\documentclass{aa}  
\usepackage{natbib}                     
\usepackage{graphicx}                   
\usepackage[varg]{txfonts}              
\usepackage[colorlinks,citecolor=blue,linkcolor=red,urlcolor=cyan]{hyperref}
\usepackage{placeins}

\newcommand{\teff}  {T$_\mathrm{eff}$}
\newcommand{\logg}  {$\log g$}
\newcommand{\kms}{\hbox{${\rm km}\:{\rm s}^{-1}$}}
\newcommand{\ha}  {H$\alpha$}
\newcommand{\ms}{m\,s$^{-1}$}

\begin{document}

  \title{Planets around evolved intermediate-mass stars\thanks{
Based on observations collected at the La Silla Observatory (Chile) with the CORALIE spectrograph mounted on the 1.2 m Swiss telescope (program 713) and with ESO-HARPS/3.6m (runs ID 075.C-0140, 076.C-0429, 078.C-0133, 079.C-0329, 080.C-0071, 081.C-0119, 082.C-0333, 083.C-0413, 091.C-0438, 092.C-0282, 094.C-0297, 099.C-0304, 0100.C-0888, 0101.C-0274, 0102.C-0812, 0104.C-0358, 105.20AZ.001, 106.21DH, 108.22LE.001) and with ESO-UVES/VLT 
at the Cerro Paranal Observatory (run 079.C-0131)}
\thanks{RV tables are only available in electronic form
at the CDS via anonymous ftp to cdsarc.cds.unistra.fr (130.79.128.5)
or via https://cdsarc.cds.unistra.fr/cgi-bin/qcat?J/A+A/}}
\subtitle{III. Planet candidates and long-term activity signals in six open clusters}

  \author{E. Delgado Mena\inst{1}
          \and J. Gomes da Silva\inst{1}
          \and J. P. Faria\inst{1}
          \and N. C. Santos\inst{1,2}
          \and J. H. Martins\inst{1}
          \and M. Tsantaki\inst{3}
          \and A. Mortier\inst{4}
          \and S. G. Sousa\inst{1}
          \and C. Lovis\inst{5}
        }

 \institute{
    Instituto de Astrof\'isica e Ci\^encias do Espa\c{c}o, Universidade do Porto, CAUP, Rua das Estrelas, 4150-762 Porto, Portugal
    \email{Elisa.Delgado@astro.up.pt}
   \and
    Departamento de F\'isica e Astronomia, Faculdade de Ci\^encias, Universidade do Porto, Rua do Campo Alegre, 4169-007 Porto, Portugal
    \and
    INAF – Osservatorio Astrofisico di Arcetri, Largo E. Fermi 5, 50125 Firenze, Italy
    \and
    School of Physics \& Astronomy, University of Birmingham, Edgbaston, Birmingham, B15 2TT, UK
    \and
    Observatoire de Gen\'eve, Université de Gen\'eve, 51 ch. des Maillettes, 1290 Sauverny, Switzerland
}

  \date{Received date / Accepted date }
 
  \abstract
  {We carried out a long-term campaign spanning 17 years to obtain high-precision radial velocities (RVs) with the HARPS spectrograph for a large sample of evolved stars in open clusters.}
  {The aim of this work is to search for planets around evolved stars, with a special focus on stars more massive than 2\,M$_\odot$ in light of previous findings that show a drop in planet occurrence around stars above this mass.}
  {We used \texttt{kima} ---a package for Bayesian modelling of RV and activity data with Gaussian process capability and Nested sampling for model comparison--- to find the Keplerian orbits most capable of explaining the periodic signals observed in RV data, which have semiamplitudes of between 75 and 500\,\ms. We also studied the variation of stellar activity indicators and photometry in order to discard stellar signals mimicking the presence of planets.}
  {We present a planet candidate in the open cluster NGC3680 that orbits the 1.64\,M$_\odot$ star  No. 41. The planet has a minimum mass of 5.13M\,$_{J}$ and a period of 1155 days. We also present periodic and large-amplitude RV signals of probable stellar origin in two more massive stars (5.84 and 3.05\,M$_\odot$ in the clusters NGC2345 and NGC3532). Finally, using new data, we revise the RV signals of the three stars analysed in our previous paper. We confirm the stellar origin of the signals observed in NGC2423 No. 3 and NGC4349 No. 127. On the other hand, the new data collected for IC4651 No. 9122 (1.79\,M$_\odot$) seem to support the presence of a bona fide planet of 6.22M\,$_{J}$ at a period of 744 days, although more data will be needed to discard a possible correlation with the CCF-FWHM.}
  {The targets presented in this work showcase the difficulties in interpreting RV data for evolved massive stars. The use of several activity indicators (CCF-FWHM, CCF-BIS, \ha), photometry, and long-term observations (covering several orbital and stellar rotational periods) is required to discern the true nature of the signals. However, in some cases, all this information is insufficient, and the inclusion of additional data ---such as the determination of magnetic field variability or RV points in the near-infrared--- will be necessary to identify the nature of the discovered signals.}
  

  \maketitle

\section{Introduction}\label{sec:Introduction}

In the last 30 years, more than 5000 planets have been discovered, mainly around main sequence (MS) solar-type stars\footnote{\url{exoplanets.eu}}, that is, FGK dwarf stars and M dwarfs. However, the number of planets detected orbiting intermediate-mass stars is still low despite the increasing frequency of giant planets with stellar mass \citep[e.g.][]{johnson10}. Moreover, several studies point to a sharp decrease in the planet occurrence around stars more massive than $\sim$2\,M$_\odot$ \citep{reffert15,wolthoff22}. The search for planets around these more massive stars is crucial to our understanding of the limits of planet formation and survival, but we are still limited by the applicability of different detection techniques to this kind of star. For example, the most successful planet-detection techniques (based on photometric transits and radial velocity (RV)) are not optimal for large, massive, and fast-rotating stars, such as early-type MS stars. Nevertheless, both methods have been applied to late-A and early-F stars, revealing a handful of substellar companions \citep[e.g.][]{desort08,borgniet19,grandjean23,collier-cameron10,sebastian22,vowell23}. 

The largest fraction of planets around intermediate-mass stars were detected by successfully applying the above-mentioned techniques to K giants, the evolved counterparts of early-type MS stars. Photometric data from the Kepler and TESS  missions led to the discovery of a good number of planets around subgiants and red giant branch (RGB) stars \citep[e.g.][]{lillobox14,grunblatt22}. On the other hand, several long-term RV surveys are being carried out by different teams, such as the Lick survey \citep{frink01}, the Okayama Planet Search Program \citep{sato05}, the Tautenburg Observatory Planet Search \citep{hatzes05}, and the PTPS-TAPAS program \citep{niedzielski15}. We refer the reader to Table 1 in \citet{ottoni22} for a exhaustive compilation of the current RV surveys around giant stars. Nevertheless, the occurrence rates found by some of these surveys can also be affected by selection effects. Because redder stars tend to show larger RV jitter \citep[e.g.][]{frink01}, it is common to apply a cutoff for a maximum $B-V$. As a consequence, the most metal-rich stars with low \logg\ values are left out of such surveys \citep{mortier13_giants}.

Alternatively, the direct imaging technique can be applied to early-type MS stars, especially those with larger masses and young ages. However, due to the inherent biases affecting the direct imaging technique, only substellar companions at large distances can be detected. The largest surveys to date, SHINE (with SPHERE$@$VLT) and GPIES (with GPI$@$Gemini South), only found a few substellar companions around A stars \citep[][ respectively]{vigan21,nielsen19}. Indeed, most of the above-mentioned surveys (either around dwarf or evolved stars) only found detected planets around stars below $\sim$2.5\,M$_\odot$ (i.e. spectral type later than A0-A1). Some theoretical studies predict that planet formation for stars more massive than 3\,M$_\odot$ is very difficult because in such stars irradiation overcomes accretion and the snowlines lie at greater distances, hindering the formation of cores \citep{kennedy08}. However, the very recent results of the BEAST  survey, wherein the direct imaging technique is applied to B-type stars (M\,$\gtrsim$2.5\,M$_\odot$), show that substellar companions (most likely brown dwarfs) can form around this kind of star, which is contradictory to the low occurrence rates found in RV surveys of evolved stars. A brown dwarf was reported in an orbit of 290\,AU around the 9\,M$_\odot$ star $\mu^{2}$ Sco \citep{squicciarini22} and a 11\,M$_{J}$ planet candidate was detected in a 556\,AU orbit around the 6-10\,M$_\odot$ binary b Cen AB \citep{janson21}.  

Given the importance of correctly determining the planet-occurrence rates for intermediate-mass stars and the difficulty in obtaining accurate masses for evolved stars, we began an RV survey around giant stars in open clusters. The advantage of open clusters is that the ages and masses of their stars can be much better constrained, meaning the planetary characterisation will be much more reliable. In the first paper of the survey, \citet[][hereafter Paper I]{lovis07} presented the discovery of a planet and a brown dwarf candidate orbiting a 2.3\,M$_\odot$ red giant in the open cluster NGC2423 and a 3.8\,M$_\odot$ red giant in NGC4349, respectively. In a subsequent work, \citet[][hereafter Paper II]{delgado18} presented a planet candidate around a 2.1\,M$_\odot$ giant in IC4651 that was found to show a suspicious signal in one activity indicator with a slightly shorter period  than that of the RV variations. In addition, the analysis of new data obtained after the publication of Paper I pointed to a probable stellar origin of the RV signals found in NGC2423 and NGC4349. 

These results made evident the complexity in analysing these noisy stars, which have long rotational periods that are compatible in many cases with the orbital periods of the candidate substellar companions. Interestingly, the large-amplitude RV signals found in those three objects had periods of close to 700 days, as is also the true for other suspicious cases in the literature such as Aldebaran \citep{hatzes15,reichert19} or $\gamma$ Draconis \citep{hatzes18}. Indeed, the planet-occurrence rate from three combined large RV surveys analysed by \citet{wolthoff22} shows a maximum at 720 days, which might be caused by the accumulation of planets with orbital periods around 600 days found orbiting the more massive stars in the surveys (those with M\,$>$1.4\,M$_\odot$). One of the explanations for this accumulation might be contamination by false positives, which only  are starting to be revealed after years of observations.

The aim of this work is to present new results of our RV survey for three stars that show periodic RV variations and may host substellar companions. The outline of the paper is as follows: in Sect. \ref{sec:observations} we present the data and the stellar parameters. A planet candidate in NGC3680 whose signal might be of stellar origin is presented in Sect. \ref{sec:NGC3680}. In Sects. \ref{sec:NGC2345} and \ref{sec:NGC3532}, we show the cases of two massive stars\footnote{We note that this definition of a massive star is put forward in the context of planet search surveys where very few planets have been found around stars with M\,$>$\,2\,M$_\odot$. In stellar physics, massive stars are usually defined as those with M\,$>$\,8-10\,M$_\odot.$} in NGC2345 and NGC3532, which show large-amplitude RV signals that are likely caused by modulation of stellar magnetic activity. However, these stars also present secondary RV signals that might be caused by a brown dwarf or a planet. In Sect. \ref{sec:others}, we present additional data for three clusters stars with RV variations mimicking substellar bodies already discussed in Paper II. In Sect. \ref{sec:discussion}, we discuss the possible origin of all the signals presented in the previous sections. Finally, in Sect. \ref{sec:conclusions}, we present our conclusions.

\begin{figure}
\centering
\includegraphics[width=1.2\linewidth]{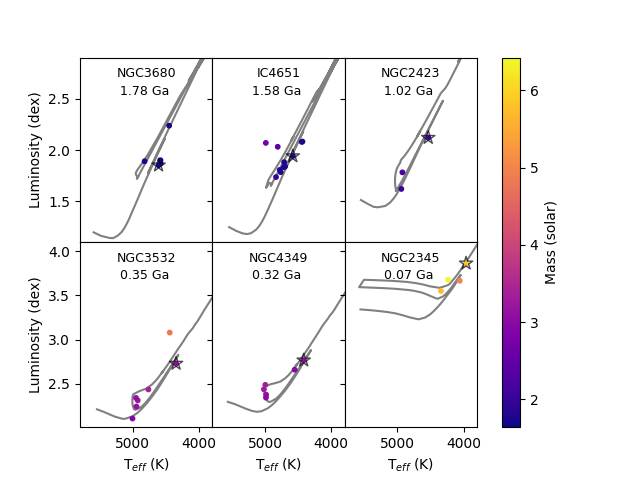}
\caption{Hertzsprung-Russell diagram for the six open clusters analysed in this work. The upper row contains the older clusters (with less massive stars, as shown in the colour scale) and the bottom row contains the younger clusters, with more massive stars. The objects with periodic RV variations are depicted with a star symbol.} 
\label{HR_diagram}
\end{figure}

\begin{center}
\begin{table*}
\caption{Stellar characteristics of the analysed planet-host candidates (the first three stars are presented here for the first time). Stellar parameters (above the horizontal line) were derived by \citet{tsantaki23}. Distances and magnitudes are extracted from WEBDA and SIMBAD, respectively. }
\centering
\begin{tabular}{llccc|ccc}
\hline
\noalign{\smallskip} 
 & &  NGC3680 No. 41 &  NGC2345 No. 50 & NGC3532 No. 670 &  IC4651 No. 9122 & NGC2423 No. 3 & NGC4349 No. 127 \\  
\noalign{\smallskip} 
\hline
\hline
\noalign{\smallskip} 
\teff& K & 4612 $\pm$ 12 & 3962 $\pm$ 10 & 4347 $\pm$ 11 & 4582 $\pm$ 12 & 4534 $\pm$ 12 & 4417 $\pm$ 12      \\
\logg&(cm\,s$^{-2}$) &2.45 $\pm$ 0.04 & 0.87 $\pm$ 0.07 & 1.75 $\pm$ 0.05 & 2.43 $\pm$ 0.04 & 2.23 $\pm$ 0.04 & 1.78 $\pm$ 0.05   \\
{[Fe/H]} & & --0.16 $\pm$ 0.02 & --0.25 $\pm$ 0.02 & --0.11 $\pm$ 0.02 & --0.03 $\pm$ 0.01 &--0.08 $\pm$ 0.01 & --0.17 $\pm$ 0.02  \\
\textit{v} sin \textit{i} & (km\,s$^{-1}$)  &  --  &  5.27  & 4.64  & 0.68  & 2.19  & 4.81  \\
$M$ & M$_\odot$ &  1.64 $\pm$ 0.06 &   5.84 $\pm$ 0.61  &  3.05 $\pm$ 0.23 & 1.79 $\pm$ 0.09& 2.03 $\pm$ 0.14 &  3.01 $\pm$ 0.24   \\
$R$ & R$_\odot$ & 11.44 $\pm$ 0.57 & 152.28 $\pm$ 17.78 & 40.95 $\pm$ 2.37 &13.36 $\pm$ 0.72&17.71 $\pm$ 1.04 & 37.97 $\pm$ 2.56   \\
log(L) & L$_\odot$ & 1.85 & 3.86 & 2.73  & 1.94 & 2.12 & 2.76  \\
Age      &Ga  & 1.78 & 0.07 & 0.35  & 1.58 & 1.02 & 0.32   \\
\noalign{\smallskip} 
\hline
\noalign{\smallskip} 
Distance &pc   & 938  &  2251 & 486  & 888 &  766 & 2176   \\
$V$ &mag& 10.88 & 10.40& 6.98 & 10.91 & 10.04 $\pm$ 0.04&10.82 $\pm$ 0.08  \\    
$\alpha$ && 11:25:48.5 & 07:08:27.0 & 11:07:57.3& 17:24:50.1& 07:37:09.2 &12:24:35.5   \\     
$\delta$ && --43:09:52.5& --13:12:32.9 &--58:17:26.3& --49:56:56.1 & --13:54:24.0&--61:49:11.7  \\      
\noalign{\medskip} %
\hline
\noalign{\medskip} %
\end{tabular}
\label{tab:stellar}
\end{table*}
\end{center}

\begin{center}
\begin{table}
\caption{Gaia DR3 identification for each target.}
\centering
\begin{tabular}{cc}
\hline
\noalign{\smallskip} 
  Star name & Gaia DR3 \\
\noalign{\smallskip} 
\hline
\noalign{\smallskip} 
\hline
\noalign{\smallskip} 
NGC3680 No. 41   & 5382186662956114304 \\ 
NGC2345 No. 50   & 3044666242714331008 \\ 
NGC3532 No. 670  & 5340186143451130112 \\ 
IC4651 No. 9122  & 5949553973093167104 \\ 
NGC2423 No. 3    & 3030262468592291072 \\ 
NGC4349 No. 127  & 6054914812271954176 \\     
\noalign{\smallskip} 
\hline
\end{tabular}
\label{tab:gaia_dr}
\end{table}
\end{center}

\section{Observations and stellar parameters} \label{sec:observations}

The RV survey on which this work is based is fully described in Paper I. The objects analysed in this survey were first observed over a time-period of nearly 5 years (from March 2005 to October 2009, ESO periods 75-83, PI: Lovis). In total, 142 stars were monitored within 17 open clusters using the HARPS spectrograph \citep{mayor03} at the ESO 3.6m telescope (\textit{R}\,$\sim$\,115000), and some of them were also observed with CORALIE
\citep[1.2 m-Swiss Telescope, La Silla][]{queloz00,udry00} in previous years. For those stars showing large RV variations, we collected more observations between March 2017 and March 2022 (ESO periods 99-108, PI: Delgado Mena). In addition, we obtained a few RV points for a number of stars in the ESO archive (periods 91-94, PI: Alves, Canto-Martins). The RV values are provided in the online tables associated with this publication. The observations were made in \textit{objA} mode (no simultaneous RV calibration) and the exposures times were estimated in order to have individual spectra with a signal-to-noise ratio (S/N) of at least \,$\sim$\,30 at $\sim$\,6000 \AA\,. This gives a typical RV photon-noise of $\sim$\,3.5\,\ms, which is sufficient to detect massive planets around the surveyed stars (with $V$ magnitudes between 7 and 12). 

We note that we applied a small negative offset to the RV points taken after May 2015, when an upgrade of the HARPS fibres took place \citep{locurto15}. The shift was calculated by extrapolating (with a linear fit) the measurements of Table 3 in \citet{locurto15} to the average full width at half maximum (FWHM) of the cross-correlation function (CCF) for each star (see Appendix A). This offset is larger for stars with larger FWHM. The offset values for the stars\footnote{For homogeneity with our previous works we will name the stars with a No. indicating the number system by Mermilliod. For example, he full name of NGC3680 No. 41 in Simbad is identified as Cl*NGC3680MMU41.} discussed in this paper are 22.3\,\ms (NGC3680 No. 41) and 27.8\,\ms (NGC3532 No. 670). No offset is required for NGC2345 No. 50 because we only have data taken after 2015. In addition, the FWHM needs to be corrected due to an instrumental focus drift affecting the measurements obtained before the fibre change in 2015. The equations for this correction are provided by \citet{gomesdasilva12}. After the fibre change, the cause of the focus drift was corrected but both the FWHM and the bisector inverse slope \citep[BIS,][]{queloz01} of the CCF\footnote{Hereafter we call these measurements as simply the FWHM and BIS.} present an offset with respect to previous values. As opposed to the RV offset, there is no published data with which to calculate the offset in the FWHM and BIS values. We refer the reader to Appendix A for more details about our attempts to correct this offset.

The stellar parameters of effective temperature (\teff), surface gravity (\logg), metallicity ([Fe/H]), and  microturbulence ($\xi_{t}$) for most of our target stars in these clusters were presented by \citet{santos09,santos12} and were improved by \citet{delgado16}, who also derived stellar ages, masses, radii, and Li abundances. With the addition of more observations and some new targets, we presented a new set of more precise parameters for the complete sample of stars \citep{tsantaki23}. In this latter analysis, we derived stellar parameters with the spectral synthesis method and made use of MARCS model atmospheres, which are more appropriate for giant stars. In addition, the masses and radii of the stars were revised using Gaia DR2 parallaxes. We refer the reader to \citet{tsantaki23} for further information.

\section{Radial velocity, stellar activity, and photometry analysis} \label{sec:analysis}

The analysis of the RV data was carried out with \texttt{kima} \citep{faria18}. This code facilitates a Keplerian fit where the number of planets is a free parameter to be estimated from the data, using the Diffusive Nested Sampling algorithm \citep{brewer_dnest} to sample from the posterior distribution of the model parameter. \texttt{kima} also allows the inclusion of Gaussian processes to model stellar activity, although we did not consider this component for the analysis of the stars presented here (as in many cases the signals show coherence). In the present work, we use \texttt{kima} by inputting wide and uninformative priors for the orbital parameters, that is, the orbital period (log-uniform prior from 0.2 to 2000 days), the semi-amplitudes (modified log-uniform prior from 1 m/s to 1 km/s), and the eccentricities (uniform prior between 0 and 1). These priors, and indeed the whole analysis with kima, are not informed by the periodogram of the RVs or the activity indicators.

As a double check, we performed a second analysis of each dataset with the \textit{yorbit} algorithm \citep{segransan11} in order to fit the whole dataset with a model composed of a Keplerian function. \textit{Yorbit} uses a hybrid method based on a fast linear algorithm (Levenberg-Marquardt) and genetic operators (breeding,  mutations, crossover), and is optimised to explore the parameter space for Keplerian fitting of RV datasets. In our case, the global search for the best-fitting orbital parameters was made with a genetic algorithm. As we show in the following subsections, the photon noise \textit{RV} errors of our observations are typically below 3\,\ms. However, the \textit{RV} variations are dominated by the stellar jitter, which is much higher than the photon noise in this kind of star \citep[typically of $\pm$\,15-20\,\ms, as discussed in ][]{hekker08,delgado18,lovis07}. Therefore, when using \textit{Yorbit,} we added in quadrature a 15\,\ms \ noise to the photon noise before fitting the data. In the case of \texttt{kima,} the RV jitter is a free parameter (with a uniform prior) that is already considered during the fitting process.

To gain insight into possible activity modulations that could interfere with RV, we used ACTIN\footnote{\url{http://github.com/gomesdasilva/ACTIN2}} \citep{gomesdasilva18} to obtain the activity indices  from the spectra based on the \ha, \ion{Na}{i} D$_1$ and D$_2$, and  \ion{He}{i} D$_3$ lines. The line parameters for the \ion{Na}{i} and \ion{He}{i} indices are described in \citet{gomesdasilva11}. For the \ha\ index, we used two versions with different bandpasses: \ha16, using a 1.6 square passband (as used in our previous work) and \ha06 using a 0.6 passband \citep[optimal for FGK dwarfs; see ][]{gomesdasilva22}. A description of the ACTIN flux and indices determination is available at \citet[][Appendix A]{gomesdasilva21}. In general, we could not use the \ion{Ca}{ii} H\&K for any of the stars due to the very low S/N ($<$\,3) in the spectral orders containing those lines. In addition, the \ion{Na}{i} and \ion{He}{i} indicators do not show any signal for our stars and therefore we do not show them in the figures.

To further investigate RV variations caused by stellar atmospheric phenomena, we also used the FWHM. These values and their errors are provided by the HARPS pipeline. Moreover, we also analysed the BIS \citep{queloz01}, but we note that this diagnostic of line asymmetry loses sensitivity for slow rotators (as some of the stars we study here) \citep[e.g.][]{saar98,santos03_planet,queloz09,santos14}. Nevertheless, we have already observed significant correlations between RV and BIS (e.g. see the case of NGC2423 No. 3 in Paper II). We note that we only obtain the periodograms for activity indicators of HARPS data because the number of observations per star with CORALIE is too low for meaningful results.

Most of the stars in our survey were observed within the All Sky Automated Survey (ASAS) from the las Campanas Observatory (Chile) with observations available between the end of 2000 and 2009 \citep{pojmanski02}. We downloaded the light curves in V magnitude from The ASAS-3 Photometric V-band Catalogue\footnote{\url{http://www.astrouw.edu.pl/asas/?page=aasc}} and performed periodograms to detect any possible variability with the same period as the RV variability.

\section{Is the RV variation of NGC3680 No. 41 due to the presence of a planet?} \label{sec:NGC3680}

\subsection{Parent-star characteristics} 

Our sample contains five giant stars in the open cluster NGC3680 (Age\,=\,1.78\,Ga) with an average metallicity of [Fe/H]\,=\,-0.15\,$\pm$\,0.02 dex \citep[see Tables 1 and 5 of][]{tsantaki23}. One of these stars, NGC3680 No. 41 with \teff\,=\,4612\,$\pm$\,12\,K, \logg\,=\,2.45\,$\pm$\,0.04\,dex, M$_{*}$\,=\,1.64\,$\pm$\,0.05\,M$_\odot$, and R$_{*}$\,=\,11.44\,$\pm$\,0.47\,R$_\odot$ (see Table \ref{tab:stellar}) seems to be on the first ascent of the RGB, and close to the luminosity bump (see Fig. \ref{HR_diagram}). The mean RV of the giants in this cluster is 1.23\,$\pm$\,0.65\,km\,s$^{-1}$ while the mean RV of NGC3680 No. 41 is 1.59\,km\,s$^{-1}$, and therefore this star is likely a cluster member. This is also supported by its parallax value. In order to obtain an estimation of the maximum rotational period of this star (which is relevant for the interpretation of the observed signals), we can use the projected rotational velocity (\textit{v} sin \textit{i}) and the radius. The \textit{v} sin \textit{i} of the sample stars were estimated by \citet{tsantaki23} using spectral synthesis and by considering a fixed  macroturbulence velocity \citep[with the empirical formula for giants by][]{hekker07}, which in this case has a value of 5.04\,\kms. For this star, the \textit{v} sin \textit{i} is very low and below the detectability threshold (i.e. the value of the macroturbulence velocity is large and can perfectly account for the broadening of the spectra). Therefore, we cannot provide a reliable estimation of the stellar rotational period but we can safely claim that this star is a slow rotator.

\begin{figure}
\centering
\includegraphics[width=1.0\linewidth]{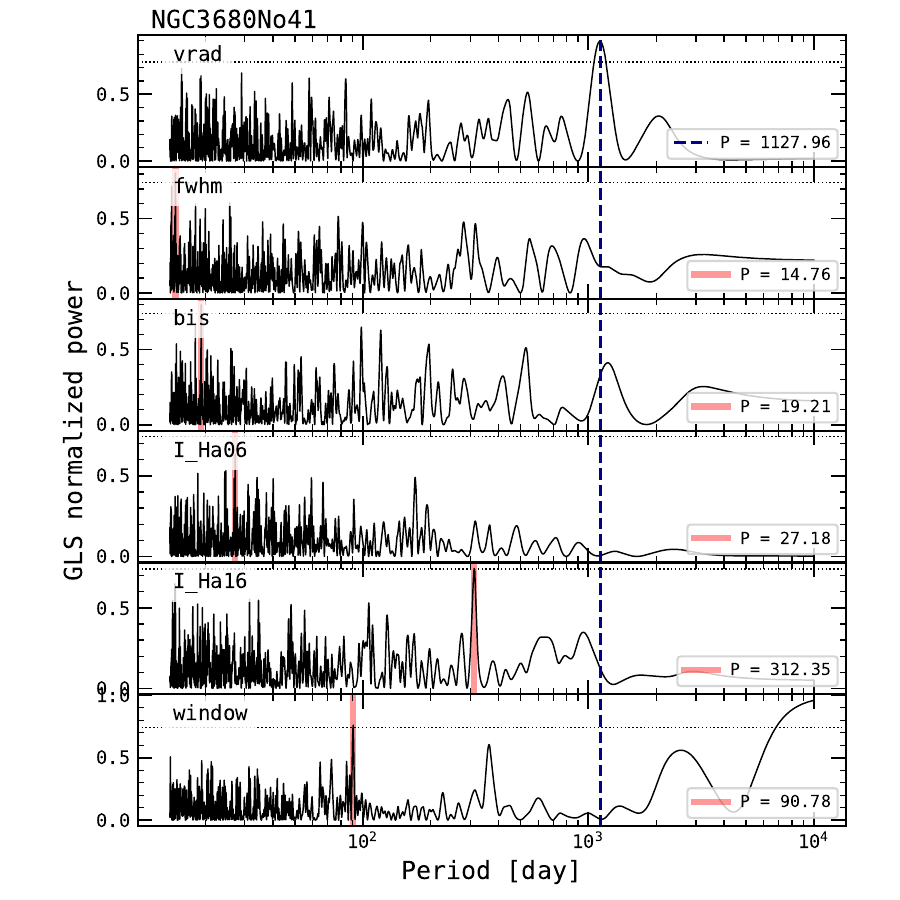}
\caption{GLS periodogram of the RV, FWHM, BIS, and stellar activity indexes for NGC3680 No. 41 HARPS data. The blue dashed line marks the period of the planet candidate. The period of the strongest peak in each periodogram is marked with a red solid line. The horizontal line marks the 1\% FAP level. The last panel shows the periodogram for the window function.} 
\label{per_NGC3680No41}
\end{figure}

\begin{figure}
\centering
\includegraphics[width=1.0\linewidth]{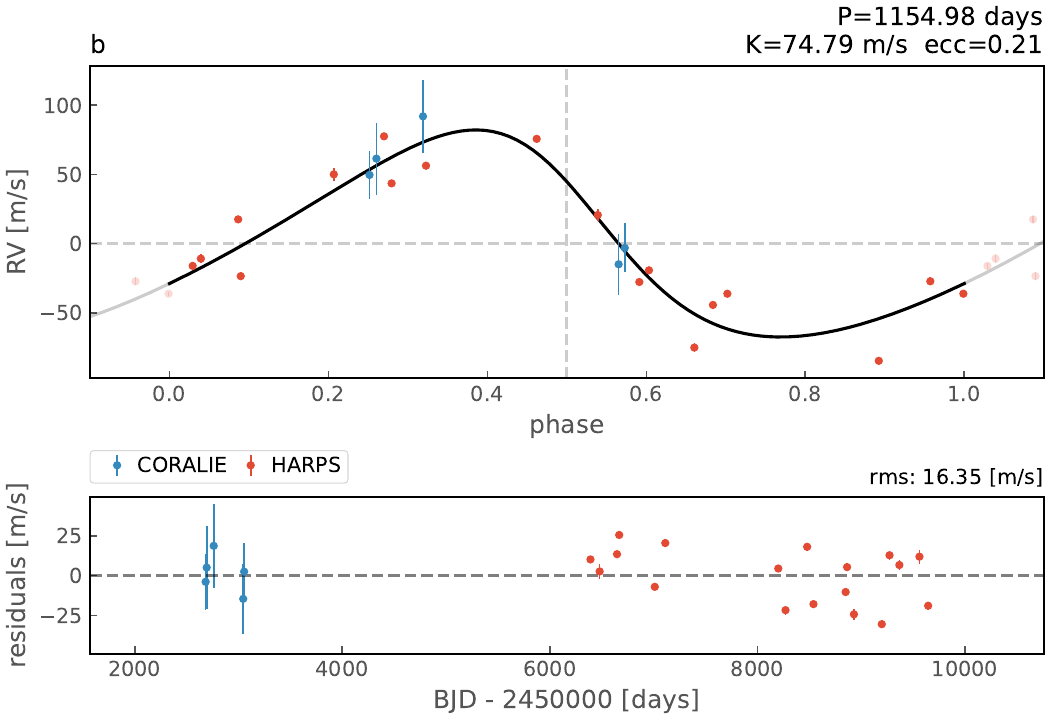}
\includegraphics[width=1.0\linewidth]{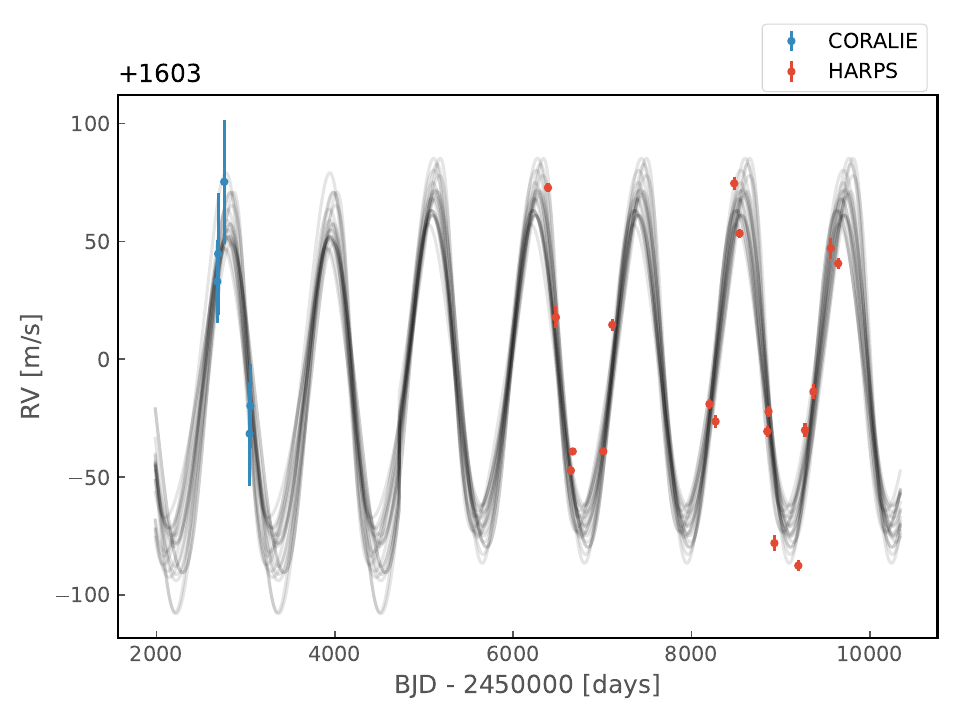}
\caption{One Keplerian fit for NGC3680No41 using \texttt{kima}. The top panel shows the phase curve for the signal, with the residuals shown below, highlighting the rms of the residual RVs. The bottom panel shows the RV data from HARPS (red points) and CORALIE (blue points) and representative samples from the posterior distribution. The systemic RV has been subtracted and is shown at the top.} 
\label{NGC3680No41_kima}
\end{figure}

\subsection{Radial-velocity analysis} 
For NGC3680 No. 41, we have 18 RV measurements obtained with HARPS over a time period of 9 years (between 2013 and 2022) and 5 measurements with CORALIE obtained over one year (2003-2004), therefore with a large time gap between the two data series. The average photon noise \textit{RV} uncertainty is 2.6 and 2.1\,\ms\ for HARPS and CORALIE data, respectively. 

An analysis using generalised Lomb-Scargle (GLS) periodograms \citep[e.g.][]{zechmeister09} was performed for RV (see upper panel of Fig.\,\ref{per_NGC3680No41}). For each star, the minimum period to search in the periodogram was defined as the minimum cadence of observations when these are done with a separation of more than 10 days. Otherwise, the minimum period is set at 10 days. The false-alarm probability (FAP) was computed by bootstrapping the data and statistically significant peaks were considered for values above FAP = 1\%. A significant signal with a period of $\sim$\,1127 days can be clearly observed in the periodogram of HARPS data only. Using \texttt{kima,} the two RV sets are well fitted by a Keplerian function with $P$\,=\,1155 days, $K$\,=\,74.79\,\ms, and $e$\,=\,0.21 (see Table \ref{tab:orbital_parameters}) with a 16.35\,\ms\ dispersion of the residuals. The posterior distribution for the fitted extra white noise is centred at 20$\pm 5$\,\ms, which is in agreement with the added jitter when using \textit{yorbit} (as explained in Sect. \ref{sec:analysis}). Considering the mass of the star, these values correspond to the expected signal induced by a planet with 5.13\,M$_{J}$ and a 2.53 AU semi-major axis. The phase curve of the best-fit solution can be seen in Fig.\,\ref{NGC3680No41_kima}. 

The results using \textit{yorbit} are very similar and the data can be best fit with a single Keplerian with a period of 1154 days, $K$\,=\,74.7\,\ms, and $e$\,=\,0.19 (figures from \textit{yorbit} can be seen in the Appendix, Fig. \ref{NGC3680No41_fit}) The dispersion of the residuals is 16\,\ms\ and the reduced $\chi^{2}$ is 1.37. In the following subsections, we assess whether or not this signal could be of stellar origin by evaluating different stellar activity or variability indicators.

\subsection{Photometry} 
\begin{figure}
\centering
\includegraphics[width=1\linewidth]{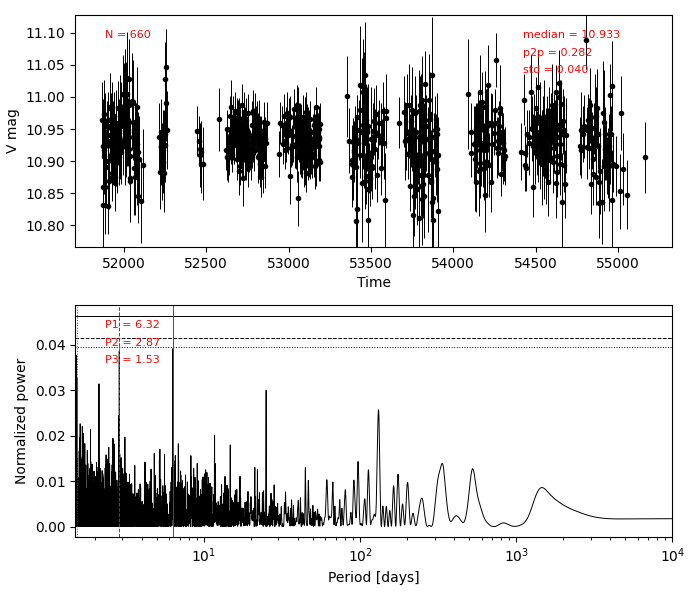}
\caption{GLS of V magnitude for NGC3680 No. 41 using ASAS-3 data. The horizontal lines indicate the FAP at levels of 0.1\%, 0.5\%, and 1\%. The vertical red lines show the periods of the three signals with the highest significance.} 
\label{NGC3680_phot}
\end{figure}

We found 664 photometric measurements classified as good quality (these are given the grade A or B, with average errors of 0.045 mag), which show a $\sim$\,0.3 peak-to-peak variability in V magnitude. In Fig. \ref{NGC3680_phot}, we can see that the GLS periodogram shows no signals above the FAP level and therefore the RV variations are likely not linked to photometric variability.

\begin{figure}
\centering
\includegraphics[width=0.8\linewidth]{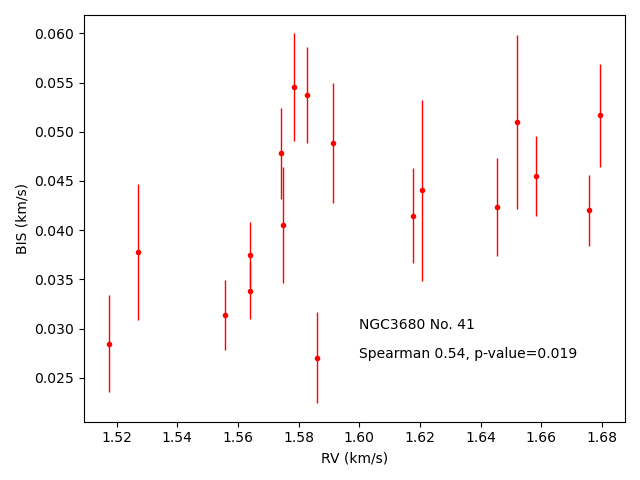}
\caption{RV versus BIS correlation for NGC3680 No. 41} 
\label{NGC3680No41_correlation}
\end{figure}

\subsection{Stellar activity and line profile analysis} 

The GLS periodograms of the FWHM and the BIS do not show any significant period (see Fig. \ref{per_NGC3680No41}). However, the RV variations from HARPS seem to be correlated with the BIS (a Pearson's coefficient value of 0.49 and Spearman rank coefficient of 0.54; see Table. \ref{correlations} and Fig. \ref{NGC3680No41_correlation}), although these correlations are not significant (with a p-value above 0.01). On the other hand, for the CORALIE data, the RV-BIS correlation is much stronger (Pearson's coefficient value of 0.85, though this is based on five points and is not significant, with a p-value of 0.08). In addition, the analysis of the \ha06 index does not show any periodic behaviour but the GLS of the \ha16 index has a peak at the 1\% FAP level for a period of 312 days. As we cannot tightly constrain the rotational period of the star, it is difficult to confidently discern whether or not this activity signal is a manifestation of the rotation of the star. As a further test, we detrended the RV points based on a linear fit to the RV versus BIS correlation. When performing a periodogram of the detrended RVs, the peak at 1127 days still shows a strong power of close to 0.8, but below the 1\% FAP line. Although the activity indicators do not exhibit any periodic behaviour with the same period as the RV, the possible RV--BIS correlation mentioned above casts doubt on the planetary hypothesis and is further discussed in Sect. \ref{sec:discussion}. Furthermore, we note that the correlation between the BIS and the RV residuals of the Keplerian fit becomes weaker (with a Pearson's coefficient value of 0.06), which could point to a stellar origin of the variation.

\begin{figure}
\centering
\includegraphics[width=1.0\linewidth]{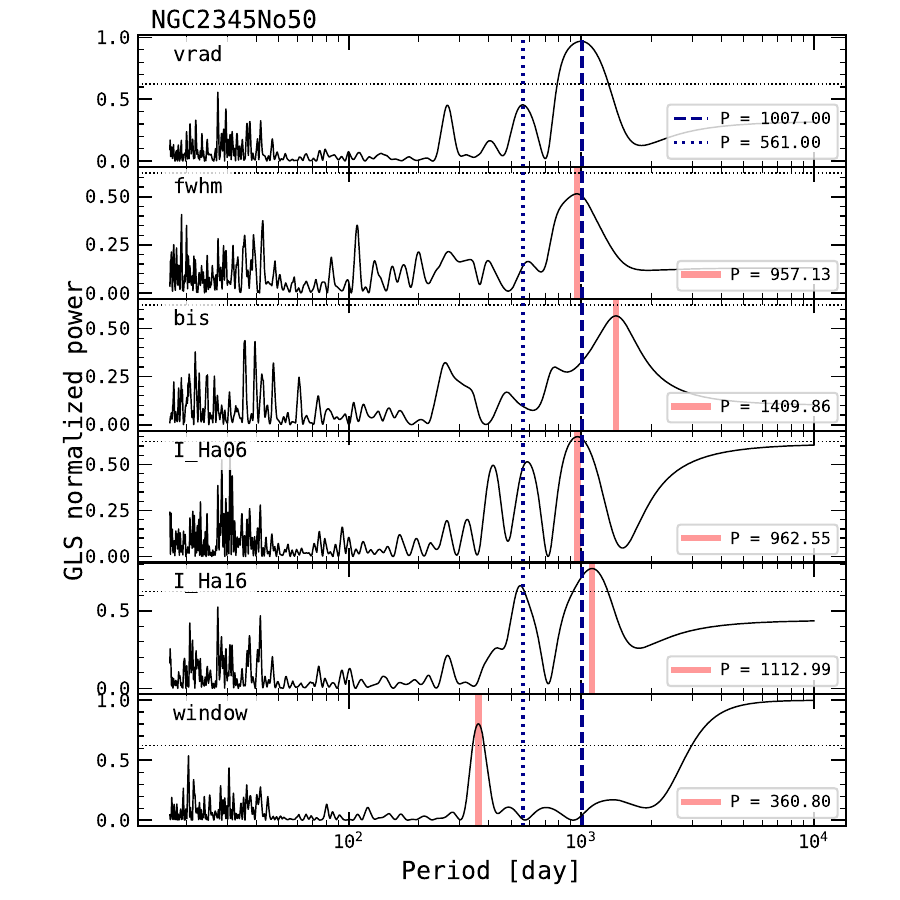}
\caption{GLS periodogram of the RV, FWHM, BIS, and stellar activity indexes for NGC2345 No. 50 HARPS data. The blue dashed line marks the period of the planet candidate. The period of the strongest peak in each periodogram is marked with a red solid line. The horizontal line marks the 1\% FAP level. The last panel shows the periodogram for the window function.}
\label{per_NGC2345No50}
\end{figure}

\begin{figure}
\centering
\includegraphics[width=1.0\linewidth]{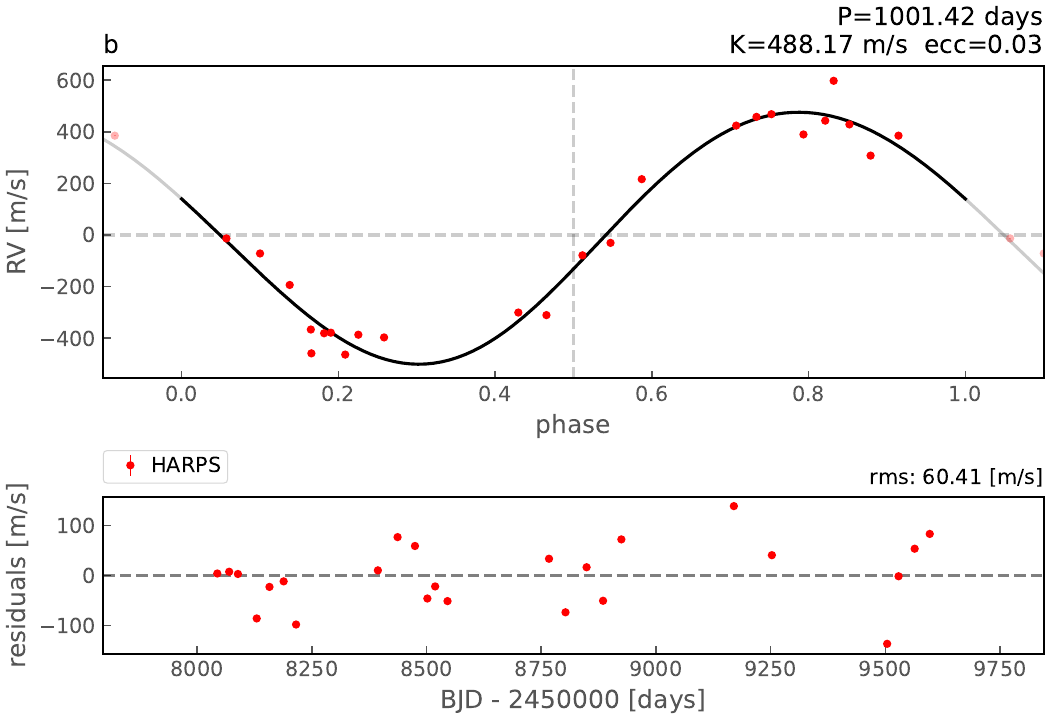}
\includegraphics[width=1.0\linewidth]{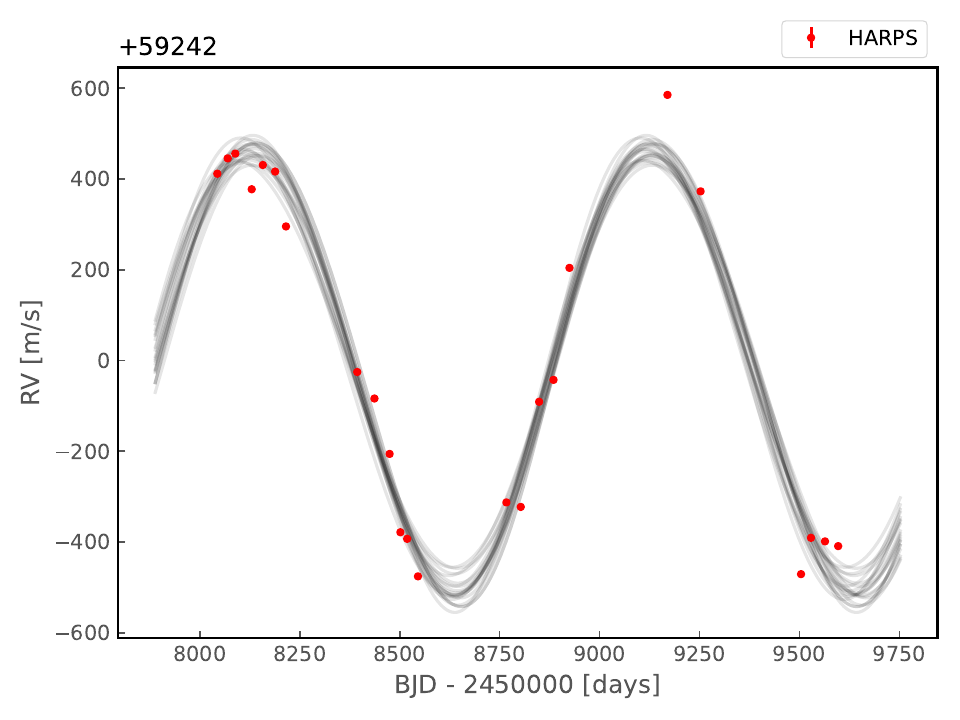}
\caption{One Keplerian fit for NGC2345No50 using \texttt{kima}. The top panel shows the phase curve for the signals, with the residuals shown below, highlighting the rms of the residual RVs. The bottom panel shows the RV data from HARPS and representative samples from the posterior distribution. The systemic RV has been subtracted and is shown at the top.} 
\label{NGC2345No50_kima}
\end{figure}

\begin{table*}
\centering
\caption{Pearson's $\rho$ (P) and Spearman rank (S) correlation coefficients, together with their respective p-values, for the correlations between RV and different activity indicators of only the HARPS data. We mark in boldface the strong correlations (those that have a coefficient with an absolute value of higher than 0.5) and the significant correlations, even if weak (those with a p-value below 10$^{-2}$). }
\tiny
\label{correlations}
\begin{tabular}{p{1.9cm}||p{0.7cm}p{0.6cm}|p{0.7cm}p{0.6cm}||p{0.6cm}p{0.6cm}|p{0.6cm}p{0.6cm}||p{0.6cm}p{0.6cm}|p{0.6cm}p{0.6cm}||p{0.6cm}p{0.6cm}|p{0.6cm}p{0.6cm}|}

\hline
\noalign{\vskip0.02\columnwidth}
\multicolumn{1}{c||}{}&\multicolumn{4}{c||}{FWHM}&\multicolumn{4}{c||}{BIS}&\multicolumn{4}{c||}{\ha06}&\multicolumn{4}{c|}{\ha16}\\
\noalign{\smallskip} 
\hline
\noalign{\smallskip} 
star & P & p & S & p & P & p & S & p & P & p & S & p & P & p & S & p \tabularnewline[0.005\columnwidth]
NGC3680No.41 & 0.26 & 0.287 & 0.32 & 0.185& 0.48 & 0.039& \textbf{0.54}& 0.019 & 0.12 & 0.638 & 0.06 & 0.797 & 0.13 & 0.602 & 0.06 & 0.797\\
NGC2345No.50 & \textbf{--0.72} & \textbf{7$\cdot$10$^{-5}$} & \textbf{--0.76} & \textbf{1$\cdot$10$^{-5}$}& \textbf{0.51} & 0.011& 0.46& 0.022 & \textbf{0.76} & \textbf{1$\cdot$10$^{-5}$} & \textbf{0.76} & \textbf{1$\cdot$10$^{-5}$} & \textbf{0.85} & \textbf{9$\cdot$10$^{-8}$} & \textbf{0.76} & \textbf{1$\cdot$10$^{-5}$}\\
NGC3532No.670 & 0.21 & 0.284 &0.17 & 0.358 & 0.42& 0.022 &  0.37 & 0.044& 0.45 & 0.013 &0.44 & 0.016 & \textbf{0.74} & \textbf{4$\cdot$10$^{-6}$} &\textbf{0.75} & \textbf{2$\cdot$10$^{-6}$} \\
\noalign{\smallskip} 
\hline 
\noalign{\smallskip} 
IC4651No.9122 & 0.00 & 0.984 & -0.06 & 0.570 & 0.28 & 0.013 & 0.34& \textbf{0.003} & 0.20 & 0.084 & 0.13 & 0.251 & 0.21 & 0.071 & 0.27 & 0.017\\
NGC2423No.3 & 0.20 & 0.117 & 0.17 & 0.180 & \textbf{0.61} & \textbf{2$\cdot$10$^{-7}$} & \textbf{0.54}& \textbf{7$\cdot$10$^{-6}$} & 0.12 & 0.366 & 0.16 & 0.216 & 0.22 & 0.087 & 0.22 & 0.091\\
NGC4349No.127 & -0.17 & 0.191 & -0.17 & 0.192 & 0.08 & 0.514& 0.145& 0.280 & 0.13 & 0.331 & 0.15 & 0.248 & 0.40 & \textbf{0.002} & 0.35 & \textbf{0.006}\\

\noalign{\smallskip} 
\hline 
\hline
\end{tabular}
\end{table*}

\section{Large RV variations in NGC2345 No. 50 mimicking the presence of massive companions} \label{sec:NGC2345}

\subsection{Parent star characteristics} 

Our sample contains only four giant stars in the open cluster NGC2345 (Age\,=\,0.07\,Ga) with an average metallicity of [Fe/H]\,=\,-0.22\,$\pm$\,0.02 dex \citep[see Tables 1 and 5 of][]{tsantaki23}. The star analysed here, NGC2345 No. 50, with \teff\,=\,3962\,$\pm$\,10\,K, \logg\,=\,0.87\,$\pm$\,0.07\,dex, M$_{*}$\,=\,5.84\,$\pm$\,0.61\,M$_\odot$, and R$_{*}$\,=\,152.28\,$\pm$\,17.78\,R$_\odot$ (see Table \ref{tab:stellar}), is the most evolved star in the cluster and seems to be ascending the early asymptotic giant branch (AGB; see Fig. \ref{HR_diagram}). This star is very interesting because it is the youngest and most massive in the full sample. Despite the fact that the other stars in the cluster have very few observations, we are able determine the mean RV of the cluster as 58.61\,$\pm$\,0.56\,\kms\ while the mean RV of NGC2345 No. 50 is 59.21\,\kms. Although the scatter is large due to the large RV variations observed in the star, its RV is within $\sim1\sigma$ of the RV dispersion in the cluster, and taking into consideration its parallax, this star seems to be a cluster member. We note that recent studies in this cluster by \citet{holanda19} and \citet{alonso-santiago19} also catalogue this star as a member and obtain similar stellar parameters. Both mentioned studies place this star in the RGB, as also suggested by its not very low $^{12}$C/$^{13}$C ratio. We estimated a \textit{v} sin \textit{i} of 5.27\,\kms\ in \citet{tsantaki23} for a fixed macroturbulence velocity of 6.06\,\kms\ \citep[with the empirical formula for bright giants by][]{hekker07}. Using the stellar radius, we can derive the maximum rotational period via $P_{max}=2\pi R_{*}/\textit{v} sin \textit{i}$. This leads to a maximum rotational period of $\sim$\,1463 days.

\subsection{Radial velocity analysis} 

A total of 24 RV measurements were obtained with HARPS over $\sim$5 years (between 2017 and 2022). The average photon noise error in \textit{RV} is 3.3\,\ms. The GLS periodogram was performed for RV (see upper panel of Fig.\,\ref{per_NGC2345No50}), which shows a strong signal (well above the 0.1\% FAP level) at a period of 1007 days. The analysis with \texttt{kima} reveals that the maximum likelihood is obtained for a model with one Keplerian with a period of 1001 days, $K$\,=\,488.17\,\ms\,, and $e$\,=\,0.03. Considering the mass of the star, these values would correspond to the expected signal induced by a body with 3.5 AU semi-major axis and with 79.34\,M$_{J}$, a value that places this companion at the threshold between brown dwarf and low-mass star. The phase curve of the best-fit solution can be seen in Fig.\,\ref{NGC2345No50_kima}. The posterior distribution for the fitted extra white noise is centred at 55$\pm 18$\,\ms. This value is relatively high but is to be expected in very evolved stars, as is the case for this star \citep{hekker08}. The residuals of the fit have a dispersion of 60.4\,\ms, which appears large and might be indicative of an additional signal. 

Indeed, the analysis performed by \textit{yorbit} with a single Keplerian provides very similar output parameters (P=1001 days, $K$\,=\,488.3\,\ms\, and $e$\,=\,0.036), including the large dispersion of the residuals and a large reduced $\chi^{2}$ of 22.5 (see Fig. \ref{NGC2345No50_k1}). We therefore attempted to make a two-Keplerian fit with \textit{yorbit} to model both the large-amplitude signal (probably caused by activity, as we show below) and a possible existing substellar body in the system. We obtained a solution for a system with a low-mass star and a brown dwarf, with the period of the `low-mass star' being 1007 days (close to the period of the single Keplerian fit explained above) and with a slightly larger semiamplitude ($K$\,=\,554\,\ms) corresponding to a mass of 87.36\,M$_{J}$. The brown-dwarf candidate would have a period of 561 days, a mass of 23.27\,M$_{J}$, and an eccentricity of 0.28 (see Fig. \ref{NGC2345No50_k2}). In this case, the dispersion of the residuals is lower, 32\,\ms, and the reduced $\chi^{2}$ is 8.1; this model therefore better represents the observed data than the single Keplerian fit. We also performed a GLS periodogram of the residuals, which did not show any significant peak. A complete analysis of activity and photometry signals is presented in the following subsection, with the aim being to better understand if the brown-dwarf candidate is bona fide.

\subsection{Photometry} 
\begin{figure}
\centering
\includegraphics[width=1\linewidth]{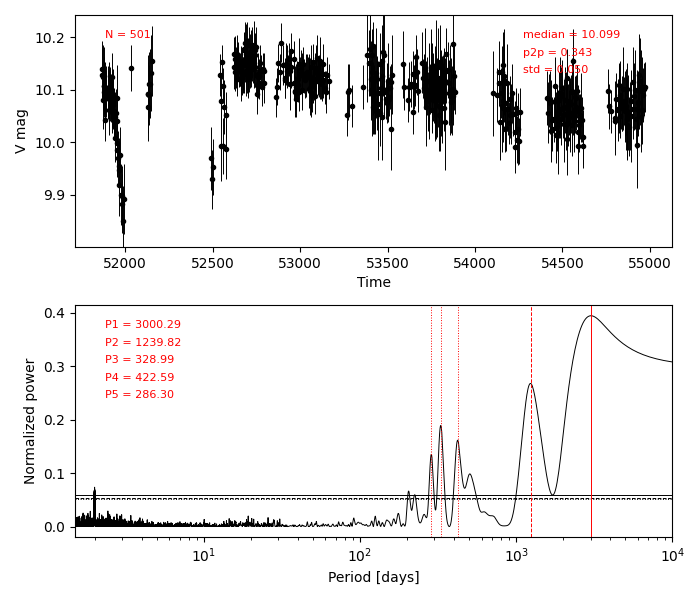}
\caption{GLS periodogram of V magnitude for NGC2345 No. 50 using ASAS-3 data. The horizontal lines indicate the FAP at the levels of 0.1\%, 0.5\%,\ and 1\%. The vertical red lines show the periods of the five signals with the highest significance. The time series are fitted by a sinusoidal with the period showing the highest significance.} 
\label{NGC2345_phot}
\end{figure}

We found 501 photometric measurements in the ASAS-3 catalogue classified as good quality (they are given a grade A or B, with average errors of 0.049 mag) that show a $\sim$\,0.2 peak-to-peak variability in V magnitude (not considering the observations in the first epoch showing a large scatter). In Fig. \ref{NGC2345_phot}, the GLS periodogram for this dataset shows several signals well above the FAP = 0.1\% line with periods of 3000, 1239, 329, 422, and 286 days. Considering the errors associated with the determination of stellar radius and especially with the determination of \textit{v} sin \textit{i} (due to the degeneracy with macroturbulence velocity), the signal at 1239 days is compatible with the maximum rotational period estimation of 1463 days. On the other hand, none of these photometric signals are close in period to the large RV variation observed at 1007 days, nor with the signal at 561 days of possible brown dwarf origin.

\begin{table*}
\centering
\caption{Summary of the strongest periods found in the different periodograms. First set of columns: Maximum stellar rotational period and periods of the most significant peak(s) in each GLS periodogram that match the most significant RV period. We highlight in blue the periods that are compatible with the estimated stellar rotation period. Second set of columns: For the stars with a possible two-Keplerian fit, we list the periods of the periodogram peaks that match the period of the hypothetical second companion in the system. The periodogram peaks that lie above the 1\% FAP level are marked in boldface.}
\tiny
\label{peaks}
\begin{tabular}{l|c|cccccc||ccccc}
\hline
\noalign{\smallskip}
Star         & $P_{max}$ & RV &  FWHM  & BIS & \ha06 & \ha16 & Phot & RV &  FWHM  & BIS & \ha06 & \ha16  \\
\hline
\noalign{\smallskip}
NGC3680No. 41  & --    & \textbf{1128}  & --  & --  & --  & --  &  & --  &  -- & --  & --  & --   \\
NGC2345No. 50  & \textcolor{blue}{1463} & \textbf{1007}  & 957  & \textcolor{blue}{1409} & \textbf{962} & \textbf{1113} & \textbf{\textcolor{blue}{1239}} & 561 & -- & -- &  586 & \textbf{545}   \\
NGC3532No. 670 & \textcolor{blue}{447}  &  \textbf{842} & \textbf{811}  & --  & --  & \textbf{889}  & \textbf{\textcolor{blue}{412}}  & 625   & \textbf{619}  & --  & --  &  574   \\
\noalign{\smallskip}
\hline 
\noalign{\smallskip}
IC4651No. 9122 & \textcolor{blue}{995} & \textbf{744}  & --  & --  & --  & \textcolor{blue}{985}  & -- & 384   & \textbf{364/393}  & --  & --  &  --   \\
NGC2423No. 3   & \textcolor{blue}{407} & \textbf{697} & -- & \textbf{697} &-- & -- & \textbf{\textcolor{blue}{417}} & \textcolor{blue}{435} & \textbf{\textcolor{blue}{380}} & -- & -- & \textbf{\textcolor{blue}{393}}  \\
NGC4349No. 127 & \textcolor{blue}{399} & \textbf{674}  & \textbf{675}  & --  & --  & --  & \textbf{\textcolor{blue}{342/428}} & --  &  -- & --  & --  & --   \\
\noalign{\smallskip}                            
\hline 
\hline
\end{tabular}
\end{table*}

\subsection{Stellar activity and line-profile analysis} 

When studying the RV signals of this star, we first noted a very strong correlation with the FWHM (with a Pearson's coefficient value of -0.72 and a p-value of 7$\cdot$10$^{-5}$, also strong and significant with the Spearman statistics; see Table \ref{correlations}), which led us to suspect the large RV variability observed. On the other hand, the correlation with the BIS is not significant. We also performed the GLS periodograms for the different activity indicators (see Fig. \ref{per_NGC2345No50}). In this case, the FWHM shows a non-significant peak at 957 days and the BIS has a non-significant peak at 1409 days. It is possible that this signal is related to the 1239 d signal observed in photometry. The most significant signals are seen in the \ha\ line. The \ha06 index shows a strong signal above the 1\% FAP level at 962 days, close to the 1007 d period of the RV. Furthermore, there are two signals at 415 and 586 days, though these are not significant. The situation is much more clear when using the \ha16 index, which shows a strong signal above the 1\% FAP level at 1113 days and another signal above the 1\% FAP level at 545 days. These two periods are suspiciously close to the 1007 d and 561 d periods of the two-Keplerian fit and therefore challenge the presence of either of the two companions in the system. Furthermore, the cadence of the observations introduces$~$yearly aliases in the periodogram (see the strong peak at 360 days in the window function periodogram). The yearly aliases of the signal at 1007 days are exactly at 286 and 573 days, with this second signal also being observed in the activity indicators and producing the RV variability that mimics a second body. We refer the reader to Table \ref{peaks} for a summary of the different peaks in the periodograms. 

In addition, we checked whether the Gaia data reveal the presence of a low-mass star companion through elevated astrometric noise. We consulted the renormalised unit weight error (RUWE\footnote{\url{https://gea.esac.esa.int/archive/documentation/GDR2/Gaia_archive/chap_datamodel/sec_dm_main_tables/ssec_dm_ruwe.html}}) statistic, which has a value of 1.05 for this star (well below 1.4) and therefore does not show evidence of a companion. This fact further supports the conclusion that the large-amplitude signal is of stellar origin and not caused by a low-mass star companion.

\begin{figure}
\centering
\includegraphics[width=1.0\linewidth]{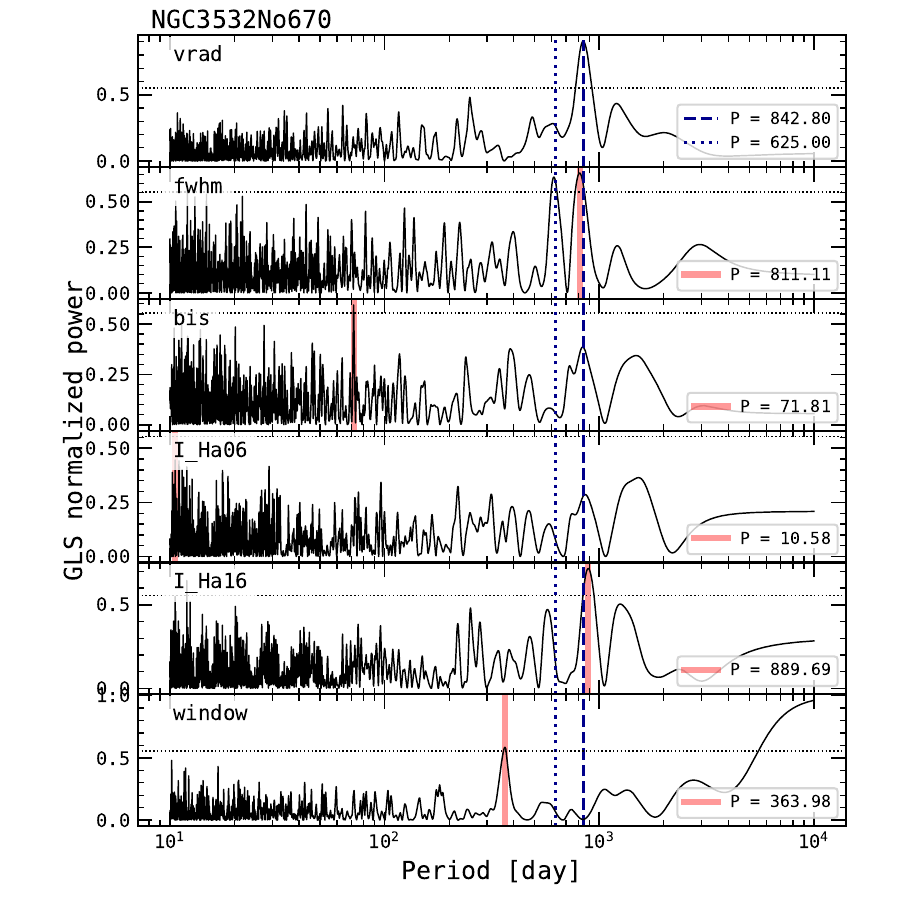}
\caption{GLS periodogram of the RV, FWHM, BIS, and stellar activity indexes for NGC3532 No. 670 HARPS data. The blue dashed line marks the period of the planet candidate. The period of the strongest peak in each periodogram is marked with a solid red line. The horizontal line marks the 1\% FAP level. The last panel shows the periodogram for the window function.} 
\label{per_NGC3532No670}
\end{figure}

\section{A planet candidate around NGC3532 No. 670 hidden in stellar activity?} \label{sec:NGC3532}

\subsection{Parent star characteristics} 

Our sample contains seven giant stars in the open cluster NGC3532 (Age\,=\,350\,Ma) with an average metallicity of  [Fe/H]\,=\,-0.08\,$\pm$\,0.03 dex \citep[see Tables 1 and 5 of][]{tsantaki23}. One of its members, NGC3532 No. 670, with \teff\,=\,4347\,$\pm$\,11\,K, \logg\,=\,1.75\,$\pm$\,0.05\,dex, M$_{*}$\,=\,3.05\,$\pm$\,0.23\,M$_\odot$, and R$_{*}$\,=\,40.95\,$\pm$\,2.37\,R$_\odot$ (see Table \ref{tab:stellar}), seems to be at the tip of the RGB but it is not clear whether it is approaching the tip of the RGB (see Fig. \ref{HR_diagram}). The mean RV of the giants in this cluster is 4.03\,$\pm$\,0.48\,km\,s$^{-1}$ while the mean RV of NGC3532 No. 670 is 4.20\,km\,s$^{-1}$. This similar value, together with its parallax, suggests that this star is likely a cluster member. We estimated a \textit{v} sin \textit{i} of 4.64\,kms\ in \citet{tsantaki23} for a fixed macroturbulence velocity of 4.52\,kms\ \citep[with the empirical formula by][for giant stars]{hekker07}. This leads to a maximum rotational period of $\sim$\,447 days.

\begin{figure}
\centering
\includegraphics[width=1.0\linewidth]{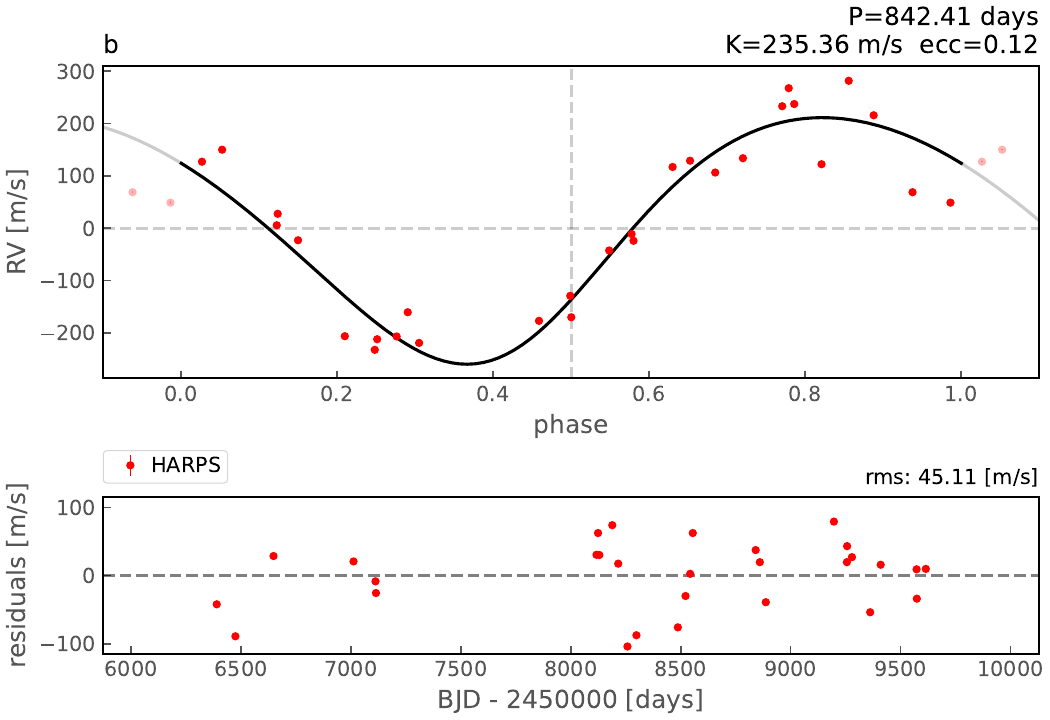}
\includegraphics[width=1.0\linewidth]{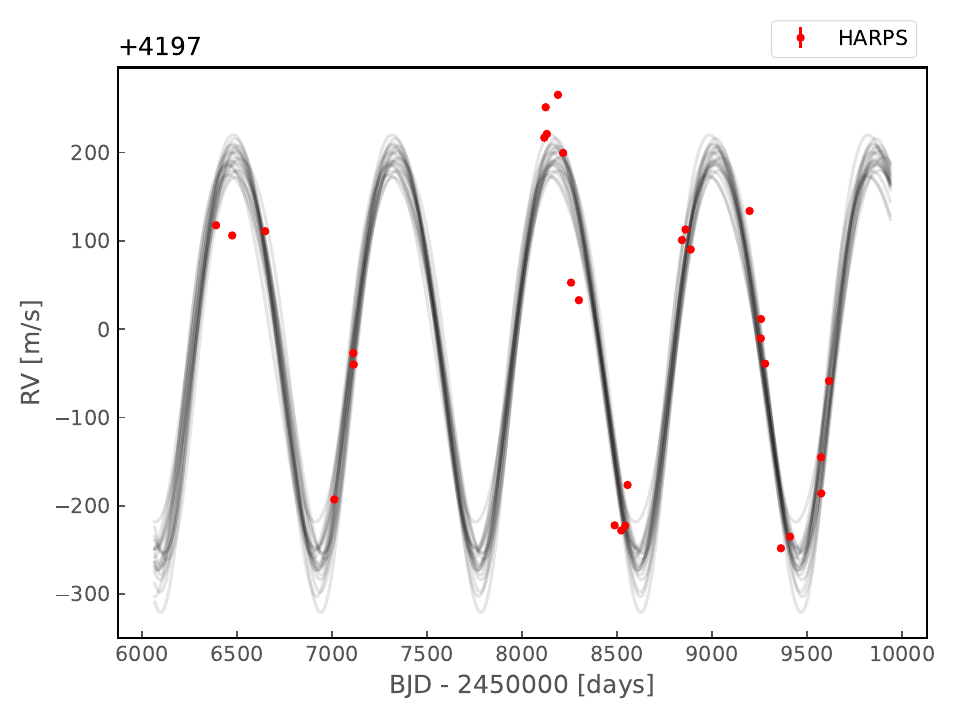}
\caption{One Keplerian fit for NGC3532 No. 670 using \texttt{kima}. The top panel shows the phase curve for the signals, with the residuals shown below, highlighting the rms of the residual RVs. The bottom panel shows the RV data from HARPS and representative samples from the posterior distribution. The systemic RV has been subtracted and is shown at the top.} 
\label{NGC3532No670_kima}
\end{figure}

\subsection{Radial velocity analysis}

For this star, we collected a total of 29 RV measurements with HARPS over $\sim$9 years (between 2013 and 2022) although there is a gap of 2.5 years without observations. The average photon noise error in \textit{RV} is 1.3\,\ms, which is due to the fact that this star is considerably brighter than the previous two described in Sections \ref{sec:NGC3680} and \ref{sec:NGC2345}, respectively. The time series of the RV show a clear variation of large amplitude. This is also demonstrated by the GLS periodogram of the RV (see upper panel of Fig.\,\ref{per_NGC3532No670}), which presents a strong signal (well above the 1\% FAP level) at a period of 842 days. The most probable solution obtained with \texttt{kima} is that with a single brown dwarf of 22.7\,M$_{J}$ with a period of 842.4 days and an eccentricity of 0.12 ($K$\,=\,235.3\,\ms; see Fig. \ref{NGC3532No670_kima} for the phase-folded orbit and the \textit{RV} time series). The posterior distribution for the fitted extra white noise is centred at 31$\pm 9$\,\ms, and the residuals of the fit show a dispersion of 45.1\,\ms, which might be indicative of another signal that cannot be fitted with a Keplerian.

The one-Keplerian fit with \textit{yorbit} provides almost identical parameters (see Fig.\,\ref{NGC3532No670_k1}). The dispersion of the residuals is 48.8\,\ms\ and the reduced $\chi^{2}$ is 13.24. Considering this dispersion and the hints of variability in the residuals, we also tried to perform a two-Keplerian fit with \textit{yorbit}. In this case, the largest-amplitude RV signal has the same period (844 days), and slightly larger semiamplitude ($K$\,=\,246.2\,\ms) and eccentricity than the single Keplerian fit. A second signal with a period of 625 days, $K$\,=\,98.6\,\ms,  and $e$\,=\,0.29 could be explained by a planet candidate of 8.3\,M$_{J}$ and a 2.07 AU semi-major axis (see Fig.\,\ref{NGC3532No670_k2}).

\subsection{Photometry} 
\begin{figure}
\centering
\includegraphics[width=1\linewidth]{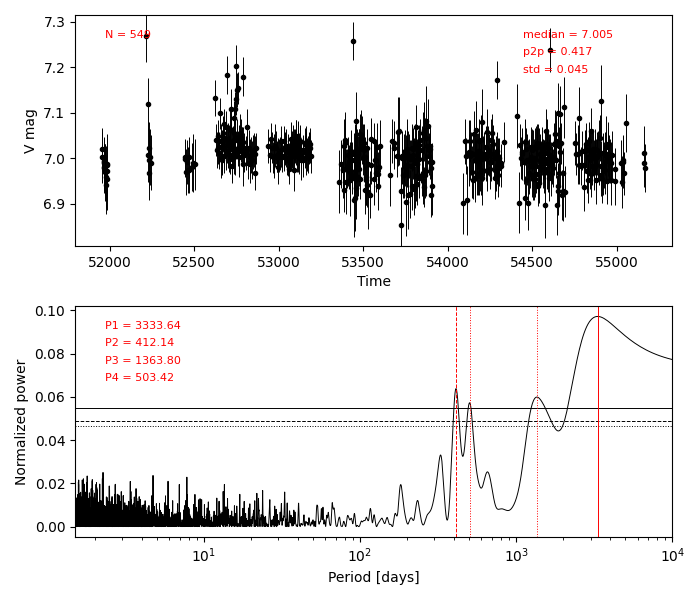}
\caption{GLS periodogram of V magnitude for NGC3532 No. 670 using ASAS-3 data. The horizontal lines indicate the FAP at levels of 0.1\%, 0.5\%,\ and 1\%. The vertical red lines show the periods of the four signals with the highest significance. The time series are fitted by a sinusoidal with the period showing the highest significance.} 
\label{NGC3532_phot}
\end{figure}

We found 549 photometric measurements classified as good quality (they are given a grade A or B, with average errors of 0.046 mag) that show a $\sim$\,0.3 peak-to-peak variability in V magnitude (excluding the probable outlier points above V\,=\,7.2). The GLS periodogram depicted in Fig. \ref{NGC3532_phot} shows four signals above the FAP = 0.1\% line with periods of 3333, 412, 1363, and 503 days. None of these periods match the signal at 842 days found in the RV time series or the planet candidate signal in the two-Keplerian fit. However, the period at 412 days is compatible with the estimated rotational period of the star and this would suggest that the RV signal at 842 days is a harmonic of the rotational period.

\subsection{Stellar activity and line profile analysis} 
For this star, we observed that the RV variations are not strongly correlated with the BIS, nor with the FWHM (see Table \ref{correlations} for the Spearman and Pearson coefficient and p-values), providing tentative evidence that the large-amplitude RV signal might be of true brown dwarf or planetary origin. However, as we also observed for similar stars in our previous work (e.g. NGC4349 No. 127, Paper II), a lack of correlation can be produced if these activity indicators are not in phase with the RV. This can be assessed with the GLS periodogram of the different activity indicators (shown in Fig.\,\ref{per_NGC3532No670}). The FWHM shows a signal at 811 days that is probably related to the RV signal at 842 days and a second signal at 619 days also above the 1\% FAP level. Therefore, the lower-amplitude signal for the two-Keplerian orbit seems to also be of stellar origin. Indeed, the window function has a significant peak at 363 days that is introducing yearly aliases in the periodograms. The yearly aliases of the 842 d signal fall at 254 and 641 days, whereas the yearly aliases of 625 days fall at 230 and 872 days. Therefore, these two signals are yearly aliases of each other. On the other hand, the BIS does not show any significant period. As was found for NGC2345 No. 50, the strongest signals are observed for the \ha16 indicator, which has a peak at 889 days that is well above the 1\% FAP level. Curiously, this star shows no significant signal in the \ha06 indicator.


\begin{figure}
\centering
\includegraphics[width=1.0\linewidth]{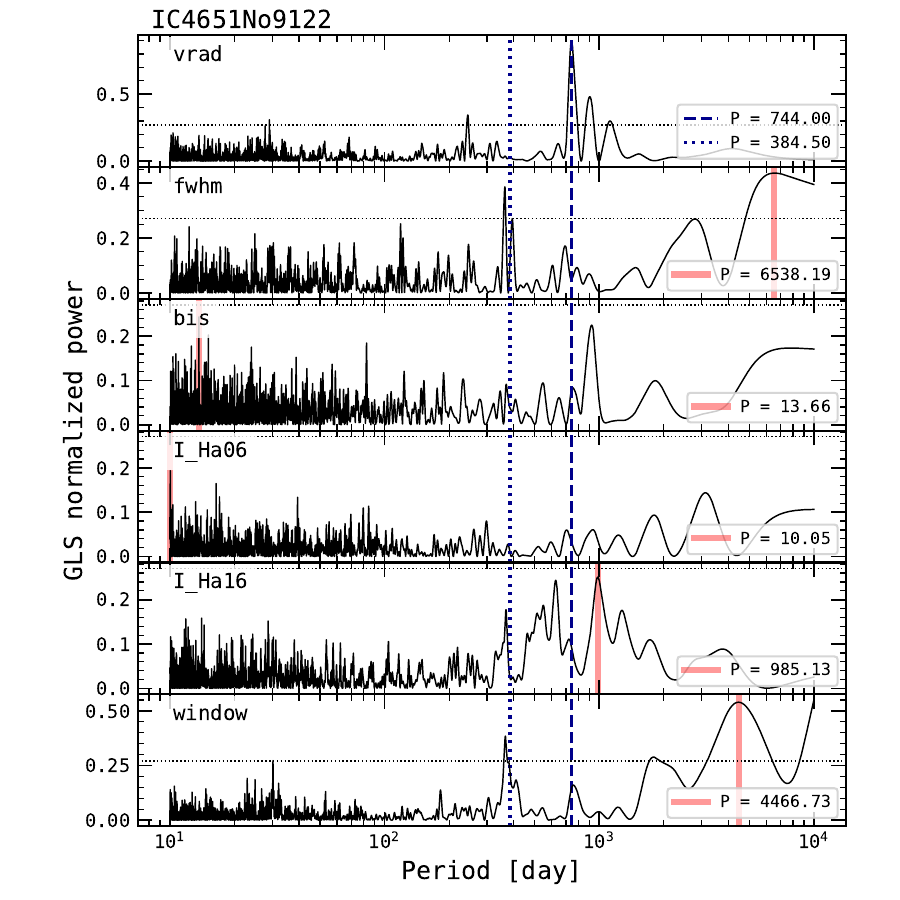}
\caption{GLS periodogram of the RV, FWHM, BIS, and stellar activity indexes for IC4651 No. 9122 HARPS data. The blue dashed line marks the period of the planet candidate. The period of the strongest peak in each periodogram is marked with a red solid line. The horizontal line marks the 1\% FAP level. The last panel shows the periodogram for the window function.} 
\label{per_IC4651No9122}
\end{figure}

\begin{figure}
\centering
\includegraphics[width=1.0\linewidth]{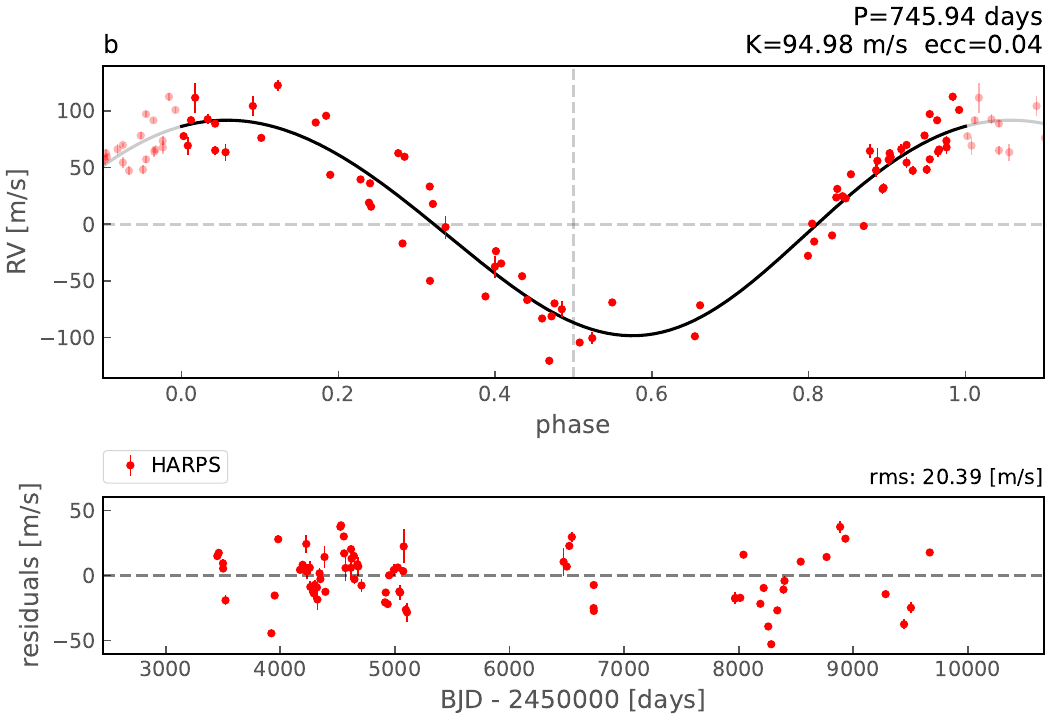}
\includegraphics[width=1.0\linewidth]{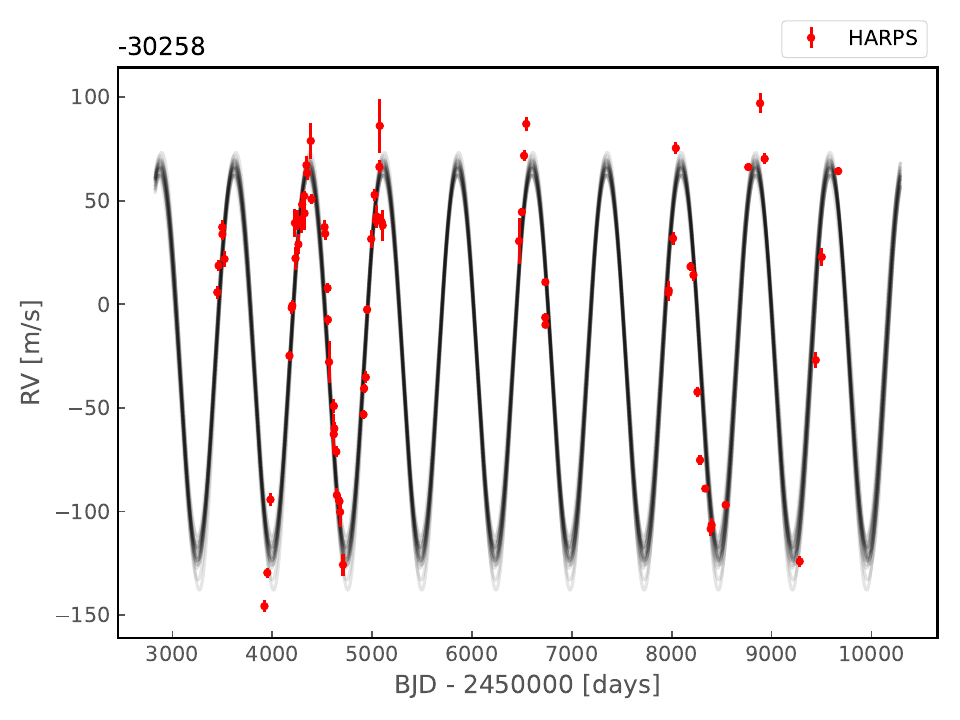}
\caption{One Keplerian fit for IC4651 No. 9122 using \texttt{kima}. The top panel shows the phase curve for the signal, with the residuals shown below, highlighting the rms of the residual RVs. The bottom panel shows the RV data from HARPS and representative samples from the posterior distribution. The systemic RV has been subtracted and is shown at the top.}
\label{IC4651No9122_kima}
\end{figure}

\begin{figure}
\centering
\includegraphics[width=1.0\linewidth]{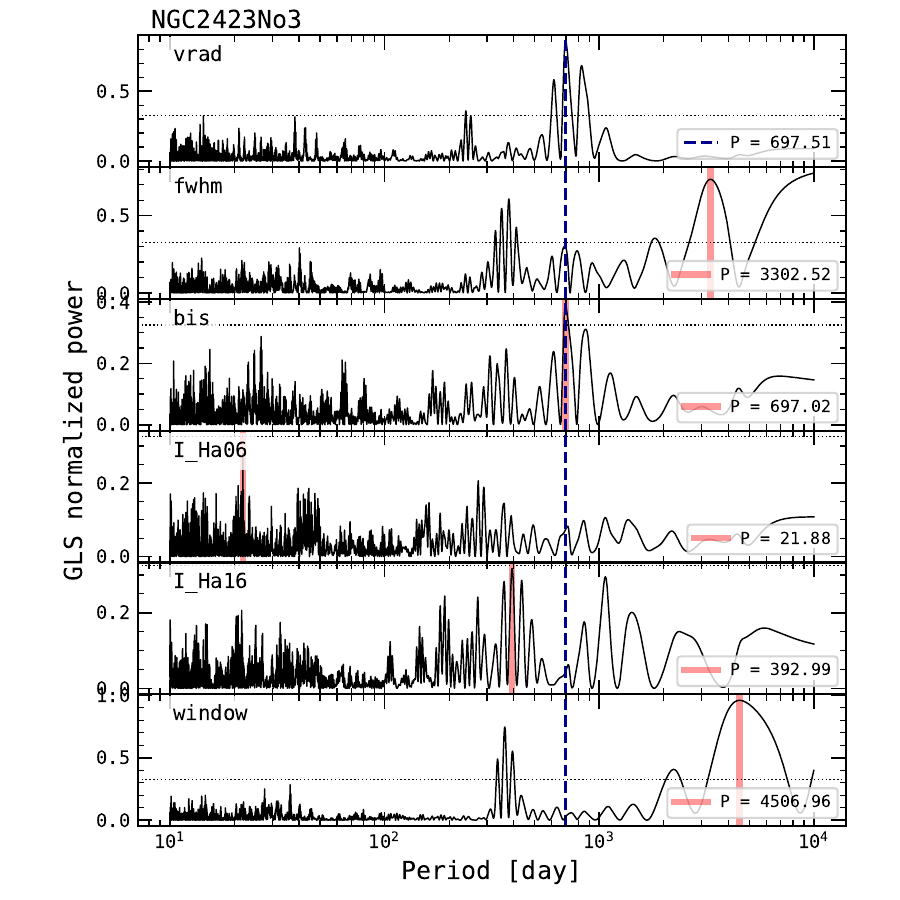}
\caption{GLS periodogram of the RV, FWHM, BIS, and stellar activity indexes for NGC2423 No. 3 HARPS data. The blue dashed line marks the period of the planet candidate. The period of the strongest peak in each periodogram is marked with a red solid line. The horizontal line marks the 1\% FAP level. The last panel shows the periodogram for the window function.} 
\label{per_NGC2423No3}
\end{figure}

\begin{figure}
\centering
\includegraphics[width=1.0\linewidth]{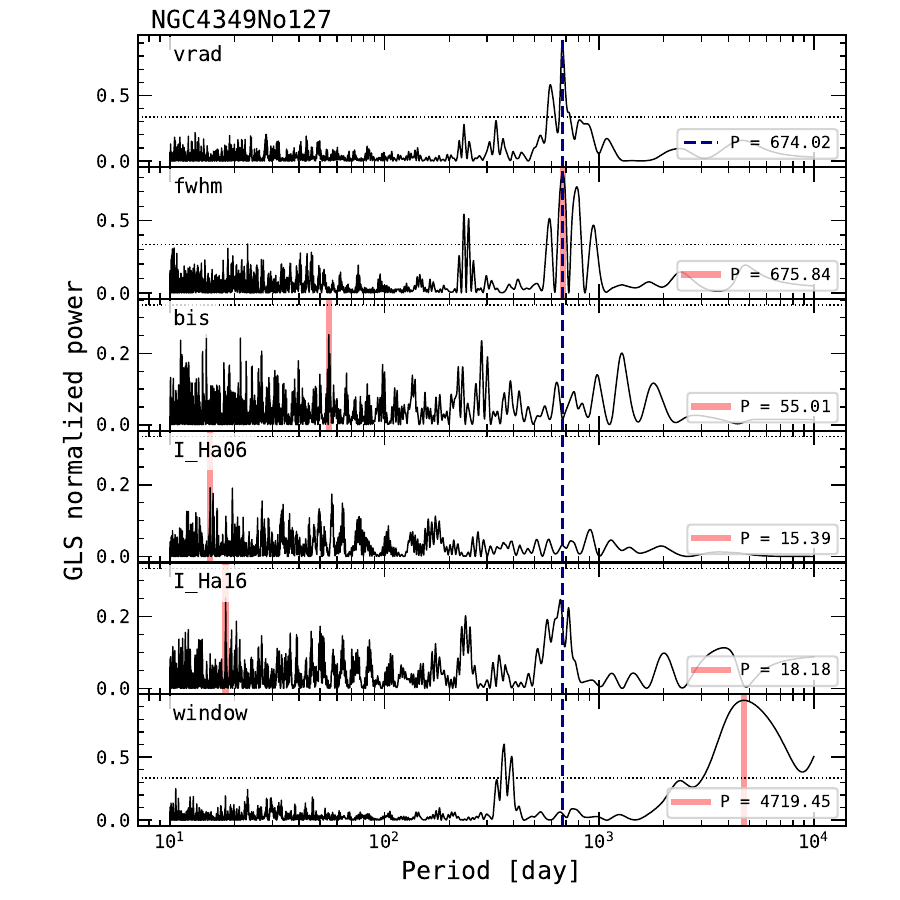}
\caption{GLS periodogram of the RV, FWHM, BIS, and stellar activity indexes for NGC4349 No. 127 HARPS data. The blue dashed line marks the period of the planet candidate. The period of the strongest peak in each periodogram is marked with a solid red line. The horizontal line marks the 1\% FAP level. The last panel shows the periodogram for the window function.} 
\label{per_NGC4349No127}
\end{figure}

\section{Revision of previous planetary candidates} \label{sec:others}

In this section, we present a summary of the updated results with the new data collected for the targets of Paper II, where we presented the cases of three stars that show large-amplitude and long-term signals disguised as substellar companions. 

\subsection{IC4651No9122}
We have collected 15 additional RV points since 2018 (for a total of 74). In paper II, this star is shown to have a strong RV signal at 747 days but with signals in FWHM and \ha16 at 714 days (above or close to the FAP level), which led us to refrain from claiming the presence of a planet around this star. However, in the GLS of the updated time series (see Fig. \ref{per_IC4651No9122}), none of the activity indicator signals has a significant peak at the same period as the most significant peak in the RV at 744 days. We also checked that the RV signal at 744 days is not an alias of other signals. The window function shows four significant peaks at 30, 365, 1785, and 4466 days. The three longer-period signals introduce significant aliases of the 744 d signal in the RV periodogram at periods of 245, 1275, and 893 days, respectively. The RV analysis with \texttt{kima} provides a two-Keplerian fit as the solution with the largest likelihood. The time series and orbital phase of the two planet candidates with periods of 743 days ($K$\,=\,99.6\,\ms and $e$\,=\,0.13) and 384 days ($K$\,=\,27.7\,\ms and $e$\,=\,{0.25}) can be seen in Fig. \ref{IC4651No9122_kima2p}. However, the period for the possible second planet in the system is very close to two significant peaks in the FWHM periodogram at 364 and 393 days. This case again highlights the importance of the FWHM as a reliable stellar indicator. Nevertheless, we note that these signals in FWHM could simply be caused by the window function, which has a peak at 366 days, and is not related to activity. On the other hand, if those are true stellar signals, then it is worth noting that they might be the first harmonic of the signal at 744 days ($P$/2\,=\,372 days), which would cause us to doubt the hypothesis that IC4651No9122 is a true planet candidate. With the data in hand, we can at least be sure that the signal fitted with the second keplerian at 384 days is either of stellar origin (reflected in the FWHM if real) or is caused by the window function or a harmonic of the 744 days signal. Therefore, we consider only the signal at 744 days as a planet candidate and repeated the fit with \texttt{kima}, fixing the number of planets to just one (see Fig. \ref{IC4651No9122_kima}). In this case, as expected, the residuals of the fit have a larger dispersion (20.39\,\ms) and the posterior distribution for the fitted extra white noise is also centred at a larger value (20.5$\pm 1.8$\,\ms). The final planet candidate has a period of 746 days ($K$\,=\,94.98\,\ms and $e$\,=\,0.04) and a minimum mass of 6.22M\,$_{J}$ (see Table \ref{tab:orbital_parameters} for the detailed parameters). 

We note that, in the present work, we are also using the updated stellar parameters from \citet{tsantaki23}, and now the estimated \textit{v} sin \textit{i} is 0.68\,kms\ for a fixed macroturbulence velocity of 4.98\,kms\ \citep[with the empirical formula by][for giant stars]{hekker07}. This leads to a maximum rotational period of $\sim$\,995 days. This period is close to the period of 985 days seen in the \ha16 periodogram; however, it is not significant and prevents us from relating them. Therefore, we cannot completely rule out the possibility that the 744 d RV signal is related to the star's rotation (if this were shorter in reality). The lack of similar periods in the activity indicator periodograms (as well as the lack of strong correlations; see Table \ref{correlations}) suggests that this is a bona fide planet but the proximity of the two significant FWHM peaks to the first harmonic of the strong RV signal warns us of the possibility of a non-planetary signal. We note that if the recent  FWHM values were corrected to take into account the observed offset (see a more detailed discussion in Appendix A), the GLS periodogram of the FWHM would show a significant peak at 687 days. This period seems safely far from the RV period at 744 days. Additionally, the peaks seen at 364 and 393 d would go below the 1\% FAP level, which would support the planet-candidate hypothesis. Nevertheless, we can see with this example how the usefulness of the FWHM indicator can be limited if corrections need to be applied. Therefore, additional data will be needed to further assess the nature of the signals observed here.

\subsection{NGC2423No3}
We collected a further 16 RV points for this star (raising the total to 92, with 32 of them from CORALIE). In our previous work, this star had a strong RV signal at 698 days but with a strong  RV--BIS correlation and a peak in the BIS periodogram close to the FAP level at a similar period. The GLS of the complete dataset is shown in Fig. \ref{per_NGC2423No3}. The RV periodogram presents the most significant peak at 697 days (with two aliases at 828 and 614 days caused by the long gap in time between the two sets of observations, specifically by the peak at 4506 days in the window function periodogram). In addition, the window function shows several significant signals around one year, which in turn produce aliases around 230 days in the RV periodogram. As opposed to previous findings, with the new data the BIS periodogram shows a significant peak at 697 days. Moreover, the RV--BIS correlation has a significant Pearson coefficient as well as Spearman rank statistics (see Table \ref{correlations}). With the new data, both \texttt{kima} and \textit{yorbit} would fit the RV time series with a two-Keplerian orbit caused by planetary bodies at 700 and 435 days (with $K$\,$\sim$\,130 and 30\,\ms, respectively, see Fig \ref{NGC2423No3_kima}). However, the long-period signal appears to be of stellar origin given the correlation with the BIS, and the second signal has a period similar to one of the peaks seen in the periodogram of the photometric data (period of 417 days, see Fig. 8 in Paper II). In addition, both the FWHM and \ha16 periodograms show significant peaks at 380 and 393 days, respectively. 

We note that with the updated stellar parameters from \citet{tsantaki23}, we estimated a \textit{v} sin \textit{i} of 2.19\,kms\ (for a fixed macroturbulence velocity of 4.88\,kms), which leads to a maximum rotational period of $\sim$\,407 days. These findings lead us to suspect that all these signals around 400 days are caused by rotational modulation of active regions, which in turn also produces the 435 d RV signal fitted in the second Keplerian orbit.

\subsection{NGC4349No127}
For this target, we previously worked with 46 HARPS measurements up to 2009 (as presented in our previous publication) and now we have 11 additional points (between 2018-2022). This star previously showed a large RV signal at 672 days but with a strong and clear signal in FWHM at 666 days (and also the periodogram of \ha16 had a close to significant peak at 689 days., see Paper II). Those findings indicated that the brown dwarf presented in Paper I was not real. We collected additional data to evaluate the stability of the RV and FWHM signals \citep[as some examples in the literature have shown changes in phase for long-term RV signals; e.g. Aldebaran][]{reichert19} and to learn more about its origin. With the new data, the RV time series can also be fitted with a single Keplerian with the same period, of namely 674 days ($K$\,$\sim$\,\textit{226}\,\ms; see Fig. \ref{NGC4349No127_kima}). However, this long-term signal is also observed in the FWHM periodogram (see Fig. \ref{per_NGC4349No127}), as was previously found to be the case, but not in the \ha\ periodograms. However, we note that we find a weak but significant correlation between RV and \ha16 (see Table \ref{correlations}). Interestingly, we do not observe a change of phase in the RV (see the orbital phase in Fig. \ref{NGC4349No127_kima}), but if we fit the FWHM with a sinusoidal with a period of 675 days (see Fig. \ref{NGC4349No127_FWHM}), there seems to be a change in the phase. In addition, the window function shows a strong peak at 4719 days caused by the long gap in between the observations, which in turn produces significant aliases of 588 days (seen in the RV and FWHM periodograms) and 785 days, which can be seen in the FWHM periodogram.

For this star, we also have updated stellar parameters from \citet{tsantaki23}, for which we estimated a \textit{v} sin \textit{i} of 4.81\,kms\ (for a fixed macroturbulence velocity of 4.66\,kms), which leads to a maximum rotational period of $\sim$\,399 days. The periodogram of the photometric data shows two significant peaks at 428 and 342 days, which may be related to the stellar rotation (see Fig. 12 in Paper II). As discussed in this latter work, we cannot discard that the estimated rotational period ($<$399 days) and the 342 d signal in the photometry are the first harmonics ($P$/2) of the RV period. Taken together, these findings point to a non-planetary origin of the signal.

\begin{center}
\begin{table*}
\caption{Orbital and physical parameters for the planet candidates.}
\centering
\begin{tabular}{llccc}
\hline
\noalign{\smallskip} 
 & & NGC3680 No.41 & IC4651 No. 9122  \\  
\noalign{\smallskip} 
\hline
\hline
\noalign{\smallskip} 
$V_{sys}$          & [km s$^{-1}$]  &  1.603 $\pm$ 0.013 & --30.258 $\pm$ 0.003     \\
$P$                & [days]        & 1154.98 $\pm$ 9.96 & 745.94 $\pm$ 1.80 \\
$K$                & [m s$^{-1}$]   & 74.79 $\pm$ 9.49  & 94.98 $\pm$ 4.38  \\
\textit{e}         &                &  0.21 $\pm$ 0.10  &  0.04 $\pm$ 0.03  \\
$\omega$           & [rad]          &  0.992         &  0.937        \\
Tp                 & [BJD-2\,400\,000]   &  51819.88      & 53004.39        \\
$m_{2}$ sin \textit{i} & [M$_{J}$]  &    5.13 $\pm$ 0.66           &      6.22 $\pm$ 0.36  \\ 
\textit{a}         & [AU]           &    2.53 $\pm$ 0.03           &      1.95 $\pm$ 0.03  \\ 
\noalign{\medskip} %
\hline                                                                                   
\noalign{\medskip} %
N$_{meas}$         &                     &    23                     &      74              \\
Span               & [years]             &    19.07                  &      17.02           \\
$\Delta$v(HARPS-Coralie) & [m s$^{-1}$]  &    14 $\pm$ 14            &    ---               \\ 
white noise        & [m s$^{-1}$]   & 20.0 $\pm$ 5(harps) -- 3.8$^{+11.3}$$_{-3.1}$ (coralie) 2 & 20.5 $\pm$ 1.8 (harps)  \\
$(O-C)       $     & [m s$^{-1}$]        &    16.35                   &    20.39              \\ \noalign{\medskip} %
\hline
\noalign{\medskip} %
\end{tabular}
\label{tab:orbital_parameters}
\end{table*}
\end{center}

\section{Discussion} \label{sec:discussion}

\subsection{The importance of each activity indicator}
The detailed analysis presented in this work (and our previous study) demonstrates the difficulty in confirming planetary signals in evolved intermediate-mass stars. Our most important finding is that there is not a single (e.g. universal) activity indicator that we can trust to unveil a spurious planetary signal; all of them have to be analysed in order to understand the origin of the different RV signals, and in some cases this might not be sufficient. We find hints that the CCF-BIS might be more useful for less massive stars (around 2\,M$_\odot$), because we find a possible RV--BIS correlation for NGC3680 No. 41 and a clear signal in the BIS periodogram (as well as a significant strong RV-BIS correlation) for NGC2423 No. 3. 

On the other hand, the CCF-FWHM and \ha\ lines seem to be more useful for tracking activity signals in more massive and evolved stars (NGC2345 No. 50, NGC3532 No. 670, and NGC4349 No. 127); see Table \ref{peaks} for a summary of the periodogram results. However, we notice that conclusions on the origin of
a given signal are highly dependent on the way the \ha\ line is measured. Meanwhile NGC2345 No. 50 has significant peaks at both the \ha16 and \ha06 indicators, NGC3532 No. 670 only shows a significant peak for \ha16, and NGC4349 No.127 shows no significant peak for any of them. Nevertheless, although not significant, the signal of \ha16 for NGC4349 No. 127 is much stronger than that of \ha06. 

We acknowledge that the sample analysed here is still not sufficiently large for us to reach clear conclusions, but there are indications that \ha16 is a more reliable indicator than \ha06 for these evolved stars \citep[the opposite is true for FGK dwarfs, for which \ha06 is more appropriate;][]{gomesdasilva22}. A thorough analysis of stellar activity indicators in evolved stars is outside the scope of the present paper but will be explored in a future work. On the contrary, measuring the \ha\ line with a very wide passband can result in spurious periods in the periodogram, because there are contaminating lines in the wings of this strong line. This is the explanation for the recent confirmation of a brown dwarf around the largest star to date \citep[HD\,18184;][]{lee23}, whose RV variations were previously deemed to be of stellar origin given the similar period in the \ha\ periodogram \citep{bang18}. However, the work by \cite{lee23} warns that a $\pm$\,1\AA\, passband around the line centre \citep[and not a $\pm$\,2\AA\, as used in][]{bang18} is needed to avoid the blending lines. We note that our \ha16 index is measured with a $\pm$\,0.8\,\AA\, passband around the line centre and is therefore not affected by nearby blending lines (see Figs. \ref{halpha1} and \ref{halpha2} in the Appendix).

Finally, we believe that the most critical indicator for this kind of star is the CCF--FWHM, because it has proven useful for detecting the probable stellar origin of the largest-amplitude signals (in the three most massive and evolved stars) but also that of the weaker RV signals (that can be fitted with a second Keplerian) found in stars with a broad range of stellar masses (the cases of NGC3532 No. 670, IC4651 No. 9122, and NGC2423 No. 3). We stress the fact that for NGC4349 No. 127, despite having nearly identical parameters to NGC3532 No. 670, only the FWHM has served to cast doubt on the planetary origin of the RV variability. Finally, the photometric data have been useful in some cases as well, but the long rotational periods of our targets make it difficult to find data of sufficient quantity to track the real period of the stars.

\subsection{Origin of RV variability}
From the six targets studied in this work, we find suggestions that it is more probable to find bona fide planets around less massive and less evolved stars. NGC3680 No. 41 and IC4651 No. 9122 both have a mass of around 1.7\,M$_\odot$ and a radius of around 12\,R$_\odot$ and seem to be in the first ascent of the RGB (see Fig. \ref{HR_diagram}). On the other hand, for stars more massive than 2\,M$_\odot$, we cannot confirm the planetary origin for any of the RV signals detected in our more massive targets. These four stars are also among the most evolved in their clusters. NGC2345 No. 50 is the most evolved of the full sample, already ascending towards the AGB. NGC 4349 No. 127 and NGC3532 No. 670 seem to be very close to the tip of the RGB and NGC2423 No. 3 is also approaching the RGB tip. This is in agreement with the decrease in planetary occurrence at M\,$\gtrsim$2\,M$_\odot$ found by the Lick \citep{reffert15} and EXPRESS surveys \citep{jones16} and the combination of both with the PPPS survey \citep{wolthoff22}. Nevertheless, the targets presented here are a fraction of the complete sample and a more in depth analysis on planet frequencies will be presented in a future work along with  results for the full sample.

In recent years, an increasing number of suspected planet candidates have been found around evolved stars with orbital periods in the range of $\sim$\,600-800 days. This is the case for NGC4349 No. 127, NGC2423 No. 3, NGC3532 No. 670 discussed here, Pollux ($\beta$ Gem) \citep{hatzes06,auriere21}, Aldebaran \citep{hatzes15,reichert19},  and $\gamma$ Draconis \citep{hatzes18}. In addition, the work by \citet{tala-pinto20} presents two evolved stars, 3CnC and 44UMa, with periodic RV variations of $\sim$\,800 days that do not show variations with a similar period in photometry or \ha\ index. However, given the similarity in stellar parameters and luminosities with the above-mentioned cases and the lack of line-shape-stability indicators such as FWHM (not available in non-stabilised spectrographs, as used in their work), \citet{tala-pinto20} classify these objects as planet candidates. The luminous red giant HD\,81817 (4.3\,M$_\odot$, 83.8\,R$_\odot$) also shows a secondary RV signal with a period of $\sim$\,630 days \citep{bang20} but the \ha\ line shows a significant peak at a similar period as well. 

The accumulation of planet candidates in that period range (and more specifically between 300 and 800 days) has also been noted by \cite{dollinger21}, who in addition points out that this only happens for stars with a radius of greater than 21\,R$_\odot$ (see their Fig. 1). These authors also warn of the possibility that the observed periodic signals are of stellar origin and propose that a large stellar radius and a low stellar metallicity should be seen as  a warning signal that  RV signals may not be of planetary nature. It is worth noting that the three objects with the clearest RV variations in our sample have large radii (they are the most evolved in each cluster) and metallicities below --0.1\,dex.

As in many cases the rotational periods of these large, slow stars are compatible with the RV periods, the simplest explanation for the RV variability would be the rotational modulation of active regions, such as spots. However, long-lived stable spots in evolved single stars have only been reported in the literature for very active stars hosting strong magnetic fields, those that are believed to be the descendants of magnetic Ap-Bp stars. This is the case for EK Eri \citep{auriere11} or $\beta$ Ceti \citep{tsvetkova13} with P$_{rot}$ of 309 and 215 days, respectively, which also present large photometric variations. 

However, the stars presented in this work seem to belong to another group of red giants; these show longer rotational periods, weak large-scale magnetic fields (sub-Gauss level), and higher-than-expected chromospheric activity \citep{auriere15}. The best studied case in the literature is Pollux, which shows RV variations with a period of 590 days ---that is stable for 25 years--- and no apparent S-index variability \citep{hatzes06}. Nevertheless, the planet candidate around Pollux was challenged by the discovery of magnetic field variations with a similar period to the RV variations \citep{auriere14}. In a more recent work with additional data, \citet{auriere21} found a longer period (660 days) for the weak magnetic field variations, but this could still compatible with the RV period due to uncertainties. Therefore, doubt remains as to the true existence of a planet orbiting Pollux, as is also true for the cases presented in the present work. In the cases of red giants with weak magnetic fields such as Pollux, if the detected RV variations are due to the magnetic field, they would not be caused by photometric spots but by magnetic plages or other magnetic structures locally reducing convection (Auri{\`e}re, \textit{priv. comm}). Therefore, the lack of photometric variation in red giants (and thus of spots) is not a guarantee that the RVs are of planetary origin. 

We do not have sufficient S/N for our sample of stars in the blue region of the spectra to be able to study the S-index variability in order to compare it with the behaviour of Pollux and other giants with weak magnetic fields like Aldebaran. However, many of our stars show \ha\ variations, which is also a sign of chromospheric activity. It would be very interesting to further study the magnetic activity of these stars, although it is very challenging to detect weak magnetic fields for relatively dull stars ($V\,\sim$\,10) with the currently available  instrumentation.

An alternative explanation for the large RV variations in evolved stars is the presence of a poorly understood class of oscillations called oscillatory convective modes \citep{saio15}. This class of oscillations is theoretically possible for 2\,M$_{\odot}$ stars with high luminosity, that is, with log\,(L/L$_{\odot}) \gtrsim$\,3\,dex; see more detailed discussion in Paper II. The clearest candidate in our sample would therefore be NGC2345 No. 50\footnote{We note that the exact minimum luminosity for these modes to appear depends on the mass and the mixing length adopted in the treatment of convection. Therefore, it is probable that for the large mass of this star (not covered by current models), the minimum needed luminosity is even higher and this star would not be a candidate to host this kind of oscillation.}, with log\,(L/L$_{\odot}) \sim$\,3.9, though NGC4349 No. 127 and NGC3532 No. 670 also have high luminosities, as do other candidates in the literature \citep[see Fig. 15 in][]{tala-pinto20}.

\section{Conclusions} \label{sec:conclusions}
In this work, we present some of the results of a long-term survey of RVs in a sample of more than 140 giant stars in 17 open clusters. Data covering such a large time span are crucial in order to find long-period planets, whereas the probability of discovering short-period planets around giant stars diminishes as the stars evolve and their radii expand. The long-term monitoring of giant stars is also essential for distinguishing RV variations caused by the rotational modulation of active regions, because the rotational periods of evolved stars are much longer than those of dwarfs (from a few hundred to more than one thousand days). From the six stars analysed in this work, we are only confident in the planetary origin of the RV signals coming from two of those systems, and the star in both of them has a mass of below 2\,M$_{\odot}$. Nevertheless, we believe that additional data are needed to fully confirm the presence of planets around those stars. On the other hand, the more massive stars in the sample show periodic variations similar to the RV in one or several stellar activity indicators (\ha, BIS and FWHM), with the FWHM of the CCF being the most critical in our opinion. We also note that in the cases where a two-Keplerian fit was a good solution to model the RV variability, the \ha16 index and especially the FWHM were decisive in ruling out the presence of a smaller body in these systems. 

The cases presented here and others in the literature demonstrate the difficulty in finding even large planets around intermediate-mass evolved stars. It is therefore of utmost importance to obtain a comprehensive picture of the different phenomena causing RV variability in evolved stars, and especially of their magnetic fields and their manifestations on  different timescales. Once we understand such phenomena, we will be in a much better position to detect giant planets but also lower-mass planets currently hidden in the stellar jitter. Near-infrared (NIR) spectra can be combined with optical spectra in order to discern between stellar activity and planetary signals, as the non-planetary RV signals are wavelength-dependent, whilst planetary signals are achromatic \citep[e.g.][]{figueira10,trifonov15,carleo20,carmona23}. In addition, the study of other stellar-activity indicators in the NIR will help us to better understand the magnetic activity of these stars and its relation with their RV variability. Therefore, as a further step towards a complete understanding of the variability in red giants, we will observe these stars with the new high-resolution NIR spectrograph NIRPS \citep{bouchy17,wildi17}  now operational on the 3.6m ESO telescope in La Silla (Chile). 

\begin{acknowledgements}
We thank Fran\c{c}ois Bouchy and Xavier Dumusque for coordinating the shared observations with HARPS and all the observers who helped collecting the data. We thank the referee for their careful review that helped to improve this paper.

E.D.M., J.G.S, J.P.F., N.C.S., J.H.M., S.G.S.  acknowledge the support from Funda\c{c}\~ao para a Ci\^encia e a Tecnologia (FCT) through national funds and from FEDER through COMPETE2020 by the following grants: UIDB/04434/2020 \& UIDP/04434/2020 and 2022.04416.PTDC. E.D.M. acknowledges the support from FCT through Stimulus FCT contract 2021.01294.CEECIND and Investigador FCT contract IF/00849/2015/CP1273/CT0003 and in the form of an exploratory project with the same reference. This research has been Co-funded by the European Union (ERC, FIERCE, 101052347). Views and opinions expressed are however those of the author(s) only and do not necessarily reflect those of the European Union or the European Research Council. Neither the European Union nor the granting authority can be held responsible for them.

This research has made use of The Extrasolar Planets Encyclopaedia, SIMBAD and WEBDA databases. This work has also made use of the IRAF facility. This work has made use of data from the European Space Agency (ESA) mission \textit{Gaia} (https://www.cosmos.esa.int/gaia), processed by the \textit{Gaia} Data Processing and Analysis Consortium (DPAC,https://www.cosmos.esa.int/web/gaia/dpac/consortium). Funding for the DPAC has been provided by national institutions, in particular the institutions participating in the {\it Gaia} Multilateral Agreement.

\end{acknowledgements}

\bibliographystyle{aa}
\bibliography{edm_bibliography_planets}

\begin{thebibliography}{73}
\expandafter\ifx\csname natexlab\endcsname\relax\def\natexlab#1{#1}\fi

\bibitem[{{Alonso-Santiago} {et~al.}(2019){Alonso-Santiago}, {Negueruela},
  {Marco}, {Tabernero}, {Gonz{\'a}lez-Fern{\'a}ndez}, \&
  {Castro}}]{alonso-santiago19}
{Alonso-Santiago}, J., {Negueruela}, I., {Marco}, A., {et~al.} 2019, \aap, 631,
  A124

\bibitem[{{Auri{\`e}re} {et~al.}(2015){Auri{\`e}re}, {Konstantinova-Antova},
  {Charbonnel}, {Wade}, {Tsvetkova}, {Petit}, {Dintrans}, {Drake}, {Decressin},
  {Lagarde}, {Donati}, {Roudier}, {Ligni{\`e}res}, {Schr{\"o}der},
  {Landstreet}, {L{\`e}bre}, {Weiss}, \& {Zahn}}]{auriere15}
{Auri{\`e}re}, M., {Konstantinova-Antova}, R., {Charbonnel}, C., {et~al.} 2015,
  \aap, 574, A90

\bibitem[{{Auri{\`e}re} {et~al.}(2014){Auri{\`e}re}, {Konstantinova-Antova},
  {Espagnet}, {Petit}, {Roudier}, {Charbonnel}, {Donati}, \&
  {Wade}}]{auriere14}
{Auri{\`e}re}, M., {Konstantinova-Antova}, R., {Espagnet}, O., {et~al.} 2014,
  in IAU Symposium, Vol. 302, Magnetic Fields throughout Stellar Evolution, ed.
  P.~{Petit}, M.~{Jardine}, \& H.~C. {Spruit}, 359--362

\bibitem[{{Auri{\`e}re} {et~al.}(2011){Auri{\`e}re}, {Konstantinova-Antova},
  {Petit}, {Roudier}, {Donati}, {Charbonnel}, {Dintrans}, {Ligni{\`e}res},
  {Wade}, {Morgenthaler}, \& {Tsvetkova}}]{auriere11}
{Auri{\`e}re}, M., {Konstantinova-Antova}, R., {Petit}, P., {et~al.} 2011,
  \aap, 534, A139

\bibitem[{{Auri{\`e}re} {et~al.}(2021){Auri{\`e}re}, {Petit}, {Mathias},
  {Konstantinova-Antova}, {Charbonnel}, {Donati}, {Espagnet}, {Folsom},
  {Roudier}, \& {Wade}}]{auriere21}
{Auri{\`e}re}, M., {Petit}, P., {Mathias}, P., {et~al.} 2021, \aap, 646, A130

\bibitem[{{Bang} {et~al.}(2018){Bang}, {Lee}, {Jeong}, {Han}, \&
  {Park}}]{bang18}
{Bang}, T.-Y., {Lee}, B.-C., {Jeong}, G.-h., {Han}, I., \& {Park}, M.-G. 2018,
  Journal of Korean Astronomical Society, 51, 17

\bibitem[{{Bang} {et~al.}(2020){Bang}, {Lee}, {Perdelwitz}, {Jeong}, {Han},
  {Oh}, \& {Park}}]{bang20}
{Bang}, T.-Y., {Lee}, B.-C., {Perdelwitz}, V., {et~al.} 2020, \aap, 638, A148

\bibitem[{{Borgniet} {et~al.}(2019){Borgniet}, {Lagrange}, {Meunier},
  {Galland}, {Arnold}, {Astudillo-Defru}, {Beuzit}, {Boisse}, {Bonfils},
  {Bouchy}, {Debondt}, {Deleuil}, {Delfosse}, {Desort}, {D{\'\i}az},
  {Eggenberger}, {Ehrenreich}, {Forveille}, {H{\'e}brard}, {Loeillet}, {Lovis},
  {Montagnier}, {Moutou}, {Pepe}, {Perrier}, {Pont}, {Queloz}, {Santerne},
  {Santos}, {S{\'e}gransan}, {da Silva}, {Sivan}, {Udry}, \&
  {Vidal-Madjar}}]{borgniet19}
{Borgniet}, S., {Lagrange}, A.~M., {Meunier}, N., {et~al.} 2019, \aap, 621, A87

\bibitem[{{Bouchy} {et~al.}(2017){Bouchy}, {Doyon}, {Artigau}, {Melo},
  {Hernandez}, {Wildi}, {Delfosse}, {Lovis}, {Figueira}, {Canto Martins},
  {Gonz{\'a}lez Hern{\'a}ndez}, {Thibault}, {Reshetov}, {Pepe}, {Santos}, {de
  Medeiros}, {Rebolo}, {Abreu}, {Adibekyan}, {Bandy}, {Benz}, {Blind},
  {Bohlender}, {Boisse}, {Bovay}, {Broeg}, {Brousseau}, {Cabral}, {Chazelas},
  {Cloutier}, {Coelho}, {Conod}, {Cumming}, {Delabre}, {Genolet}, {Hagelberg},
  {Jayawardhana}, {K{\"a}ufl}, {Lafreni{\`e}re}, {de Castro Le{\~a}o}, {Malo},
  {de Medeiros Martins}, {Matthews}, {Metchev}, {Oshagh}, {Ouellet}, {Parro},
  {Rasilla Pi{\~n}eiro}, {Santos}, {Sarajlic}, {Segovia}, {Sordet}, {Udry},
  {Valencia}, {Vall{\'e}e}, {Venn}, {Wade}, \& {Saddlemyer}}]{bouchy17}
{Bouchy}, F., {Doyon}, R., {Artigau}, {\'E}., {et~al.} 2017, The Messenger,
  169, 21

\bibitem[{{Brewer} {et~al.}(2010){Brewer}, {P{\'a}rtay}, \&
  {Cs{\'a}nyi}}]{brewer_dnest}
{Brewer}, B.~J., {P{\'a}rtay}, L.~B., \& {Cs{\'a}nyi}, G. 2010, {DNEST:
  Diffusive Nested Sampling}, Astrophysics Source Code Library, record
  ascl:1010.029

\bibitem[{{Carleo} {et~al.}(2020){Carleo}, {Malavolta}, {Lanza}, {Damasso},
  {Desidera}, {Borsa}, {Mallonn}, {Pinamonti}, {Gratton}, {Alei}, {Benatti},
  {Mancini}, {Maldonado}, {Biazzo}, {Esposito}, {Frustagli},
  {Gonz{\'a}lez-{\'A}lvarez}, {Micela}, {Scandariato}, {Sozzetti}, {Affer},
  {Bignamini}, {Bonomo}, {Claudi}, {Cosentino}, {Covino}, {Fiorenzano},
  {Giacobbe}, {Harutyunyan}, {Leto}, {Maggio}, {Molinari}, {Nascimbeni},
  {Pagano}, {Pedani}, {Piotto}, {Poretti}, {Rainer}, {Redfield}, {Baffa},
  {Baruffolo}, {Buchschacher}, {Billotti}, {Cecconi}, {Falcini}, {Fantinel},
  {Fini}, {Galli}, {Ghedina}, {Ghinassi}, {Giani}, {Gonzalez}, {Gonzalez},
  {Guerra}, {Hernandez Diaz}, {Hernandez}, {Iuzzolino}, {Lodi}, {Oliva},
  {Origlia}, {Perez Ventura}, {Puglisi}, {Riverol}, {Riverol}, {San Juan},
  {Sanna}, {Scuderi}, {Seemann}, {Sozzi}, \& {Tozzi}}]{carleo20}
{Carleo}, I., {Malavolta}, L., {Lanza}, A.~F., {et~al.} 2020, \aap, 638, A5

\bibitem[{{Carmona} {et~al.}(2023){Carmona}, {Delfosse}, {Bellotti},
  {Cort{\'e}s-Zuleta}, {Ould-Elhkim}, {Heidari}, {Mignon}, {Donati}, {Moutou},
  {Cook}, {Artigau}, {Fouqu{\'e}}, {Martioli}, {Cadieux}, {Morin}, {Forveille},
  {Boisse}, {H{\'e}brard}, {D{\'\i}az}, {Lafreni{\`e}re}, {Kiefer}, {Petit},
  {Doyon}, {Acu{\~n}a}, {Arnold}, {Bonfils}, {Bouchy}, {Bourrier}, {Dalal},
  {Deleuil}, {Demangeon}, {Dumusque}, {Hara}, {Hoyer}, {Mousis}, {Santerne},
  {S{\'e}grasan}, {Stalport}, \& {Udry}}]{carmona23}
{Carmona}, A., {Delfosse}, X., {Bellotti}, S., {et~al.} 2023, \aap, 674, A110

\bibitem[{{Collier Cameron} {et~al.}(2010){Collier Cameron}, {Guenther},
  {Smalley}, {McDonald}, {Hebb}, {Andersen}, {Augusteijn}, {Barros}, {Brown},
  {Cochran}, {Endl}, {Fossey}, {Hartmann}, {Maxted}, {Pollacco}, {Skillen},
  {Telting}, {Waldmann}, \& {West}}]{collier-cameron10}
{Collier Cameron}, A., {Guenther}, E., {Smalley}, B., {et~al.} 2010, \mnras,
  407, 507

\bibitem[{{Delgado Mena} {et~al.}(2018){Delgado Mena}, {Lovis}, {Santos}, {da
  Silva}, {Mortier}, {Tsantaki}, {Sousa}, {Figueira}, {Cunha}, {Campante},
  {Adibekyan}, {Faria}, \& {Montalto}}]{delgado18}
{Delgado Mena}, E., {Lovis}, C., {Santos}, N.~C., {et~al.} 2018, \aap, 619, A2

\bibitem[{{Delgado Mena} {et~al.}(2016){Delgado Mena}, {Tsantaki}, {Sousa},
  {Kunitomo}, {Adibekyan}, {Zaworska}, {Santos}, {Israelian}, \&
  {Lovis}}]{delgado16}
{Delgado Mena}, E., {Tsantaki}, M., {Sousa}, S.~G., {et~al.} 2016, \aap, 587,
  A66

\bibitem[{{Desort} {et~al.}(2008){Desort}, {Lagrange}, {Galland}, {Beust},
  {Udry}, {Mayor}, \& {Lo Curto}}]{desort08}
{Desort}, M., {Lagrange}, A.-M., {Galland}, F., {et~al.} 2008, \aap, 491, 883

\bibitem[{{D{\"o}llinger} \& {Hartmann}(2021)}]{dollinger21}
{D{\"o}llinger}, M.~P. \& {Hartmann}, M. 2021, \apjs, 256, 10

\bibitem[{{Faria} {et~al.}(2018){Faria}, {Santos}, {Figueira}, \&
  {Brewer}}]{faria18}
{Faria}, J.~P., {Santos}, N.~C., {Figueira}, P., \& {Brewer}, B.~J. 2018, The
  Journal of Open Source Software, 3, 487

\bibitem[{{Figueira} {et~al.}(2010){Figueira}, {Pepe}, {Melo}, {Santos},
  {Lovis}, {Mayor}, {Queloz}, {Smette}, \& {Udry}}]{figueira10}
{Figueira}, P., {Pepe}, F., {Melo}, C.~H.~F., {et~al.} 2010, \aap, 511, A55

\bibitem[{{Frink} {et~al.}(2001){Frink}, {Quirrenbach}, {Fischer}, {R{\"o}ser},
  \& {Schilbach}}]{frink01}
{Frink}, S., {Quirrenbach}, A., {Fischer}, D., {R{\"o}ser}, S., \& {Schilbach},
  E. 2001, \pasp, 113, 173

\bibitem[{{Gomes da Silva} {et~al.}(2022){Gomes da Silva}, {Bensabat},
  {Monteiro}, \& {Santos}}]{gomesdasilva22}
{Gomes da Silva}, J., {Bensabat}, A., {Monteiro}, T., \& {Santos}, N.~C. 2022,
  \aap, 668, A174

\bibitem[{{Gomes da Silva} {et~al.}(2018){Gomes da Silva}, {Figueira},
  {Santos}, \& {Faria}}]{gomesdasilva18}
{Gomes da Silva}, J., {Figueira}, P., {Santos}, N., \& {Faria}, J. 2018, The
  Journal of Open Source Software, 3, 667

\bibitem[{{Gomes da Silva} {et~al.}(2021){Gomes da Silva}, {Santos},
  {Adibekyan}, {Sousa}, {Campante}, {Figueira}, {Bossini}, {Delgado-Mena},
  {Monteiro}, {de Laverny}, {Recio-Blanco}, \& {Lovis}}]{gomesdasilva21}
{Gomes da Silva}, J., {Santos}, N.~C., {Adibekyan}, V., {et~al.} 2021, \aap,
  646, A77

\bibitem[{{Gomes da Silva} {et~al.}(2011){Gomes da Silva}, {Santos}, {Bonfils},
  {Delfosse}, {Forveille}, \& {Udry}}]{gomesdasilva11}
{Gomes da Silva}, J., {Santos}, N.~C., {Bonfils}, X., {et~al.} 2011, \aap, 534,
  A30

\bibitem[{{Gomes da Silva} {et~al.}(2012){Gomes da Silva}, {Santos}, {Bonfils},
  {Delfosse}, {Forveille}, {Udry}, {Dumusque}, \& {Lovis}}]{gomesdasilva12}
{Gomes da Silva}, J., {Santos}, N.~C., {Bonfils}, X., {et~al.} 2012, \aap, 541,
  A9

\bibitem[{{Grandjean} {et~al.}(2023){Grandjean}, {Lagrange}, {Meunier},
  {Chauvin}, {Borgniet}, {Desidera}, {Galland}, {Kiefer}, {Messina},
  {Iglesias}, {Nicholson}, {Pantoja}, {Rubini}, {Sedaghati}, {Sterzik}, \&
  {Zicher}}]{grandjean23}
{Grandjean}, A., {Lagrange}, A.~M., {Meunier}, N., {et~al.} 2023, \aap, 669,
  A12

\bibitem[{{Grunblatt} {et~al.}(2022){Grunblatt}, {Saunders}, {Sun}, {Chontos},
  {Soares-Furtado}, {Eisner}, {Pereira}, {Komacek}, {Huber}, {Collins}, {Wang},
  {Stockdale}, {Quinn}, {Tronsgaard}, {Zhou}, {Nowak}, {Deeg}, {Ciardi},
  {Boyle}, {Rice}, {Dai}, {Blunt}, {Van Zandt}, {Beard}, {Akana Murphy},
  {Dalba}, {Lubin}, {Polanski}, {Brinkman}, {Howard}, {Buchhave}, {Angus},
  {Ricker}, {Jenkins}, {Wohler}, {Goeke}, {Levine}, {Colon}, {Huang},
  {Kunimoto}, {Shporer}, {Latham}, {Seager}, {Vanderspek}, \&
  {Winn}}]{grunblatt22}
{Grunblatt}, S.~K., {Saunders}, N., {Sun}, M., {et~al.} 2022, \aj, 163, 120

\bibitem[{{Hatzes} {et~al.}(2015){Hatzes}, {Cochran}, {Endl}, {Guenther},
  {MacQueen}, {Hartmann}, {Zechmeister}, {Han}, {Lee}, {Walker}, {Yang},
  {Larson}, {Kim}, {Mkrtichian}, {D{\"o}llinger}, {Simon}, \&
  {Girardi}}]{hatzes15}
{Hatzes}, A.~P., {Cochran}, W.~D., {Endl}, M., {et~al.} 2015, \aap, 580, A31

\bibitem[{{Hatzes} {et~al.}(2006){Hatzes}, {Cochran}, {Endl}, {Guenther},
  {Saar}, {Walker}, {Yang}, {Hartmann}, {Esposito}, {Paulson}, \&
  {D{\"o}llinger}}]{hatzes06}
{Hatzes}, A.~P., {Cochran}, W.~D., {Endl}, M., {et~al.} 2006, \aap, 457, 335

\bibitem[{{Hatzes} {et~al.}(2018){Hatzes}, {Endl}, {Cochran}, {MacQueen},
  {Han}, {Lee}, {Kim}, {Mkrtichian}, {D{\"o}llinger}, {Hartmann},
  {Karjalainen}, \& {Dreizler}}]{hatzes18}
{Hatzes}, A.~P., {Endl}, M., {Cochran}, W.~D., {et~al.} 2018, \aj, 155, 120

\bibitem[{{Hatzes} {et~al.}(2005){Hatzes}, {Guenther}, {Endl}, {Cochran},
  {D{\"o}llinger}, \& {Bedalov}}]{hatzes05}
{Hatzes}, A.~P., {Guenther}, E.~W., {Endl}, M., {et~al.} 2005, \aap, 437, 743

\bibitem[{{Hekker} \& {Mel{\'e}ndez}(2007)}]{hekker07}
{Hekker}, S. \& {Mel{\'e}ndez}, J. 2007, \aap, 475, 1003

\bibitem[{{Hekker} {et~al.}(2008){Hekker}, {Snellen}, {Aerts}, {Quirrenbach},
  {Reffert}, \& {Mitchell}}]{hekker08}
{Hekker}, S., {Snellen}, I.~A.~G., {Aerts}, C., {et~al.} 2008, \aap, 480, 215

\bibitem[{{Holanda} {et~al.}(2019){Holanda}, {Pereira}, \& {Drake}}]{holanda19}
{Holanda}, N., {Pereira}, C.~B., \& {Drake}, N.~A. 2019, \mnras, 482, 5275

\bibitem[{{Janson} {et~al.}(2021){Janson}, {Gratton}, {Rodet}, {Vigan},
  {Bonnefoy}, {Delorme}, {Mamajek}, {Reffert}, {Stock}, {Marleau}, {Langlois},
  {Chauvin}, {Desidera}, {Ringqvist}, {Mayer}, {Viswanath}, {Squicciarini},
  {Meyer}, {Samland}, {Petrus}, {Helled}, {Kenworthy}, {Quanz}, {Biller},
  {Henning}, {Mesa}, {Engler}, \& {Carson}}]{janson21}
{Janson}, M., {Gratton}, R., {Rodet}, L., {et~al.} 2021, \nat, 600, 231

\bibitem[{{Johnson} {et~al.}(2010){Johnson}, {Aller}, {Howard}, \&
  {Crepp}}]{johnson10}
{Johnson}, J.~A., {Aller}, K.~M., {Howard}, A.~W., \& {Crepp}, J.~R. 2010,
  \pasp, 122, 905

\bibitem[{{Jones} {et~al.}(2016){Jones}, {Jenkins}, {Brahm}, {Wittenmyer},
  {Olivares E.}, {Melo}, {Rojo}, {Jord{\'a}n}, {Drass}, {Butler}, \&
  {Wang}}]{jones16}
{Jones}, M.~I., {Jenkins}, J.~S., {Brahm}, R., {et~al.} 2016, \aap, 590, A38

\bibitem[{{Kennedy} \& {Kenyon}(2008)}]{kennedy08}
{Kennedy}, G.~M. \& {Kenyon}, S.~J. 2008, \apj, 673, 502

\bibitem[{{Lee} {et~al.}(2023){Lee}, {Koo}, {Jeong}, {Park}, {Han}, \&
  {Choi}}]{lee23}
{Lee}, B.-C., {Koo}, J.-R., {Jeong}, G., {et~al.} 2023, Journal of Korean
  Astronomical Society, 56, 35

\bibitem[{{Lillo-Box} {et~al.}(2014){Lillo-Box}, {Barrado}, {Moya},
  {Montesinos}, {Montalb{\'a}n}, {Bayo}, {Barbieri}, {R{\'e}gulo}, {Mancini},
  {Bouy}, \& {Henning}}]{lillobox14}
{Lillo-Box}, J., {Barrado}, D., {Moya}, A., {et~al.} 2014, \aap, 562, A109

\bibitem[{{Lo Curto} {et~al.}(2015){Lo Curto}, {Pepe}, {Avila}, {Boffin},
  {Bovay}, {Chazelas}, {Coffinet}, {Fleury}, {Hughes}, {Lovis}, {Maire},
  {Manescau}, {Pasquini}, {Rihs}, {Sinclaire}, \& {Udry}}]{locurto15}
{Lo Curto}, G., {Pepe}, F., {Avila}, G., {et~al.} 2015, The Messenger, 162, 9

\bibitem[{{Lovis} \& {Mayor}(2007)}]{lovis07}
{Lovis}, C. \& {Mayor}, M. 2007, \aap, 472, 657

\bibitem[{{Mayor} {et~al.}(2003){Mayor}, {Pepe}, {Queloz}, {Bouchy},
  {Rupprecht}, {Lo Curto}, {Avila}, {Benz}, {Bertaux}, {Bonfils}, {Dall},
  {Dekker}, {Delabre}, {Eckert}, {Fleury}, {Gilliotte}, {Gojak}, {Guzman},
  {Kohler}, {Lizon}, {Longinotti}, {Lovis}, {Megevand}, {Pasquini}, {Reyes},
  {Sivan}, {Sosnowska}, {Soto}, {Udry}, {van Kesteren}, {Weber}, \&
  {Weilenmann}}]{mayor03}
{Mayor}, M., {Pepe}, F., {Queloz}, D., {et~al.} 2003, The Messenger, 114, 20

\bibitem[{{Mortier} {et~al.}(2013){Mortier}, {Santos}, {Sousa}, {Adibekyan},
  {Delgado Mena}, {Tsantaki}, {Israelian}, \& {Mayor}}]{mortier13_giants}
{Mortier}, A., {Santos}, N.~C., {Sousa}, S.~G., {et~al.} 2013, \aap, 557, A70

\bibitem[{{Niedzielski} {et~al.}(2015){Niedzielski}, {Villaver}, {Wolszczan},
  {Adam{\'o}w}, {Kowalik}, {Maciejewski}, {Nowak},
  {Garc{\'{\i}}a-Hern{\'a}ndez}, {Deka}, \& {Adamczyk}}]{niedzielski15}
{Niedzielski}, A., {Villaver}, E., {Wolszczan}, A., {et~al.} 2015, \aap, 573,
  A36

\bibitem[{{Nielsen} {et~al.}(2019){Nielsen}, {De Rosa}, {Macintosh}, {Wang},
  {Ruffio}, {Chiang}, {Marley}, {Saumon}, {Savransky}, {Ammons}, {Bailey},
  {Barman}, {Blain}, {Bulger}, {Burrows}, {Chilcote}, {Cotten}, {Czekala},
  {Doyon}, {Duch{\^e}ne}, {Esposito}, {Fabrycky}, {Fitzgerald}, {Follette},
  {Fortney}, {Gerard}, {Goodsell}, {Graham}, {Greenbaum}, {Hibon}, {Hinkley},
  {Hirsch}, {Hom}, {Hung}, {Dawson}, {Ingraham}, {Kalas}, {Konopacky},
  {Larkin}, {Lee}, {Lin}, {Maire}, {Marchis}, {Marois}, {Metchev},
  {Millar-Blanchaer}, {Morzinski}, {Oppenheimer}, {Palmer}, {Patience},
  {Perrin}, {Poyneer}, {Pueyo}, {Rafikov}, {Rajan}, {Rameau}, {Rantakyr{\"o}},
  {Ren}, {Schneider}, {Sivaramakrishnan}, {Song}, {Soummer}, {Tallis},
  {Thomas}, {Ward-Duong}, \& {Wolff}}]{nielsen19}
{Nielsen}, E.~L., {De Rosa}, R.~J., {Macintosh}, B., {et~al.} 2019, \aj, 158,
  13

\bibitem[{{Ottoni} {et~al.}(2022){Ottoni}, {Udry}, {S{\'e}gransan}, {Buldgen},
  {Lovis}, {Eggenberger}, {Pezzotti}, {Adibekyan}, {Marmier}, {Mayor},
  {Santos}, {Sousa}, {Lagarde}, \& {Charbonnel}}]{ottoni22}
{Ottoni}, G., {Udry}, S., {S{\'e}gransan}, D., {et~al.} 2022, \aap, 657, A87

\bibitem[{{Pojmanski}(2002)}]{pojmanski02}
{Pojmanski}, G. 2002, \actaa, 52, 397

\bibitem[{{Queloz} {et~al.}(2009){Queloz}, {Bouchy}, {Moutou}, {Hatzes},
  {H{\'e}brard}, {Alonso}, {Auvergne}, {Baglin}, {Barbieri}, {Barge}, {Benz},
  {Bord{\'e}}, {Deeg}, {Deleuil}, {Dvorak}, {Erikson}, {Ferraz Mello},
  {Fridlund}, {Gandolfi}, {Gillon}, {Guenther}, {Guillot}, {Jorda}, {Hartmann},
  {Lammer}, {L{\'e}ger}, {Llebaria}, {Lovis}, {Magain}, {Mayor}, {Mazeh},
  {Ollivier}, {P{\"a}tzold}, {Pepe}, {Rauer}, {Rouan}, {Schneider},
  {Segransan}, {Udry}, \& {Wuchterl}}]{queloz09}
{Queloz}, D., {Bouchy}, F., {Moutou}, C., {et~al.} 2009, \aap, 506, 303

\bibitem[{{Queloz} {et~al.}(2001){Queloz}, {Henry}, {Sivan}, {Baliunas},
  {Beuzit}, {Donahue}, {Mayor}, {Naef}, {Perrier}, \& {Udry}}]{queloz01}
{Queloz}, D., {Henry}, G.~W., {Sivan}, J.~P., {et~al.} 2001, \aap, 379, 279

\bibitem[{{Queloz} {et~al.}(2000){Queloz}, {Mayor}, {Weber}, {Bl{\'e}cha},
  {Burnet}, {Confino}, {Naef}, {Pepe}, {Santos}, \& {Udry}}]{queloz00}
{Queloz}, D., {Mayor}, M., {Weber}, L., {et~al.} 2000, \aap, 354, 99

\bibitem[{{Reffert} {et~al.}(2015){Reffert}, {Bergmann}, {Quirrenbach},
  {Trifonov}, \& {K{\"u}nstler}}]{reffert15}
{Reffert}, S., {Bergmann}, C., {Quirrenbach}, A., {Trifonov}, T., \&
  {K{\"u}nstler}, A. 2015, \aap, 574, A116

\bibitem[{{Reichert} {et~al.}(2019){Reichert}, {Reffert}, {Stock}, {Trifonov},
  \& {Quirrenbach}}]{reichert19}
{Reichert}, K., {Reffert}, S., {Stock}, S., {Trifonov}, T., \& {Quirrenbach},
  A. 2019, \aap, 625, A22

\bibitem[{{Saar} {et~al.}(1998){Saar}, {Butler}, \& {Marcy}}]{saar98}
{Saar}, S.~H., {Butler}, R.~P., \& {Marcy}, G.~W. 1998, \apjl, 498, L153

\bibitem[{{Saio} {et~al.}(2015){Saio}, {Wood}, {Takayama}, \& {Ita}}]{saio15}
{Saio}, H., {Wood}, P.~R., {Takayama}, M., \& {Ita}, Y. 2015, \mnras, 452, 3863

\bibitem[{{Santos} {et~al.}(2012){Santos}, {Lovis}, {Melendez}, {Montalto},
  {Naef}, \& {Pace}}]{santos12}
{Santos}, N.~C., {Lovis}, C., {Melendez}, J., {et~al.} 2012, \aap, 538, A151

\bibitem[{{Santos} {et~al.}(2009){Santos}, {Lovis}, {Pace}, {Melendez}, \&
  {Naef}}]{santos09}
{Santos}, N.~C., {Lovis}, C., {Pace}, G., {Melendez}, J., \& {Naef}, D. 2009,
  \aap, 493, 309

\bibitem[{{Santos} {et~al.}(2014){Santos}, {Mortier}, {Faria}, {Dumusque},
  {Adibekyan}, {Delgado-Mena}, {Figueira}, {Benamati}, {Boisse}, {Cunha},
  {Gomes da Silva}, {Lo Curto}, {Lovis}, {Martins}, {Mayor}, {Melo}, {Oshagh},
  {Pepe}, {Queloz}, {Santerne}, {S{\'e}gransan}, {Sozzetti}, {Sousa}, \&
  {Udry}}]{santos14}
{Santos}, N.~C., {Mortier}, A., {Faria}, J.~P., {et~al.} 2014, \aap, 566, A35

\bibitem[{{Santos} {et~al.}(2003){Santos}, {Udry}, {Mayor}, {Naef}, {Pepe},
  {Queloz}, {Burki}, {Cramer}, \& {Nicolet}}]{santos03_planet}
{Santos}, N.~C., {Udry}, S., {Mayor}, M., {et~al.} 2003, \aap, 406, 373

\bibitem[{{Sato}(2005)}]{sato05}
{Sato}, B. 2005, Journal of Korean Astronomical Society, 38, 315

\bibitem[{{Sebastian} {et~al.}(2022){Sebastian}, {Guenther}, {Deleuil},
  {Dorsch}, {Heber}, {Heuser}, {Gandolfi}, {Grziwa}, {Deeg}, {Alonso},
  {Bouchy}, {Csizmadia}, {Cusano}, {Fridlund}, {Geier}, {Irrgang}, {Korth},
  {Nespral}, {Rauer}, {Tal-Or}, \& {CoRoT-team}}]{sebastian22}
{Sebastian}, D., {Guenther}, E.~W., {Deleuil}, M., {et~al.} 2022, \mnras, 516,
  636

\bibitem[{{S{\'e}gransan} {et~al.}(2011){S{\'e}gransan}, {Mayor}, {Udry},
  {Lovis}, {Benz}, {Bouchy}, {Lo Curto}, {Mordasini}, {Moutou}, {Naef}, {Pepe},
  {Queloz}, \& {Santos}}]{segransan11}
{S{\'e}gransan}, D., {Mayor}, M., {Udry}, S., {et~al.} 2011, \aap, 535, A54

\bibitem[{{Squicciarini} {et~al.}(2022){Squicciarini}, {Gratton}, {Janson},
  {Mamajek}, {Chauvin}, {Delorme}, {Langlois}, {Vigan}, {Ringqvist}, {Meeus},
  {Reffert}, {Kenworthy}, {Meyer}, {Bonnefoy}, {Bonavita}, {Mesa}, {Samland},
  {Desidera}, {D'Orazi}, {Engler}, {Alecian}, {Miglio}, {Henning}, {Quanz},
  {Mayer}, {Flasseur}, \& {Marleau}}]{squicciarini22}
{Squicciarini}, V., {Gratton}, R., {Janson}, M., {et~al.} 2022, \aap, 664, A9

\bibitem[{{Tala Pinto} {et~al.}(2020){Tala Pinto}, {Reffert}, {Quirrenbach},
  {Stock}, {Trifonov}, \& {Mitchell}}]{tala-pinto20}
{Tala Pinto}, M., {Reffert}, S., {Quirrenbach}, A., {et~al.} 2020, \aap, 644,
  A1

\bibitem[{{Trifonov} {et~al.}(2015){Trifonov}, {Reffert}, {Zechmeister},
  {Reiners}, \& {Quirrenbach}}]{trifonov15}
{Trifonov}, T., {Reffert}, S., {Zechmeister}, M., {Reiners}, A., \&
  {Quirrenbach}, A. 2015, \aap, 582, A54

\bibitem[{{Tsantaki} {et~al.}(2023){Tsantaki}, {Delgado-Mena}, {Bossini},
  {Sousa}, {Pancino}, \& {Martins}}]{tsantaki23}
{Tsantaki}, M., {Delgado-Mena}, E., {Bossini}, D., {et~al.} 2023, \aap, 674,
  A157

\bibitem[{{Tsvetkova} {et~al.}(2013){Tsvetkova}, {Petit}, {Auri{\`e}re},
  {Konstantinova-Antova}, {Wade}, {Charbonnel}, {Decressin}, \&
  {Bogdanovski}}]{tsvetkova13}
{Tsvetkova}, S., {Petit}, P., {Auri{\`e}re}, M., {et~al.} 2013, \aap, 556, A43

\bibitem[{{Udry} {et~al.}(2000){Udry}, {Mayor}, {Naef}, {Pepe}, {Queloz},
  {Santos}, {Burnet}, {Confino}, \& {Melo}}]{udry00}
{Udry}, S., {Mayor}, M., {Naef}, D., {et~al.} 2000, \aap, 356, 590

\bibitem[{{Vigan} {et~al.}(2021){Vigan}, {Fontanive}, {Meyer}, {Biller},
  {Bonavita}, {Feldt}, {Desidera}, {Marleau}, {Emsenhuber}, {Galicher}, {Rice},
  {Forgan}, {Mordasini}, {Gratton}, {Le Coroller}, {Maire}, {Cantalloube},
  {Chauvin}, {Cheetham}, {Hagelberg}, {Lagrange}, {Langlois}, {Bonnefoy},
  {Beuzit}, {Boccaletti}, {D'Orazi}, {Delorme}, {Dominik}, {Henning}, {Janson},
  {Lagadec}, {Lazzoni}, {Ligi}, {Menard}, {Mesa}, {Messina}, {Moutou},
  {M{\"u}ller}, {Perrot}, {Samland}, {Schmid}, {Schmidt}, {Sissa}, {Turatto},
  {Udry}, {Zurlo}, {Abe}, {Antichi}, {Asensio-Torres}, {Baruffolo}, {Baudoz},
  {Baudrand}, {Bazzon}, {Blanchard}, {Bohn}, {Brown Sevilla}, {Carbillet},
  {Carle}, {Cascone}, {Charton}, {Claudi}, {Costille}, {De Caprio},
  {Delboulb{\'e}}, {Dohlen}, {Engler}, {Fantinel}, {Feautrier}, {Fusco},
  {Gigan}, {Girard}, {Giro}, {Gisler}, {Gluck}, {Gry}, {Hubin}, {Hugot},
  {Jaquet}, {Kasper}, {Le Mignant}, {Llored}, {Madec}, {Magnard}, {Martinez},
  {Maurel}, {M{\"o}ller-Nilsson}, {Mouillet}, {Moulin}, {Orign{\'e}}, {Pavlov},
  {Perret}, {Petit}, {Pragt}, {Puget}, {Rabou}, {Ramos}, {Rickman}, {Rigal},
  {Rochat}, {Roelfsema}, {Rousset}, {Roux}, {Salasnich}, {Sauvage}, {Sevin},
  {Soenke}, {Stadler}, {Suarez}, {Wahhaj}, {Weber}, \& {Wildi}}]{vigan21}
{Vigan}, A., {Fontanive}, C., {Meyer}, M., {et~al.} 2021, \aap, 651, A72

\bibitem[{{Vowell} {et~al.}(2023){Vowell}, {Rodriguez}, {Quinn}, {Zhou},
  {Vanderburg}, {Mann}, {Hooton}, {Stassun}, {Howard}, {Bieryla}, {Latham},
  {Howell}, {Guillot}, {Ziegler}, {Collins}, {Carmichael}, {Jenkins},
  {Shporer}, {ABE}, {Bendjoya}, {Bush}, {Buttu}, {Collins}, {Eastman},
  {Fields}, {Gasparetto}, {G{\"u}nther}, {Kostov}, {Kraus}, {Lester}, {Levine},
  {Littlefield}, {Marie-Sainte}, {M{\'e}karnia}, {Osborn}, {Rapetti}, {Ricker},
  {Seager}, {Sefako}, {Srdoc}, {Suarez}, {Torres}, {Triaud}, {Vanderspek}, \&
  {Winn}}]{vowell23}
{Vowell}, N., {Rodriguez}, J.~E., {Quinn}, S.~N., {et~al.} 2023, \aj, 165, 268

\bibitem[{{Wildi} {et~al.}(2017){Wildi}, {Blind}, {Reshetov}, {Hernandez},
  {Genolet}, {Conod}, {Sordet}, {Segovilla}, {Rasilla}, {Brousseau},
  {Thibault}, {Delabre}, {Bandy}, {Sarajlic}, {Cabral}, {Bovay}, {Vall{\'e}e},
  {Bouchy}, {Doyon}, {Artigau}, {Pepe}, {Hagelberg}, {Melo}, {Delfosse},
  {Figueira}, {Santos}, {Gonz{\'a}lez Hern{\'a}ndez}, {de Medeiros}, {Rebolo},
  {Broeg}, {Benz}, {Boisse}, {Malo}, {K{\"a}ufl}, \& {Saddlemyer}}]{wildi17}
{Wildi}, F., {Blind}, N., {Reshetov}, V., {et~al.} 2017, in Society of
  Photo-Optical Instrumentation Engineers (SPIE) Conference Series, Vol. 10400,
  Society of Photo-Optical Instrumentation Engineers (SPIE) Conference Series,
  ed. S.~{Shaklan}, 1040018

\bibitem[{{Wolthoff} {et~al.}(2022){Wolthoff}, {Reffert}, {Quirrenbach},
  {Jones}, {Wittenmyer}, \& {Jenkins}}]{wolthoff22}
{Wolthoff}, V., {Reffert}, S., {Quirrenbach}, A., {et~al.} 2022, \aap, 661, A63

\bibitem[{{Zechmeister} \& {K{\"u}rster}(2009)}]{zechmeister09}
{Zechmeister}, M. \& {K{\"u}rster}, M. 2009, \aap, 496, 577

\end{thebibliography}



\begin{appendix}

\section{RV and FWHM offset for HARPS data after the fibre change}

\begin{figure}
\centering
\includegraphics[width=1\linewidth]{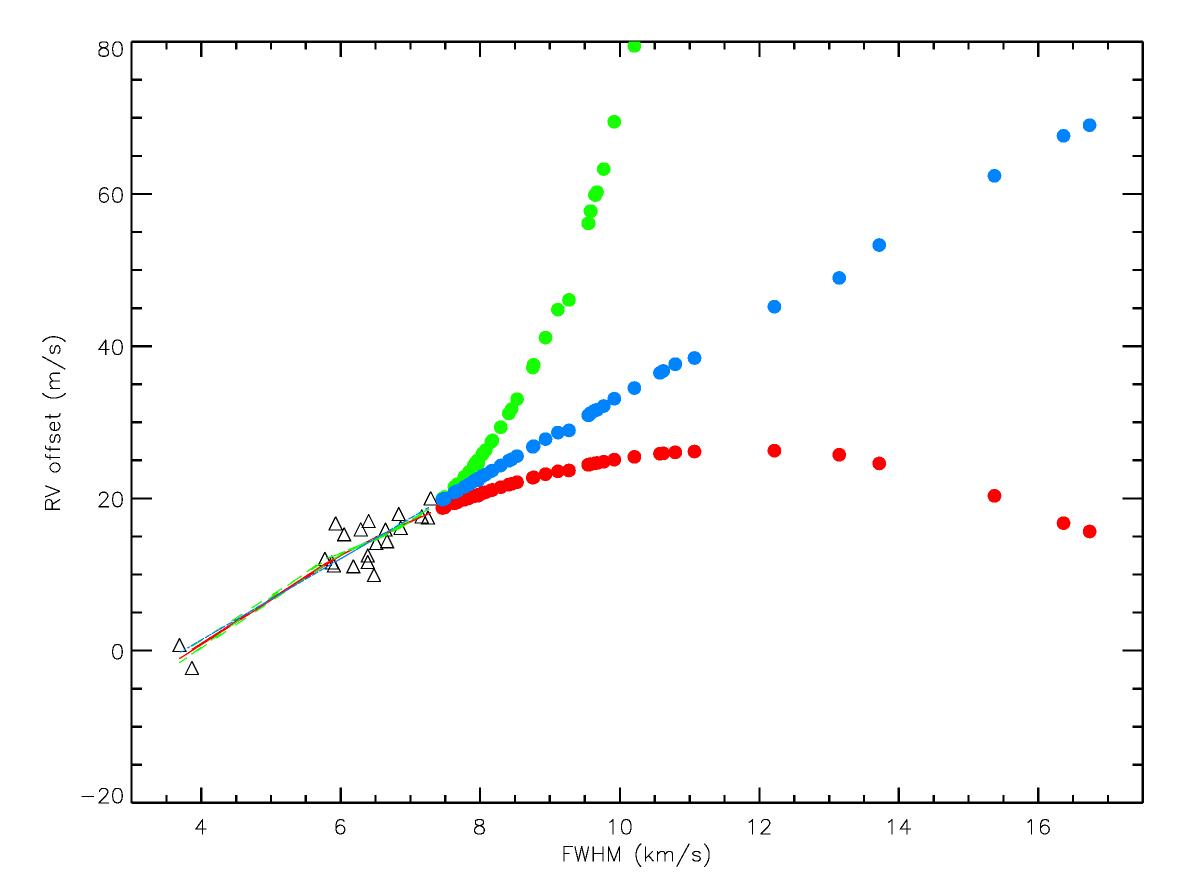}
\caption{RV offset correction after fiber change as a function of FWHM (new observations have larger RV). Triangles are observed standard stars by LoCurto. The 60 sample stars with observations pre and post fiber change are coloured in blue, red or green circles (with their mean FWHM pre fiber change and after correcting from focus drift) and extrapolated using polynomials of 1st, 2nd or 3rd order fitting the standard stars, respectively.}
\label{rv_offset}
\end{figure}

In 2015, a major upgrade of the HARPS fibres took place, which produced an offset in the RV values. Several tests were carried out by \cite{locurto15} to trace the change in RV for stars of different FWHM values. Figure \ref{rv_offset} shows the observed standard stars by LoCurto, which show a positive RV offset that is larger for stars with larger FWHM values. Almost all our data points have mean FWHM values of greater than the FWHM of the standard stars of the tests. We applied different fits to the data, with the third-order polynomial giving very high and probably unrealistic values. We note that we also tried to fit the RV series as separate datasets (pre- and post-fiber change) to calculate the offset. However, due to the low number of points per star and the fact that the RV offset is at the level of the stellar jitter, the obtained solutions were not reliable (the fitted offsets were either too low or too high in many cases). Therefore, given the lack of information, we decided to apply the most conservative linear fit to obtain the RV offset for our stars. Fortunately, the FWHM of the analysed stars in this work is below 10 \kms and the calculated RV offset does not differ significantly between the different fits. In addition, the RV offset is well below the amplitude of the detected signals, and therefore has little impact on the final results.

\begin{figure}
\centering
\includegraphics[width=0.75\linewidth]{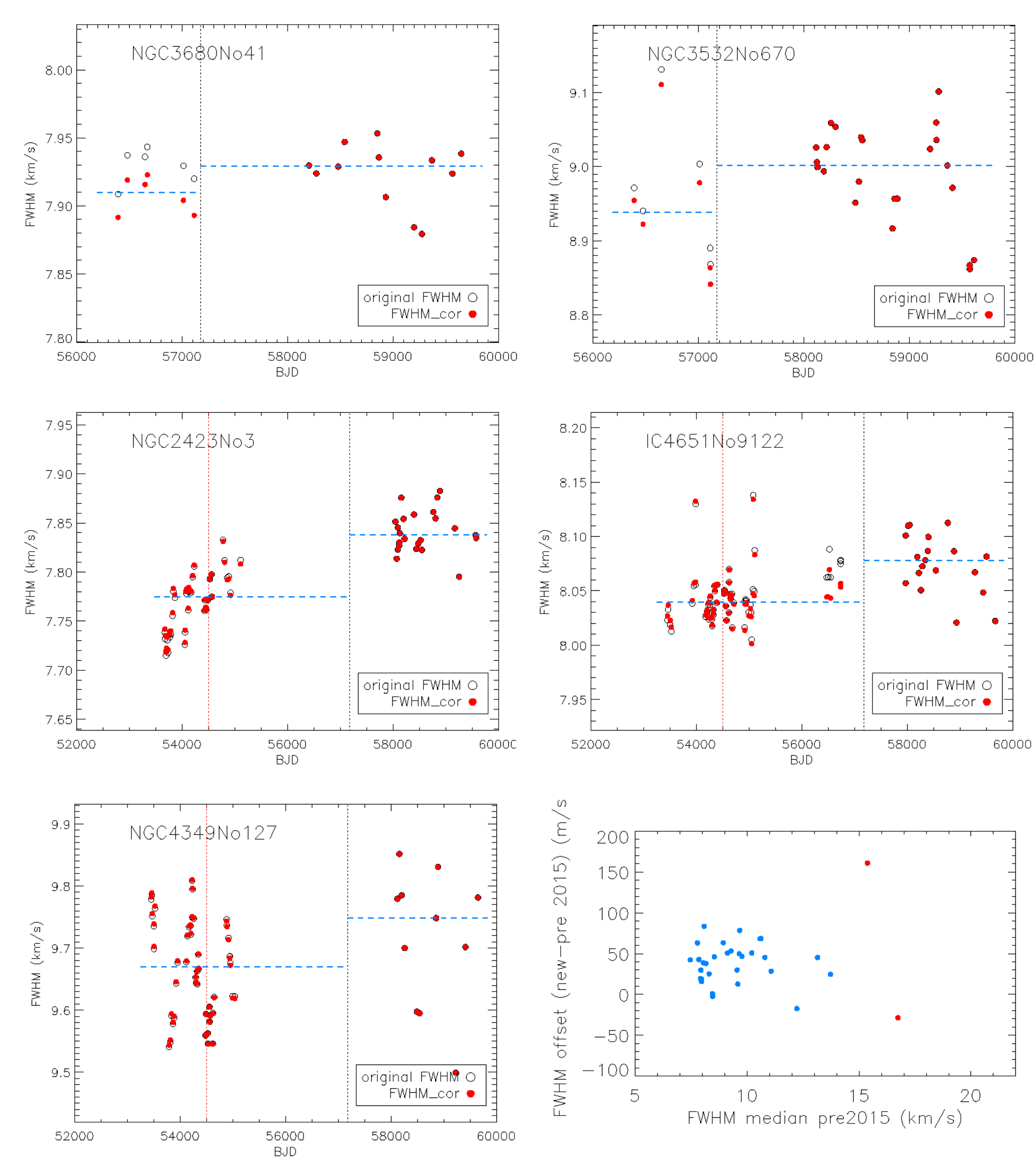}
\caption{FWHM values before and after the 2015 fiber change (marked with a black dashed line) for the 5 stars in this paper. The red points are the final FWHM values after considering the focus drift correction which only affect the data pre-2015. The horizontal blue dashed lines mark the median values before and after fiber change to calculate the observed FWHM offset. The bottom-right plot shows the FWHM offset (which is positive for most of the stars) for all the stars in our sample with more than 5 observations before and after the fiber change.}
\label{fwhm_offset}
\end{figure}

We also attempted to make a correction of the FWHM offset observed after the 2015 fibre change. In Fig. \ref{fwhm_offset}, we show the time series of FWHM values for the five stars in this paper with data taken before and after the fibre change. The blue dashed line shows the average FWHM in each period. The last panel shows the difference in average FWHM values (newer data values minus older data values) as a function of the median FWHM (calculated with the pre-2015 values) of all the stars in the sample. Only the stars with at least five points before and after the fibre change are plotted here. A correction could be applied if we were to find a correlation between the offset and the FWHM (in the same way as for the RV offset). However, neither the Pearson nor Spearman coefficients provide a strong correlation (regardless of whether or not we remove the two stars at very large FWHM depicted with red points) between the offset and the FWHM. Therefore, we decided to not correct the FWHM values of spectra taken after 2015. In a similar way, we also found that in general the BIS values are also slightly higher for the newer observations (thus a positive offset) but we were not able to find any correlation of such offsets with the FWHM either. We note that in the case of the BIS, the offsets are smaller, with a median value of $\sim$\,8\,\ms\ and we have not applied any correction.

To assess whether the non-correction of the FWHM affects the results presented in the paper, we applied a negative offset to the more recent values. This offset is just the difference between the median FWHM values pre- and post-fibre change (i.e. the blue dashed lines in Fig. \ref{fwhm_offset}). We repeated the GLS periodograms with the new FWHM time series and the periods of the most significant peaks remain mostly the same for NGC4349 No. 127 (673 days instead of 675) and for NGC3680 no. 41 (no peaks above the 1\% FAP level). For NGC3532 No. 670, the main peak lies at 828 days instead of 811 (but below the 1\% FAP level) and now there is no significant peak at 619 days. For NGC2423 No. 3, the secondary peak at 380 days is now at 310 days, decreasing the similarity to the \ha16 peak at 393 days. It is worth noting that in the case of IC4651 No. 9122, when correcting the FWHM values, there appears to be a significant peak (above the 0.1\% FAP level) at 687 days, but the peak at 363 days that was previously above the 0.1\% FAP level is now below the 1\% FAP level. 

\clearpage

\section{Figures from \textit{yorbit}: Single-keplerian fits}

In this section, we show the single-Keplerian fits done with \textit{yorbit} for the three new stars presented in this work  for comparison.

\begin{figure}[!h]
\centering
\includegraphics[width=1.0\linewidth]{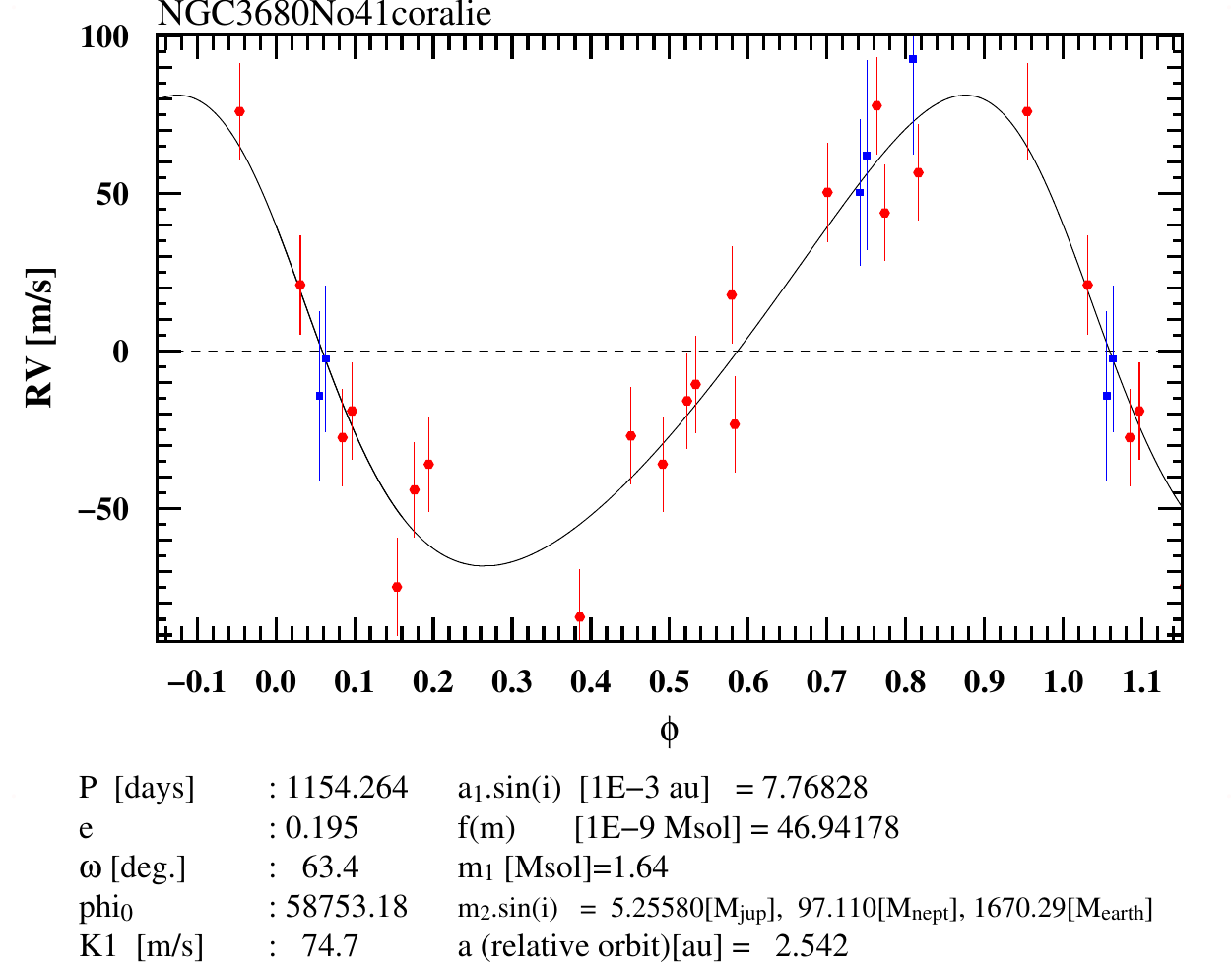}
\includegraphics[width=1.0\linewidth]{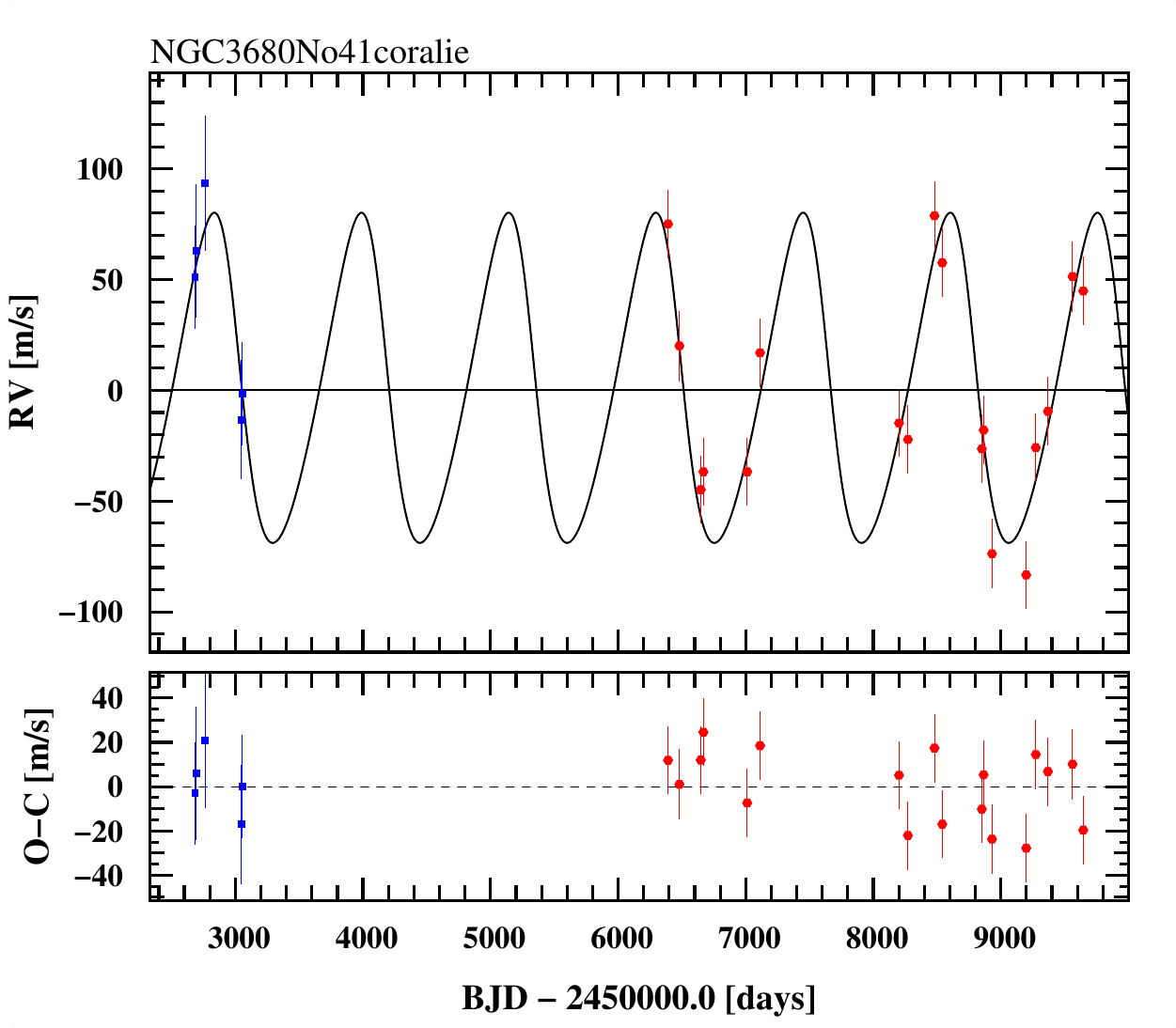}
\caption{Orbital phase curve and RV curve as a function of time for NGC3680 No.41. The data from CORALIE and HARPS are depicted with blue and red symbols, respectively.} 
\label{NGC3680No41_fit}
\end{figure}

\begin{figure}
\centering
\includegraphics[width=1.0\linewidth]{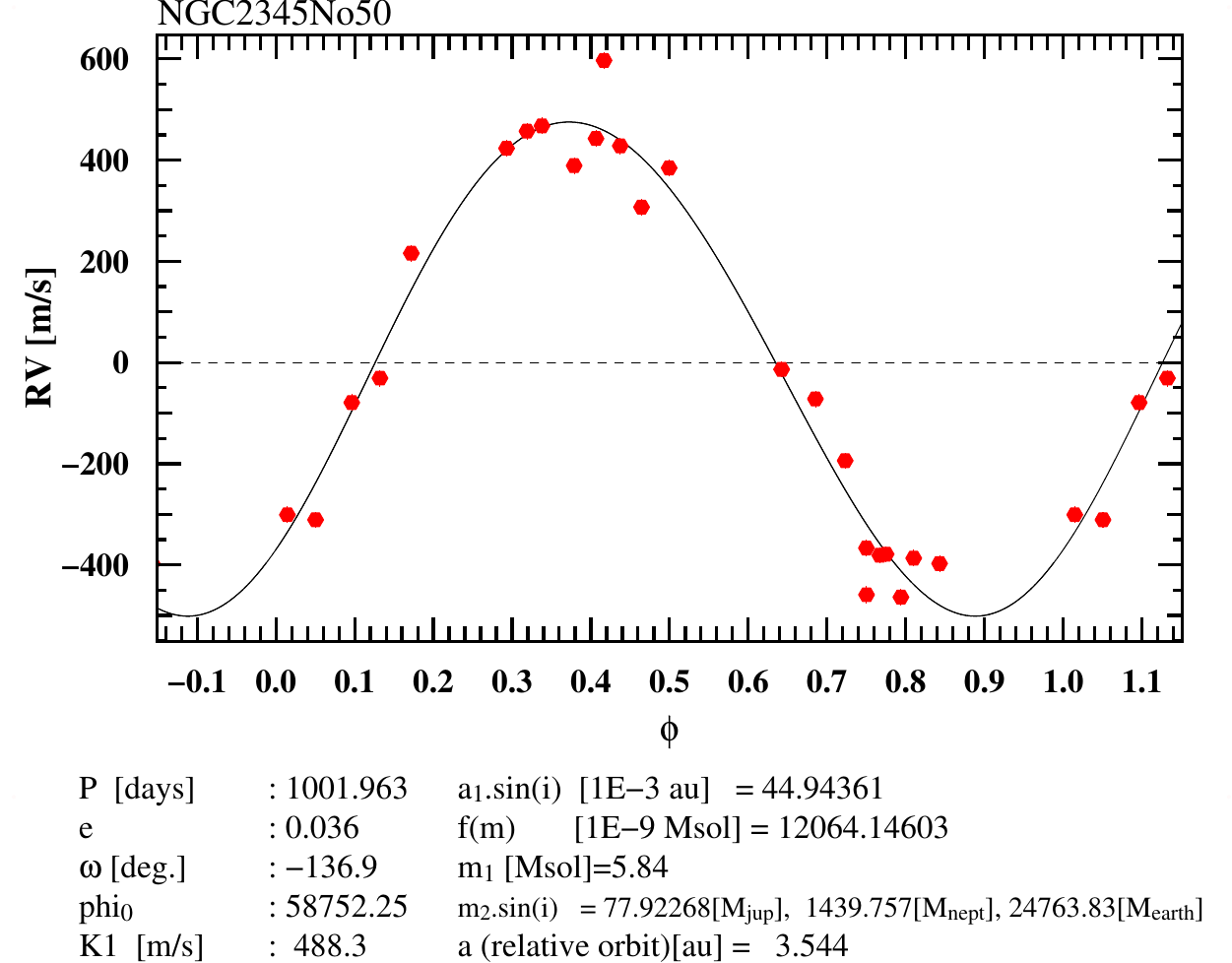}
\includegraphics[width=1.0\linewidth]{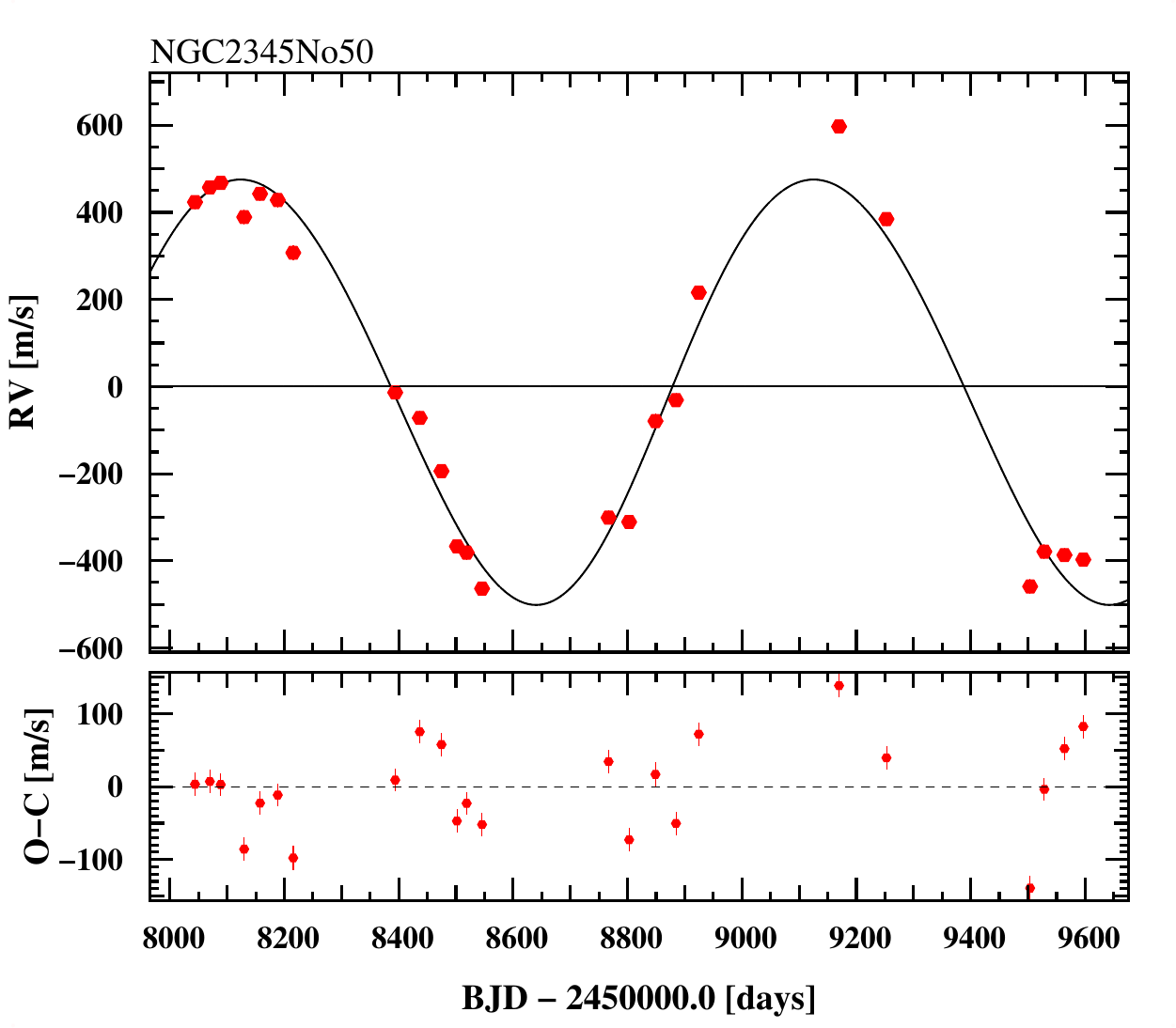}
\caption{Radial velocity curve as a function of time for a brown dwarf/binary candidate around NGC2345 No.50 using a single Keplerian fit.} 
\label{NGC2345No50_k1}
\end{figure}

\begin{figure}
\centering
\includegraphics[width=1.0\linewidth]{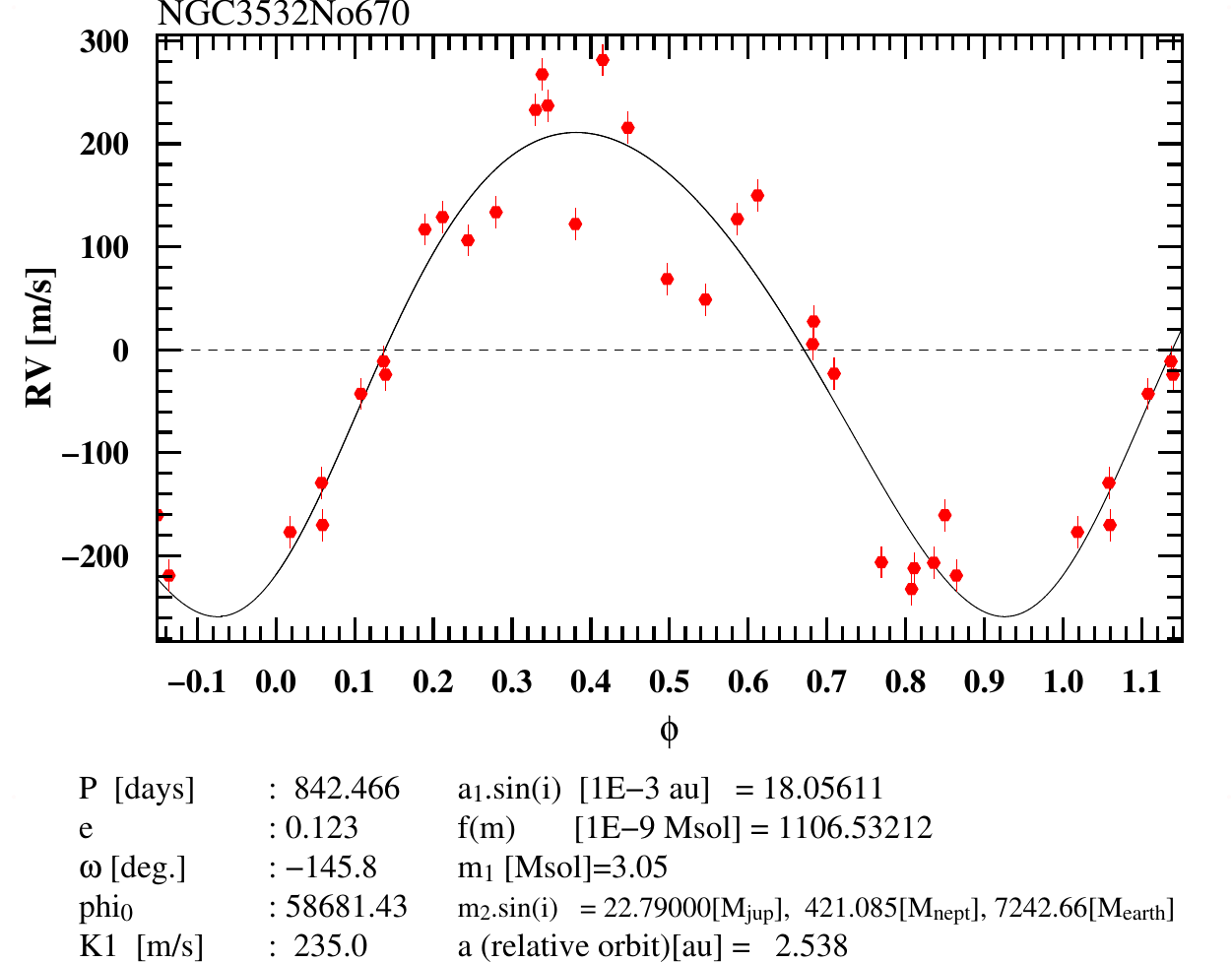}
\includegraphics[width=1.0\linewidth]{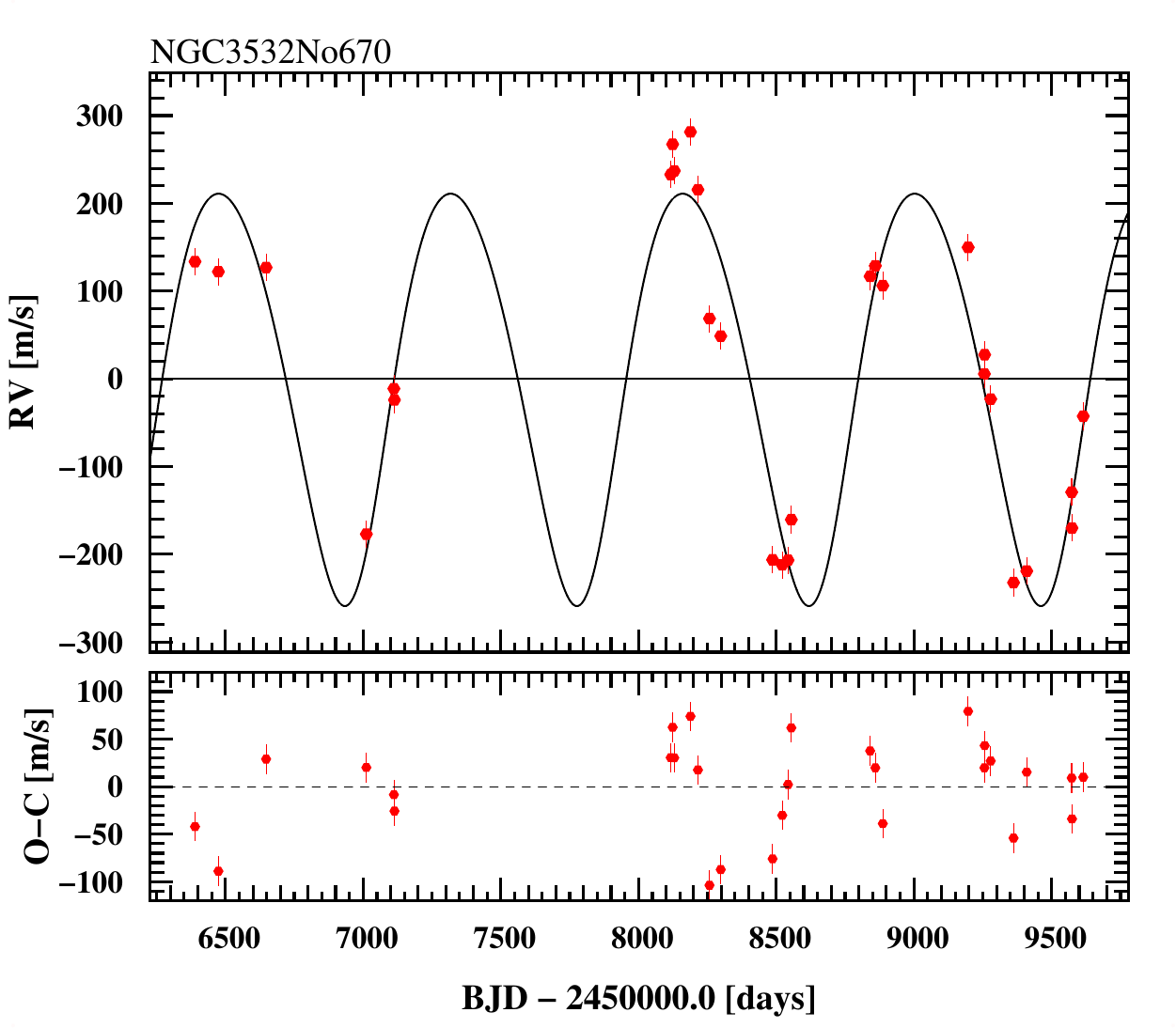}
\caption{Radial velocity curve as a function of time for NGC3532 No.670. The data have been fitted with a single-Keplerian orbit model.} 
\label{NGC3532No670_k1}
\end{figure}

\clearpage

\section{Figures from \textit{yorbit}: Two-Keplerian fit for NGC2345 No.50 and NGC3532 No.670}

In this section, we show the two-Keplerian fits done with \textit{yorbit} for NGC2345 No.50 and NGC3532 No.670  for comparison.

\begin{figure}[h!]
\centering
\includegraphics[width=0.9\linewidth]{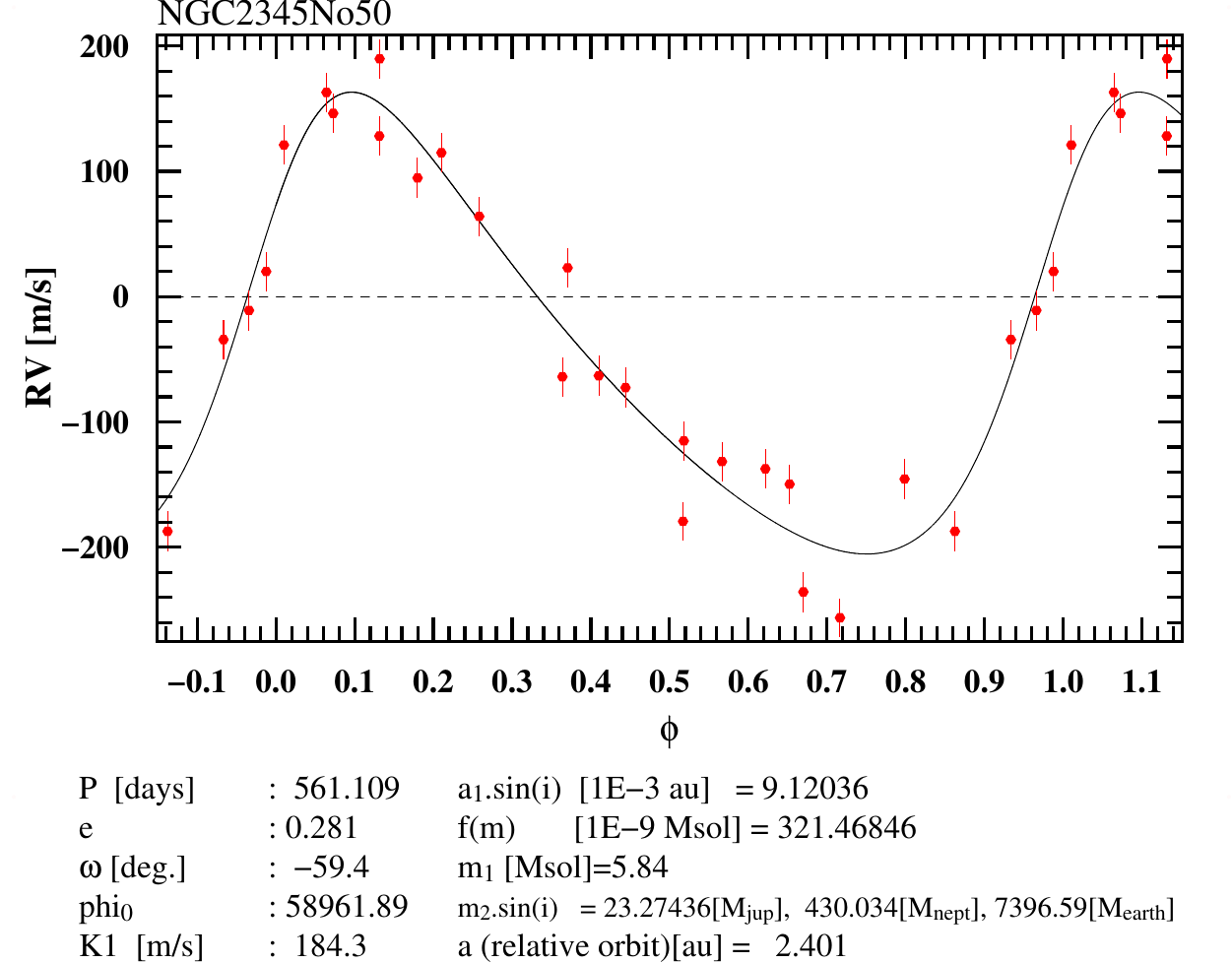}
\includegraphics[width=0.9\linewidth]{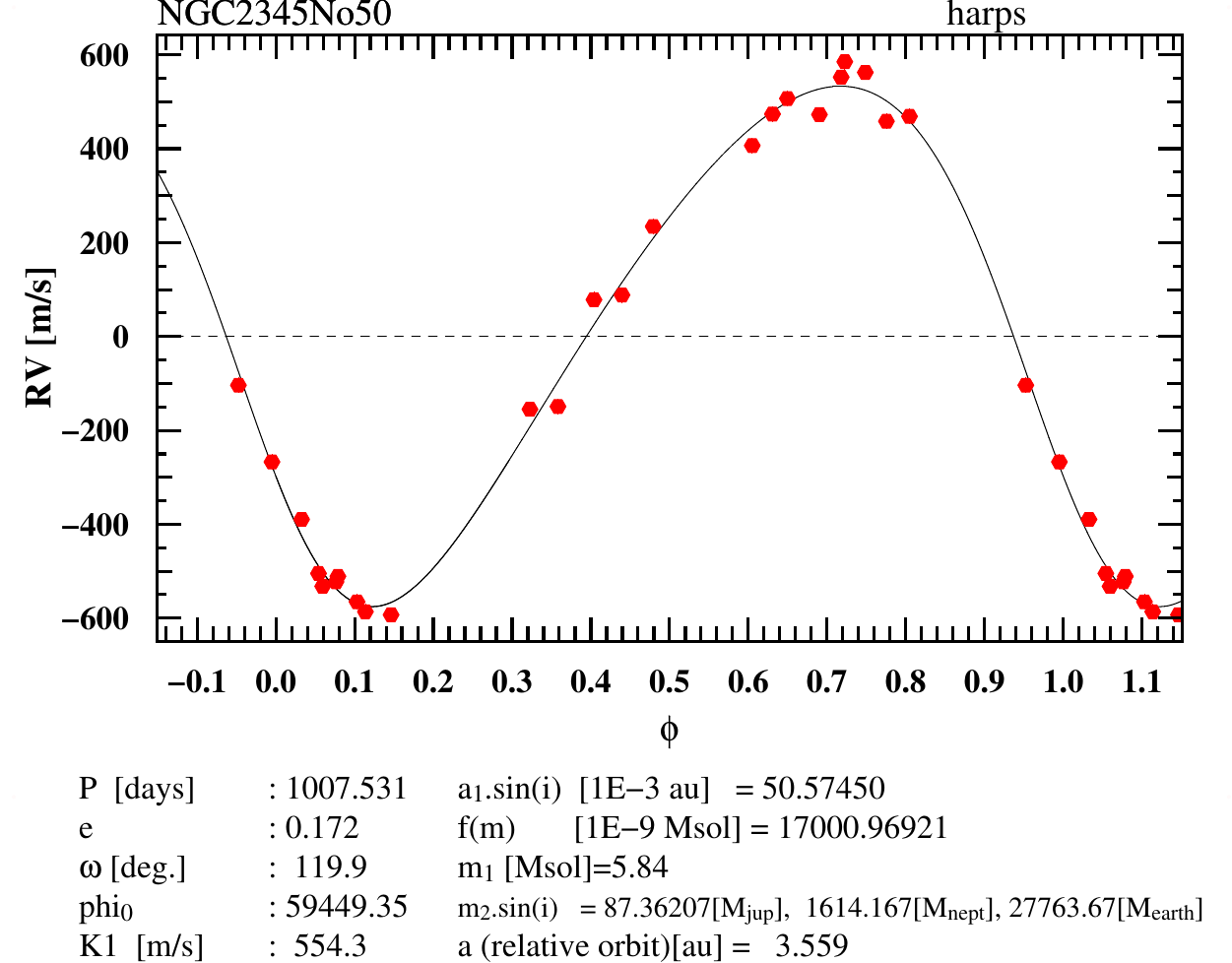}
\includegraphics[width=0.9\linewidth]{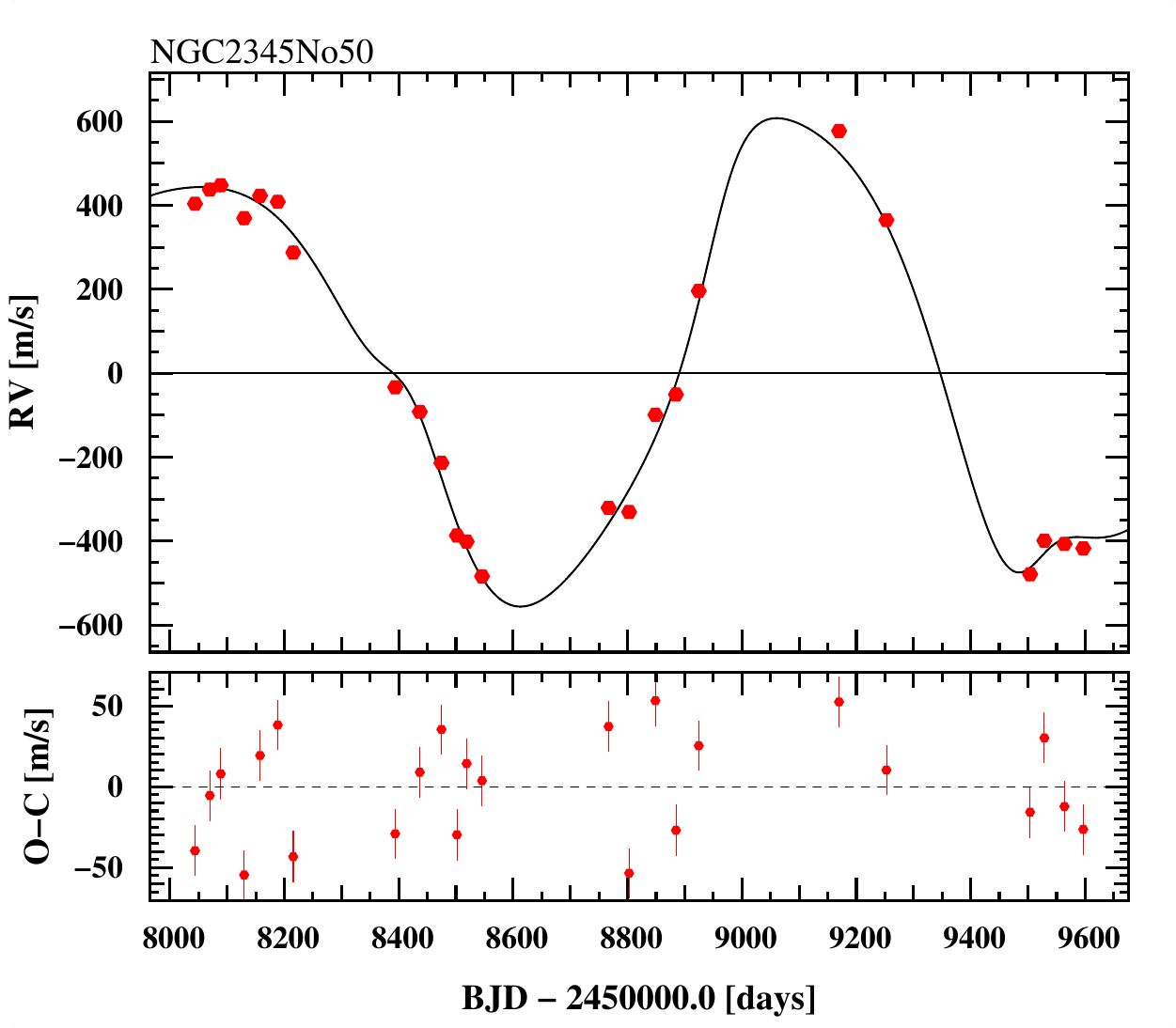}
\caption{Orbital phase curves and RV curve as a function of time for NGC2345 No.50. The data have been fitted with a two-Keplerian orbit model.} 
\label{NGC2345No50_k2}
\end{figure}

\begin{figure}[h!]
\centering
\includegraphics[width=0.9\linewidth]{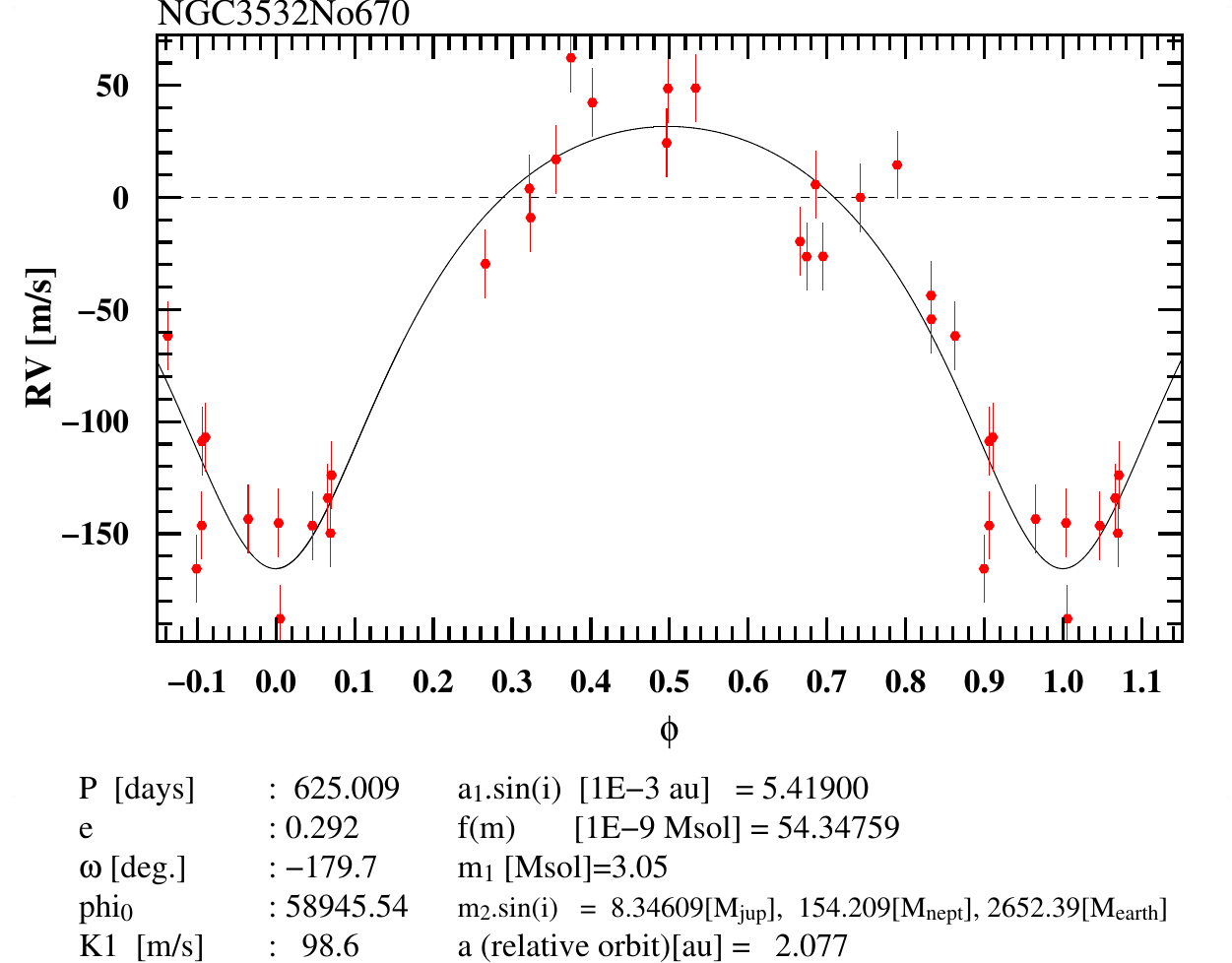}
\includegraphics[width=0.9\linewidth]{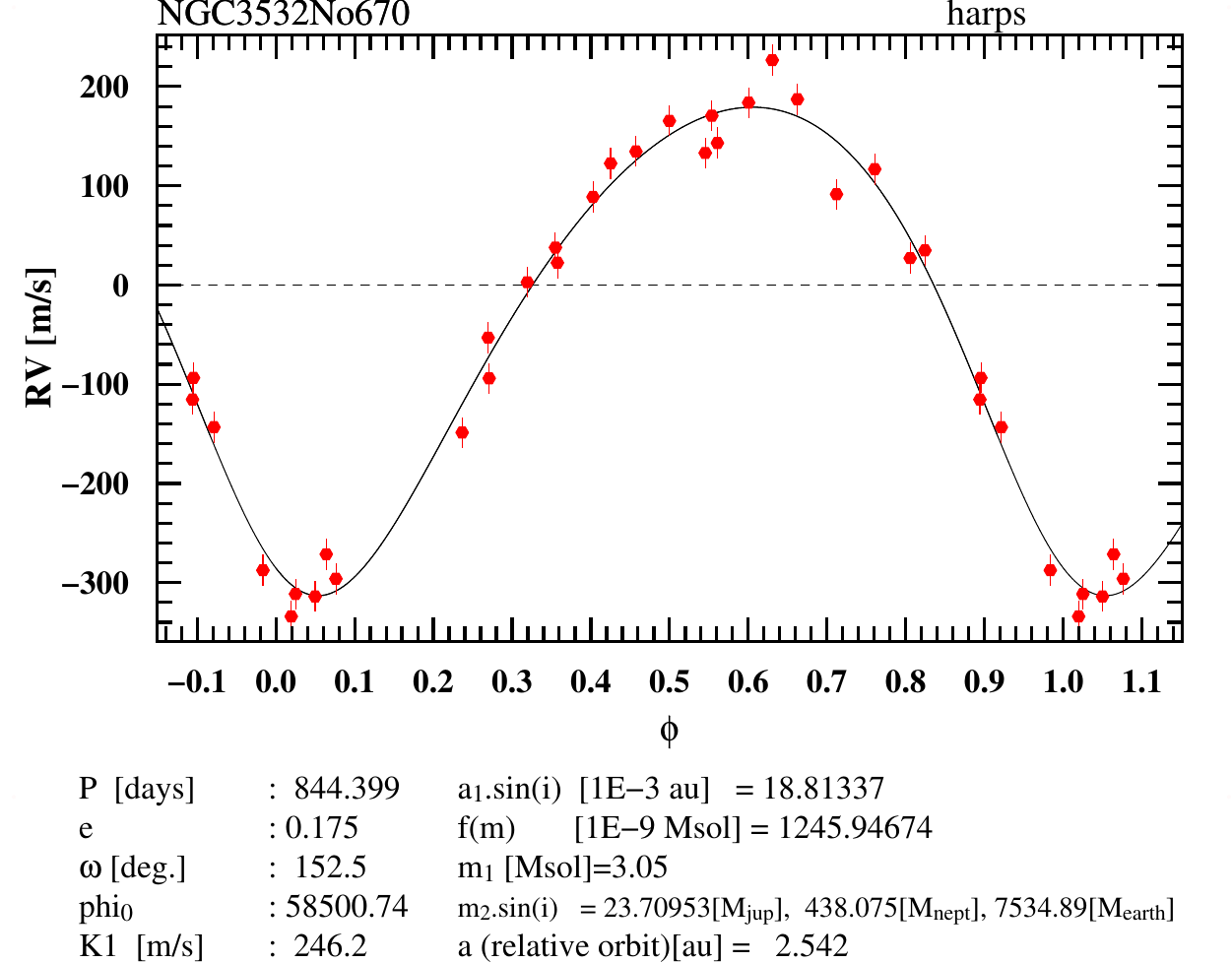}
\includegraphics[width=0.9\linewidth]{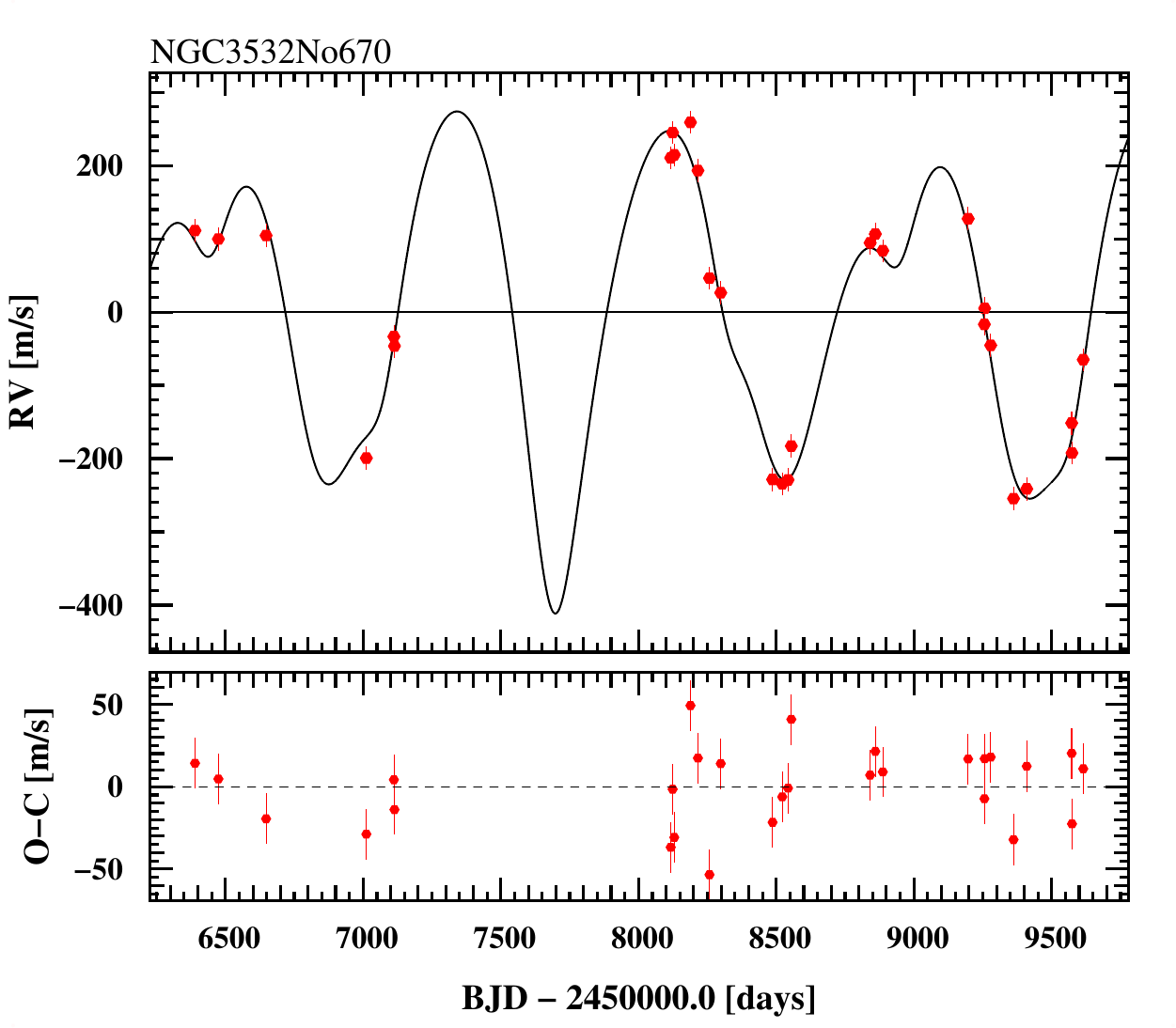}
\caption{Orbital phase curves and RV curve as a function of time for NGC3532 No.670. The data have been fitted with a two-Keplerian orbit model.} 
\label{NGC3532No670_k2}
\end{figure}

\clearpage

\section{Previous candidates}

In this section, we show the best Keplerian fits found by \texttt{kima} for the three stars discussed in the previous paper with new data presented here (see Sect. \ref{sec:others}).

\begin{figure}[h!]
\centering
\includegraphics[width=1.0\linewidth]{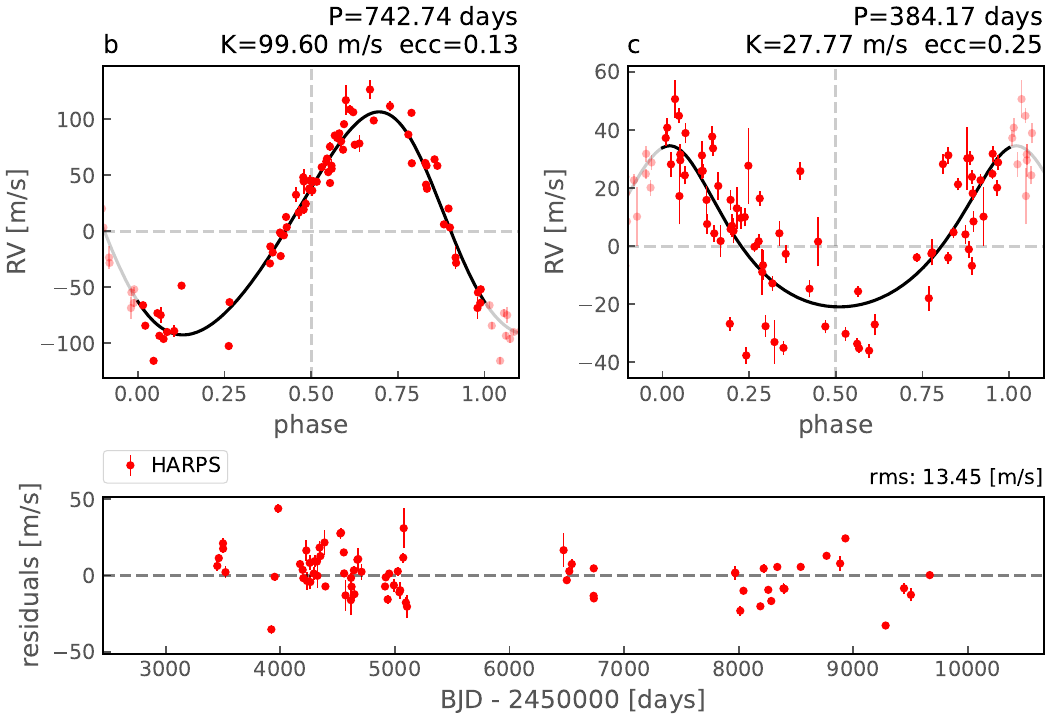}
\includegraphics[width=1.0\linewidth]{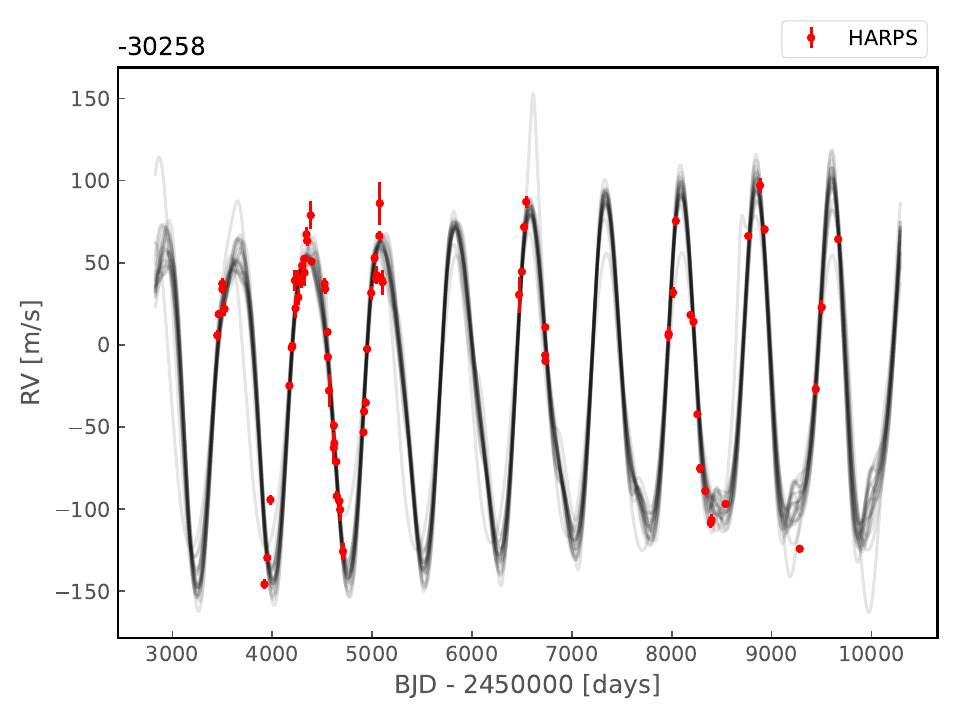}
\caption{Two-Keplerian fit for IC4651 No. 9122 using \texttt{kima}. The top panels show the phase curve for the two signals, with the residuals shown below, highlighting the rms of the residual RVs. The bottom panel shows the RV data from HARPS and representative samples from the posterior distribution. The systemic RV has been subtracted and is shown at the top.}
\label{IC4651No9122_kima2p}
\end{figure}

\begin{figure}
\centering
\includegraphics[width=1.0\linewidth]{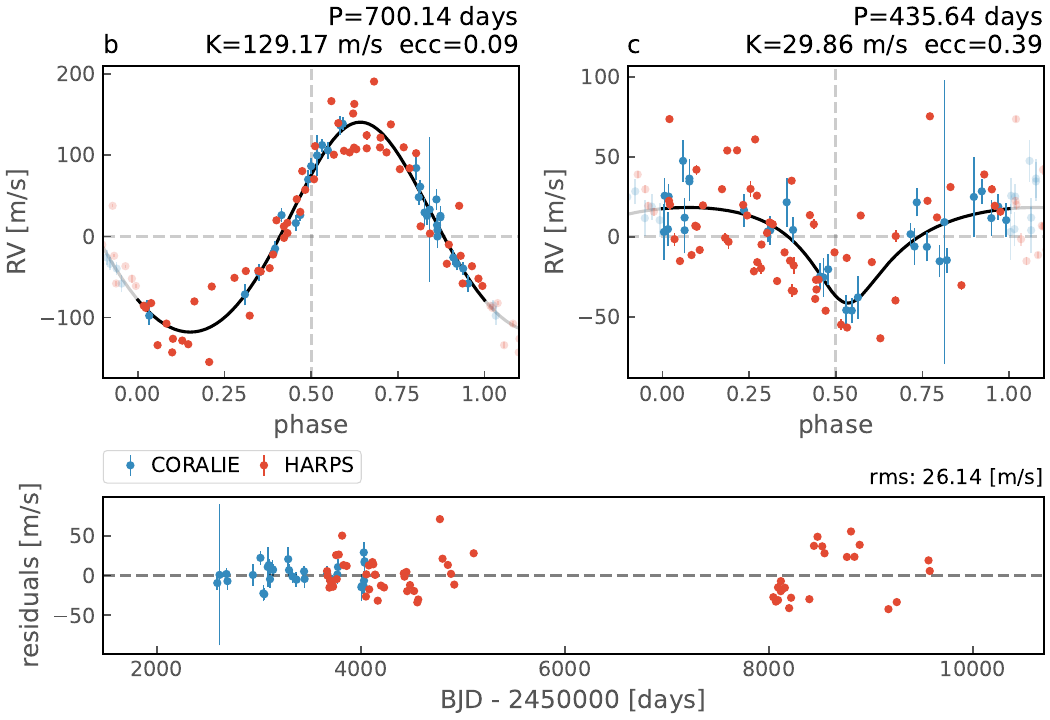}
\includegraphics[width=1.0\linewidth]{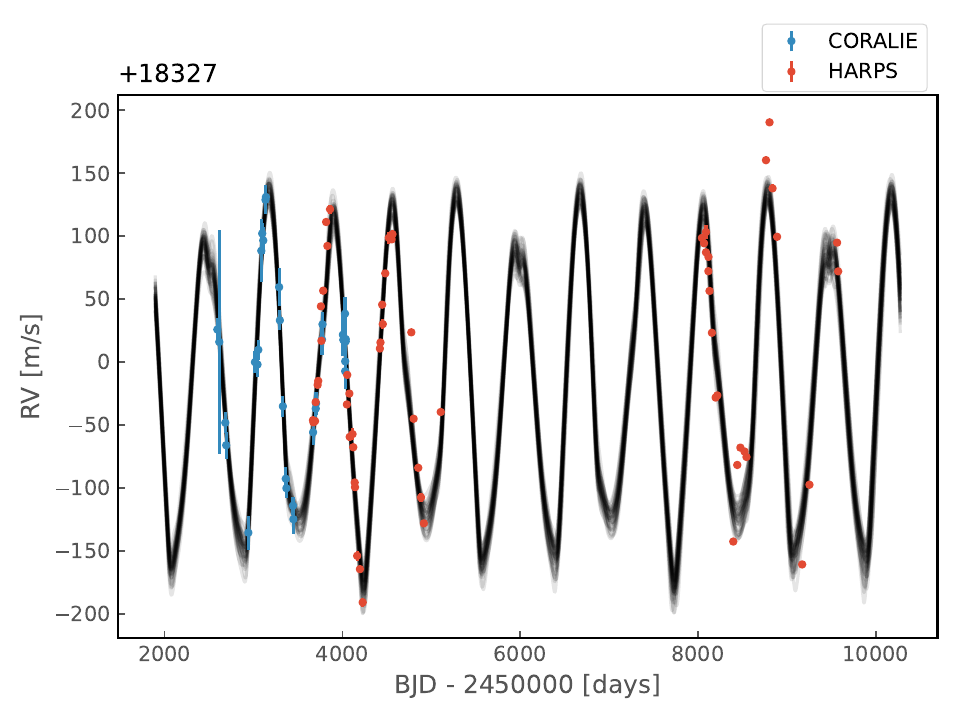}
\caption{Two-Keplerian fit for NGC2423 No. 3 using \texttt{kima}. The top panels show the phase curve for the two signals, with the residuals shown below, highlighting the rms of the residual RVs. The bottom panel shows the RV data from HARPS (red points) and CORALIE (blue points) and representative samples from the posterior distribution. The systemic RV has been subtracted and is shown at the top.}.
\label{NGC2423No3_kima}
\end{figure}

\begin{figure}
\centering
\includegraphics[width=1.0\linewidth]{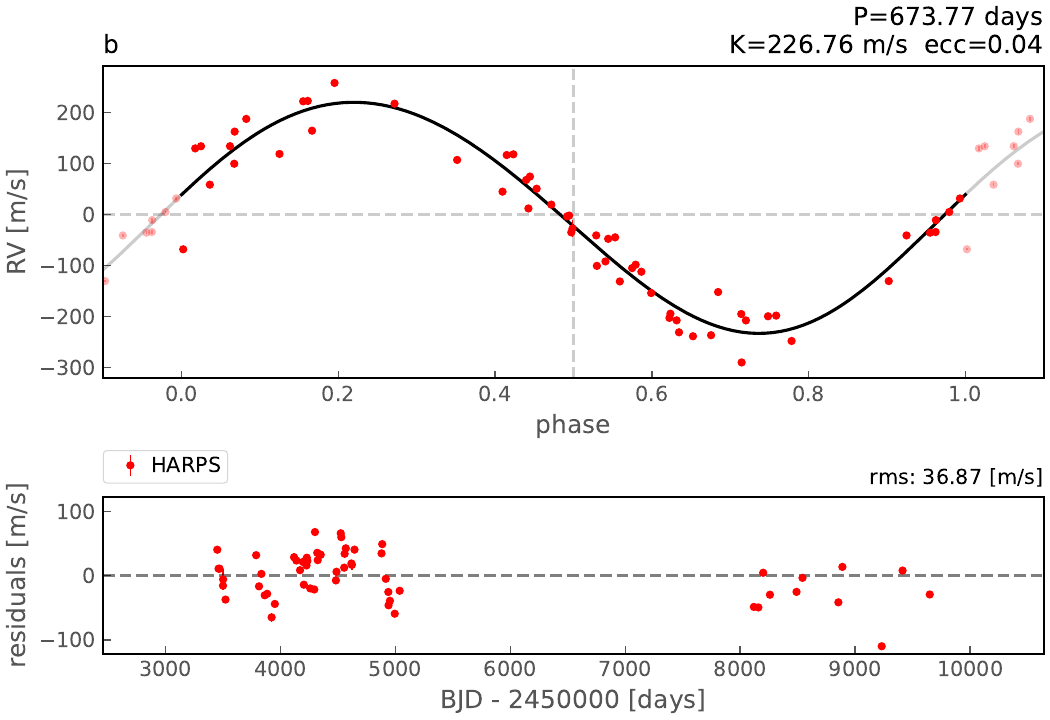}
\includegraphics[width=1.0\linewidth]{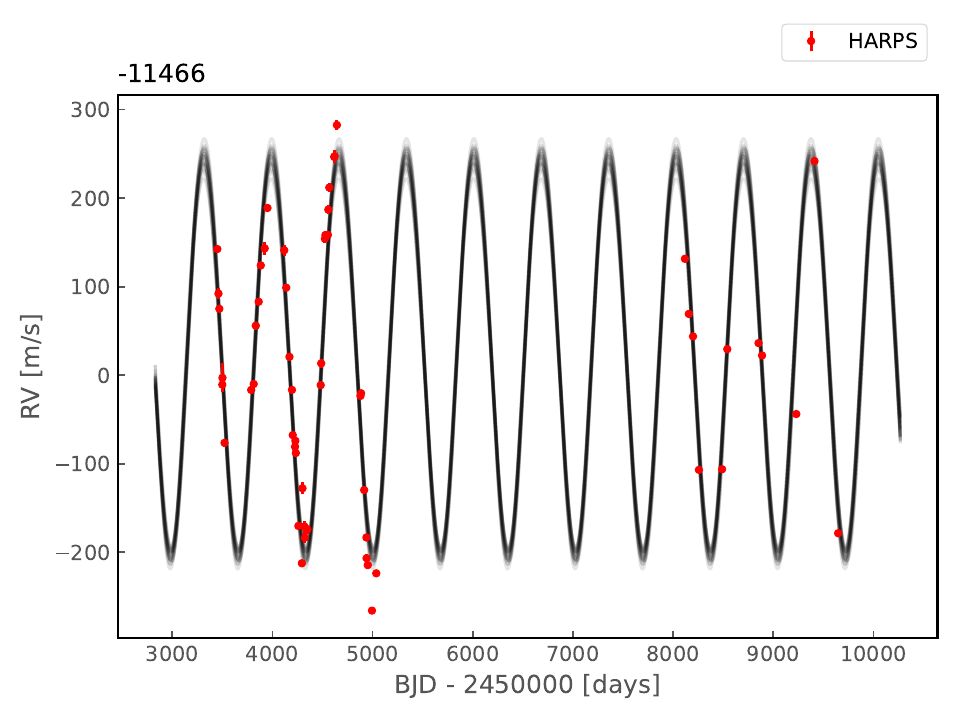}
\caption{One-Keplerian fit for NGC4349 No. 127 using \texttt{kima}. The top panel shows the phase curve for the signal, with the residuals shown below, highlighting the rms of the residual RVs. The bottom panel shows the RV data from HARPS and representative samples from the posterior distribution. The systemic RV has been subtracted and is shown at the top.}
\label{NGC4349No127_kima}
\end{figure}

\begin{figure}
\centering
\includegraphics[width=1.0\linewidth]{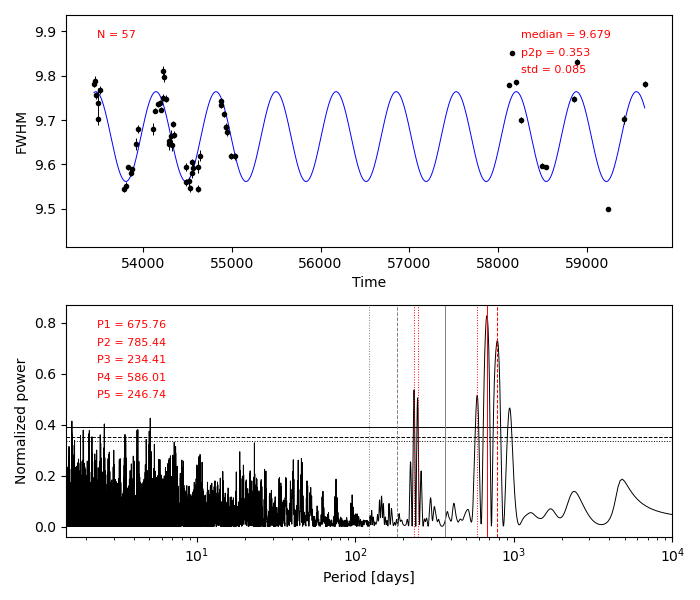}
\caption{Time series for the FWHM of NGC4349 No. 127 with a sinusoidal function fit with period 675 days.}
\label{NGC4349No127_FWHM}
\end{figure}

\clearpage

\section{\ha\ index measurements}
Examples of \ha\ measurements with the two bandpasses implemented in ACTIN.
\FloatBarrier


\begin{figure}[h!]
\centering
\includegraphics[width=0.7\linewidth]{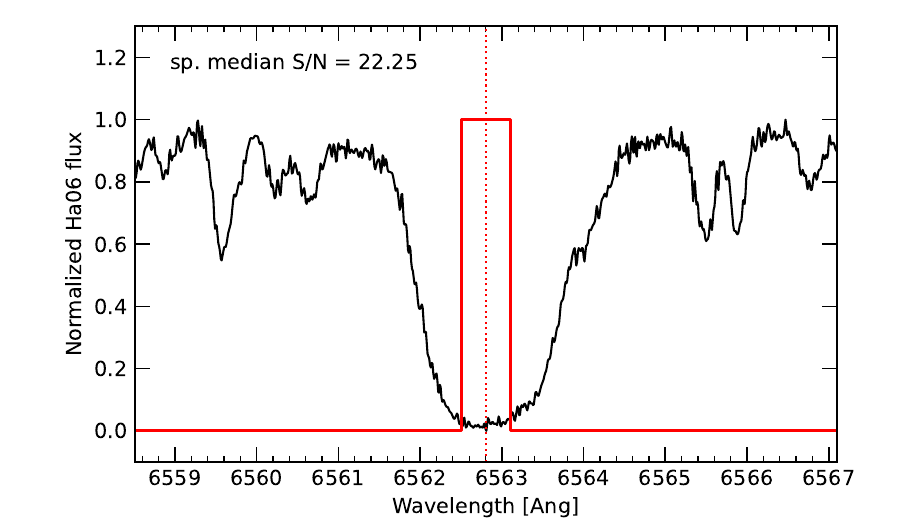}
\includegraphics[width=0.7\linewidth]{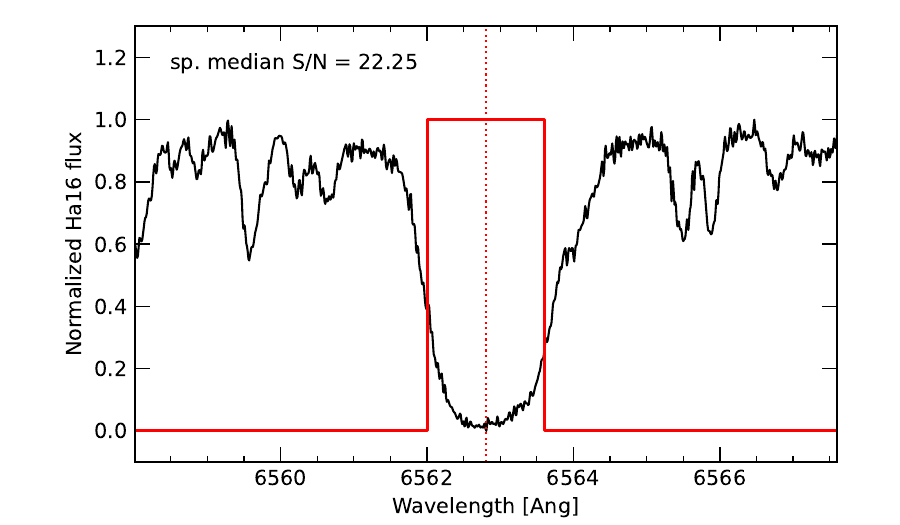}
\includegraphics[width=0.7\linewidth]{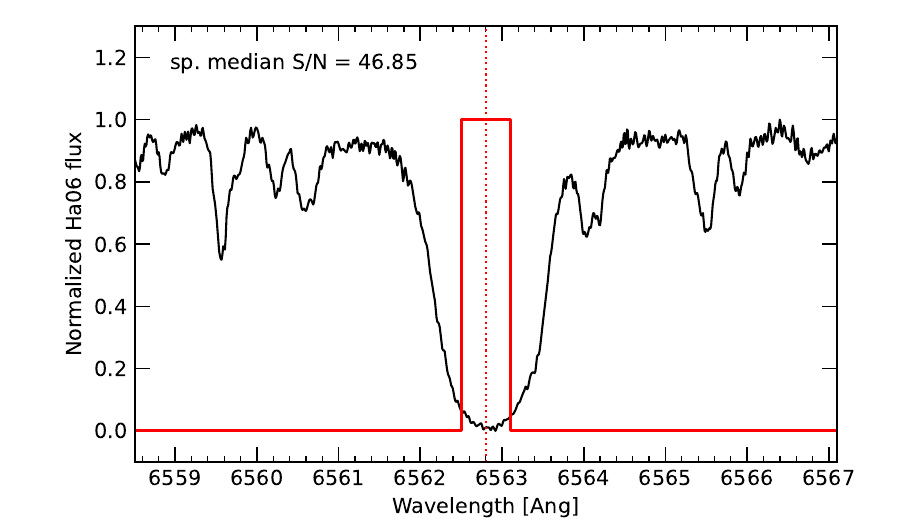}
\includegraphics[width=0.7\linewidth]{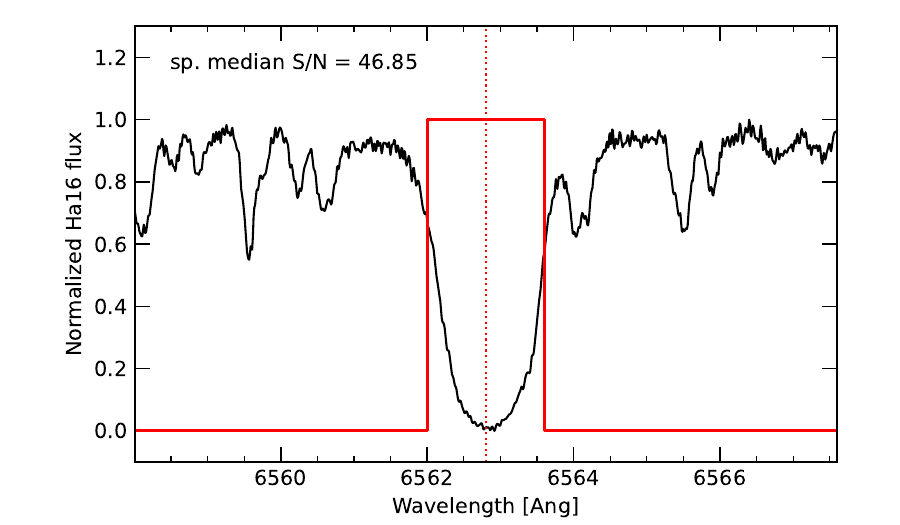}
\caption{\ha\ index measured with the 0.7 passband ($\pm$\,0.3\AA\, around the center line) or with the 1.6 passband ($\pm$\,0.8\AA\, around the centre line). Upper figures (NGC2345 No. 50), lower figures (NGC3532 No. 670).}
\label{halpha1}
\end{figure}

\begin{figure}
\centering
\includegraphics[width=0.7\linewidth]{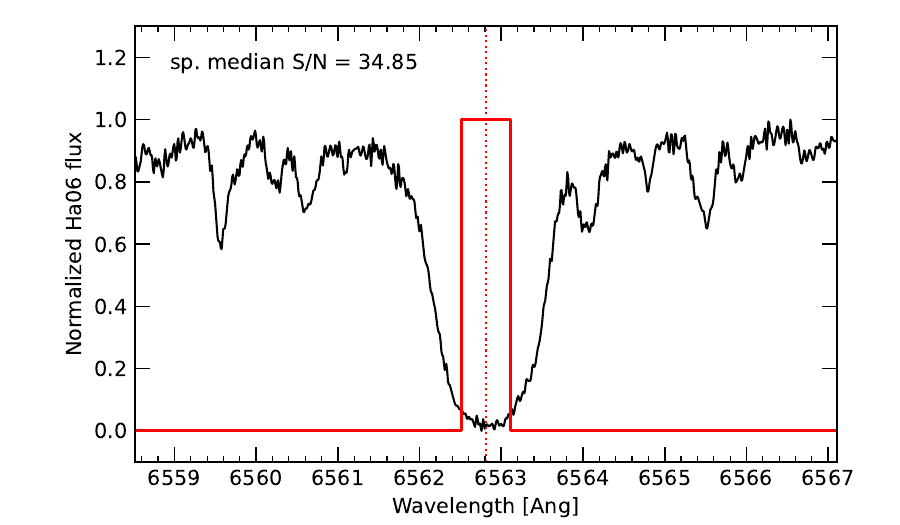}
\includegraphics[width=0.7\linewidth]{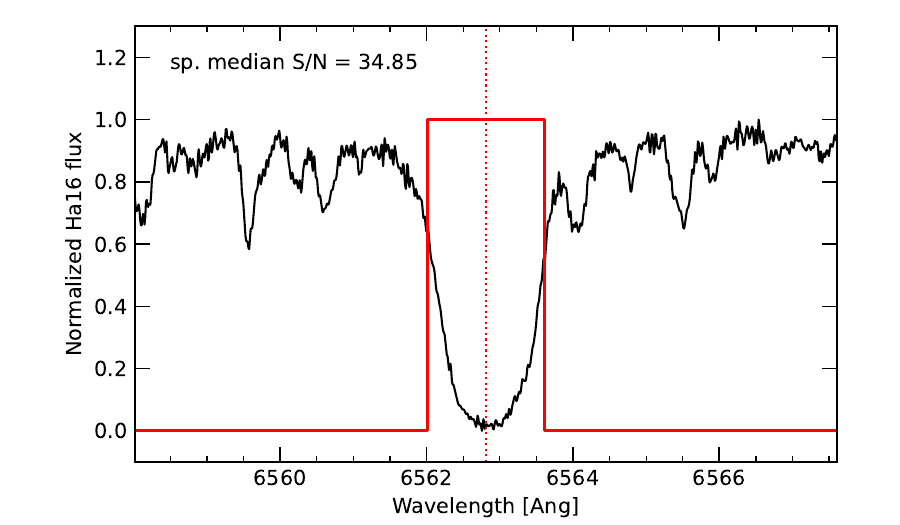}
\includegraphics[width=0.7\linewidth]{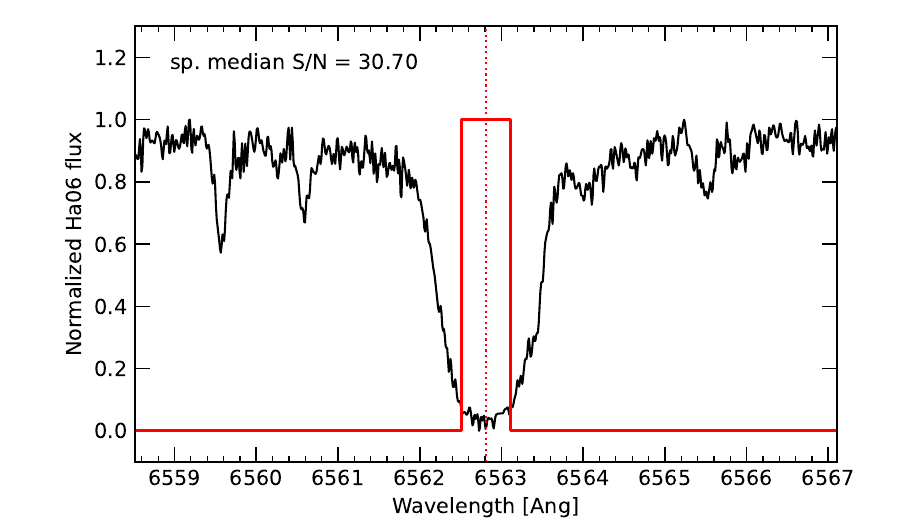}    
\includegraphics[width=0.7\linewidth]{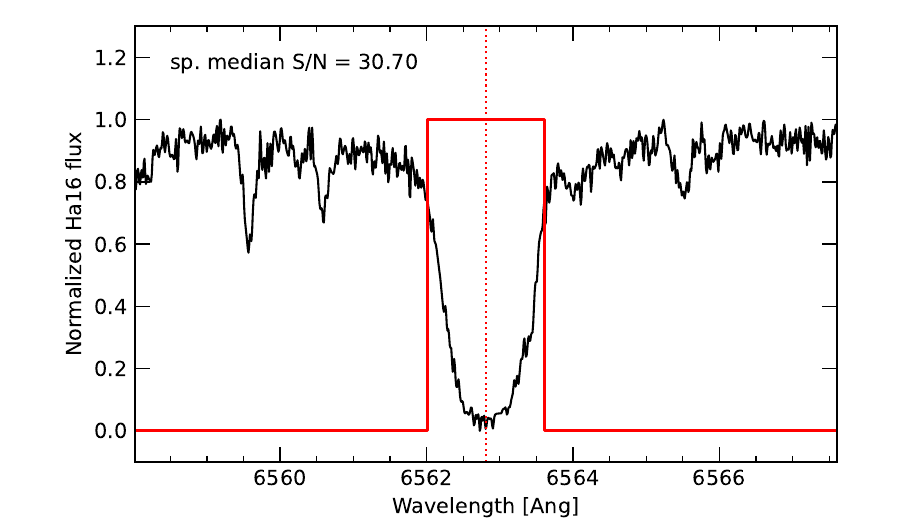}
\caption{\ha\ index measured with the 0.7 passband or with the 1.6 passband. Upper figures (NGC4349 No. 127), lower figures (NGC3680 No. 41).}
\label{halpha2}
\end{figure}

\end{appendix}

\end{document}